\documentclass[aps,pre,groupedaddress]{revtex4-2}

\usepackage{xcolor}
\usepackage{hyperref}
\usepackage{graphicx}
\usepackage{amsmath}
\usepackage{amssymb}
\usepackage{bm}
\usepackage{comment}
\usepackage{float}
\usepackage{makecell}

\def\N{{N}}

\def\ffunc{{\sf f}}
\def\gfunc{{\sf g}}
\def\hfunc{{\sf h}}
\def\vc{{v_{\rm c}}}
\def\v1{{v_1}}
\def\C{{\sf C}}
\def\x{{\bf x}}
\def\w{{\bf w}}
\def\taum{{\tau_{\rm m}}}
\def\tauc{{\tau_{\rm c}}}
\def\sigmac{{\sigma_{\rm c}}}
\def\sigmai{{\sigma_{\rm I}}}
\def\mui{{\mu_{\rm I}}}
\def\alphaV{{\alpha_{\rm V}}}
\def\nsrV{{\nu_{\rm V}}}
\def\sigmaV{{\sigma_{\rm V}}}
\def\vV{{v_{\rm V}}}
\def\xiC{{\xi_{\rm C}}}
\def\fc{{v}}
\def\rro{{r_{\rm R}}}
\def\rq{{r_{\rm q}}}
\def\gq{{g_{\rm q}}}
\def\vq{{v_{\rm q}}}
\def\uq{{u_{\rm q}}}
\def\rw{{r_{\rm w}}}
\def\vw{{v_{\rm w}}}
\def\tw{{t_{\rm w}}}
\def\vro{{v_{\rm R}}}
\def\rhoro{{\varrho}}
\def\nro{{n_{\rm R}}}
\def\sigmacref{{\sigma_{\rm c}^{({\rm ref})}}}
\def\taucref{{\tau_{\rm c}^{({\rm ref})}}}
\def\Tc{{T_{\rm c}}}
\def\nis{{n_{\rm is}}}
\def\ntrials{{n_{\rm tr}}}

\def\muip{{\mu_{\rm I}^{(+)}}}
\def\muim{{\mu_{\rm I}^{(-)}}}
\def\rrop{{r_{\rm R}^{(+)}}}
\def\rrom{{r_{\rm R}^{(-)}}}

\def\<{\left<}
\def\>{\right>}

\def\d{{\rm d}}
\begin{document}


\title{Theory of  enhanced-by-coincidence neural information transmission}


\author{Miguel Ib\'a\~nez-Berganza}
\email[]{miguel.ibanezberganza@imtlucca.it}
\affiliation{IMT School for Advanced Studies Lucca, Piazza San Francesco 19, 50100, Lucca, Italy}
\affiliation{INdAM-GNAMPA Istituto Nazionale di Alta Matematica ‘Francesco Severi’, P.le Aldo Moro 5, 00185 Rome, Italy}
\author{Giulio Bondanelli}
\affiliation{Neural Computation Laboratory, Istituto Italiano di Tecnologia, Genoa, Italy}
\affiliation{Institute for Neural Information Processing, Center for Molecular Neurobiology, University Medical Center Hamburg-Eppendorf, 20251, Hamburg, Germany}
\author{Stefano Panzeri}
\affiliation{Institute for Neural Information Processing, Center for Molecular Neurobiology, University Medical Center Hamburg-Eppendorf, 20251, Hamburg, Germany}


\date{\today}

\begin{abstract}
The activity of neurons within brain circuits has been ubiquitously reported to be correlated. The impact of these correlations on brain function has been  extensively investigated. Correlations can in principle increase or decrease the information that neural populations carry about sensory stimuli, but experiments in cortical areas have mostly reported information-limiting correlations, which decrease the information encoded in the population. However, a second stream of evidence suggests that temporal correlations between the spiking activity of different neurons may increase the impact of neural activity downstream, implying that  temporal correlations affect both the encoding of information and its downstream readout.  The principle of how encoding and readout combine are still unclear. Here, we consider a model of  transmission of stimulus information encoded in pre-synaptic input spike trains with information-limiting time-correlations to the output firing of a post-synaptic biologicaly-plausible leaky integrate and fire (LIF) readout neuron. We derive an analytical solution of the model in the diffusion approximation, in which the encoding spiking activity is treated as a continuous-time stochastic variable. An ansatz based on a separation of timescales allows us compute the stimulus information transmitted to the readout over a broad range of  parameters. Our analytical results reveal that, for sufficiently low input firing rates, large enough difference in input stimulus-specific activity, and moderately large input temporal correlations, the  stimulus discriminability of the firing of the LIF readout neuron can be enhanced by the presence of input time  correlations, despite they decrease the stimulus information encoded in its inputs.
\end{abstract}


\maketitle

\section{Introduction \label{sec:intro}}

Since the first extensive simultaneous recordings of the activity of multiple neurons in the brain {\em in vivo} were performed decades ago \cite{Vaadia1995, Wilson1993, Gray1989, Mastronarde1983b, Urai2022, Stringer2024}, neuroscientists have consistently observed that the  times at which different neurons fire are correlated. That is, the timing at which a neuron fires does not depend only on externally-measurable correlates as the sensory stimulus or the animal's decision, but also on the timing of firing of other neurons. When considering the encoding of sensory stimuli, correlations are usually measured as so called
 {\it noise correlations}, defined as correlations between the timing of spikes of different neurons at a fixed value of the sensory stimulus \cite{averbeck2006,panzeri2022}. 

Previous work has shown that noise correlations between the spike timing of different neurons is related to their anatomical connectivity  \cite{Kuan2024}. However, despite being ubiquitously reported across brain activity recordings {\em in vivo}, their possible contribution to brain function remains intensively debated. Theoretical work has shown that  temporally correlated inputs can profoundly affect  the dynamics of spiking neuron models \cite{brunel1998,kempter1998,feng2000,brunel2001,salinas2000,salinas2002,fourcaud2002,moreno2002,middleton2003,morenobote2004,lindner2004,schwalger2008,lindner2009}. It has been proposed that one function of such correlations is to shape the amount of information that can be encoded  in a neural population \cite{averbeck2006,kohn2016,panzeri2022,lindner2009,salinas2001,abbott1999}. In particular, these studies have shown that noise correlations can have both an information-enhancing effect or an information-limiting effect on the information carried by a population of neurons \cite{zohary1994,abbott1999,Panzeri1999,panzerischultz2001,averbeck2006,moreno2014}. However, the majority of empirical studies analyzing neural recordings  have reported  
information-limiting correlations \cite{moreno2014}, defined here as correlations that decrease the amount of information in a neuronal population \cite{zohary1994, Rumyantsev2020,bartolo2020,valente2021}. 
This has led to the widespread view that noise correlations eventually limit the ability of an organism to discriminate sensory stimuli, thus impairing perceptual discrimination and decision-making \cite{zohary1994,moreno2014,bartolo2020,gold2001}. 

However, recent experimental findings, reviewed in the next section and sketched in Fig.\ref{fig:cartoon_valente},  have challenged this notion, reporting that  noise correlations are stronger during correct than during incorrect perceptual judgments \cite{runyan2017,Francis2022,valente2021, Safaai2023, BalaguerBallester2020, Shahidi2019}, suggesting that noise correlations benefit perceptual abilities. This has been reported even in experiments in which noise correlations are information-limiting, ruling out that this effect may be due to enhanced encoded information. While some of these results can be understood also neglecting neural dynamics and considering only time-averaged responses \cite{zylberberg2017}, they have a natural explanation considering the time-dependent nature of neural correlations and neural information encoding and transmission. Temporally-correlated presynaptic activity can elicit stronger responses in postsynaptic neurons with short integration time constants via the non-linear mechanism of coincidence detection \cite{salinas2001,koch1996brief}. Under such conditions,   noise correlations can amplify responses in downstream neural populations and thus  increase the  amount of sensory information that is transmitted downstream, despite reducing the \textit{encoded} input information. In particular, stronger information-limiting noise correlations may make different neurons more redundant, which enhances the consistency of stimulus-encoding neural responses across time and neurons. A downstream neuron with short integration time constants may exploit the enhanced consistency in its input to outweigh the information-limiting effects of noise correlations, ultimately increasing the stimulus information that is read out downstream \cite{valente2021, Koren2023}.

Understanding the feasibility of such an enhanced-by-coincidence information transmission between pre- and post-synaptic activity requires models of neurons as physical systems that integrate both stimulus encoding and readout mechanisms. To understand how correlated input at the pre-synaptic stage enhance information transmission to the post-synaptic stage, we propose a biologically plausible  Neural Encoding and Readout Model (NERM)\cite{valente2021}. The model consists of a population of pre-synaptic neurons encoding a binary stimulus, which subsequently drives a single post-synaptic readout neuron. The encoding neurons generate spike trains with a certain strength and time scales of correlation across neurons. These correlations are designed to be information-limiting, that is to decrease the information encoded in the input when the strength of correlation increases. 
The readout neuron is modeled as a leaky integrate-and-fire (LIF) neuron and integrates afferent currents as a  weighted sum of spike trains with a given membrane integration time constant.
Numerical solutions of this model reveal that the stimulus discriminability  of the readout neuron (as quantified by the signal-to-noise ratio between the difference in stimulus-specific mean firing rates and its variance, SNR) can be, seemingly paradoxically, enhanced by stronger information-limiting correlations for sufficiently short readout  time constants and sufficiently low input firing rate. In these conditions correlations amplify the stimulus information transmitted to the readout even when decreasing input information.

Our previous study left open important questions about the conditions needed for noise correlations to overcome their information-limiting impact at the readout stage. The study of the model was  limited to only $N=2$ input encoding neurons, while information transmission in the brain involves larger neural populations. Moreover, we did not provide an analytical understanding of the interplay between encoding and readout of information in neural populations.  To overcome these issues and allow an understanding of these neural operations as a statistical physical system, here we develop an analytical framework to characterize the SNR enhancement effect beyond numerical simulations. This framework allows to study encoding in a population with an arbitrary number $N$ of pre-synaptic input neurons  with time-dependent correlations, and study how this information is transmitted to a readout LIF neuron. The ability to obtain solutions for arbitrary input population sizes allows us to study the interplay between the spatial dimension of population size (including the large $N$ limit) and the time scales of input correlations and of readout integration.
Our approach employs a diffusion approximation, where the readout neuron membrane potential obeys a second-order stochastic differential equation. In this framework, the inter-spike interval (ISI) of the readout neuron corresponds to the first passage time of an \emph{inertial} stochastic process. We derive analytical expressions for the mean and variance of the readout neuron firing rate, and its temporal correlation structure.  Treating this problem within the theory of stochastic processes involves analyzing first-passage times of non-Markovian random walks, which typically requires solving coupled Fokker-Planck equations. By introducing a novel analytical approach —the \textit{quenched-noise approximation}— we provide tractable expressions for the firing rate mean and variance under the assumption that the noise correlations timescale $\tauc$ is much larger than the membrane time constant of the readout neuron $\taum$. The quenched-noise approximation exploits classical expressions for the mean rate of the LIF neuron with white noise input \cite{burkitt2006}. Our results extend previous expressions for the mean rate of LIF neurons with colored noise input with $\taum\ll\tauc$ \cite{moreno2002}, which were recovered in the limit of weak correlation strength. Moreover, we are able to study analytically the opposite limit $\tauc\ll\taum$ using a Poisson approximation.
These results provide the necessary conditions for noise correlations to enhance the readout SNR. Notably, our analysis demonstrates that SNR enhancement can occur for a wide range of noise correlation timescales $\tauc$, regardless of whether they exceed or fall below the membrane time constant $\taum$.

The  article is structured as follows. We first review experimental work and define the resulting NERM (Sec. \ref{sec:model}). We then (Sec. \ref{sec:auxiliary})  investigate how a readout LIF neuron with or without time-correlated inputs discriminates between binary stimuli $s$ by computing the SNR of its firing activity. In Sec. \ref{sec:diffapp}, we approximate analytically the SNR using the mean and variance of the firing rate of a LIF neuron with input current satisfying a scalar Ornstein-Uhlenbeck equation, within the quenched-noise approximation in the $\tauc\gg\taum$ regime (Secs. \ref{sec:quenchedmean}, \ref{sec:quenchedvar}). In Sec. \ref{sec:intuitive}, we explain intuitively why and when  noise correlations  enhance downstream stimulus information transmission  under the quenched-noise approximation. In Sec. \ref{sec:valentemodelsolution} we describe analytically how noise correlations influence the readout SNR. We  finally draw  conclusions in Sec. \ref{sec:discussion}.


\section{Definition of the Neural Encoding-Readout Model (NERM) \label{sec:model}}

\begin{figure}
    \centering
    \includegraphics[width=1\linewidth]{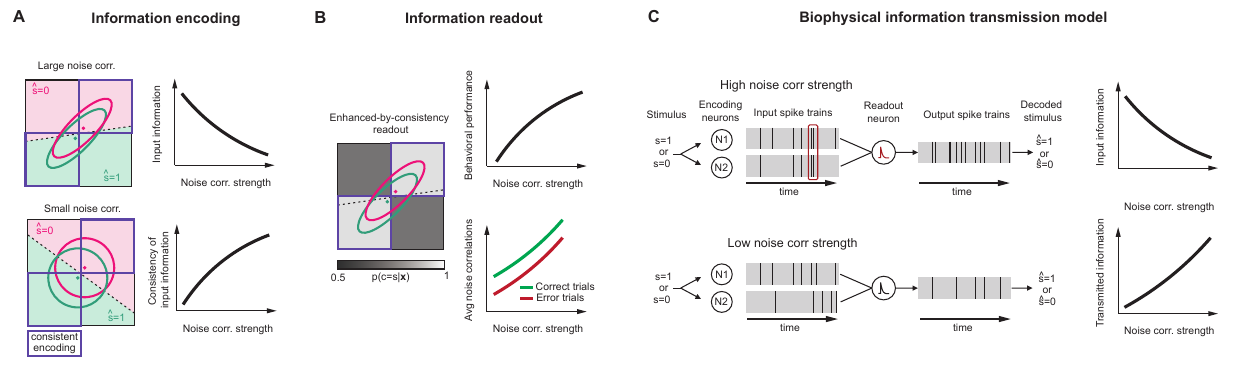}
    \caption{{\it Summary of recent neuro-biological results on the role of neural correlations in information encoding and readout. } A. Left: Single-trial neural activity of two neurons in the case of large and small noise correlations. Right: Input information decreases with the strength of noise correlations, while the consistency of input information (fraction of trials where the two neurons signal the same stimulus) increases with the noise correlation strength. B. Schematics of the enhanced-by-consistency readout model in \cite{valente2021} in which the probability that the choice equals the stimulus is higher for trials with consistent input information. This readout accounts for experimental observations \cite{valente2021} of enhanced behavioral performance in presence of correlations and higher noise correlation strength in correct than in error trials. C. Biophysical information-transmission model that implements an enhanced-by-consistency readout. When the readout unit has a sufficiently short membrane time-constant correlated input activity can drive the readout to spike. Under this condition the readout transmits more stimulus information when input noise correlations are strong, so that increasing noise correlations increases the amount of transmitted information, even when decreasing the amount of input stimulus information.}
    \label{fig:cartoon_valente}
\end{figure}
\begin{figure}
\includegraphics[width=0.5\textwidth]{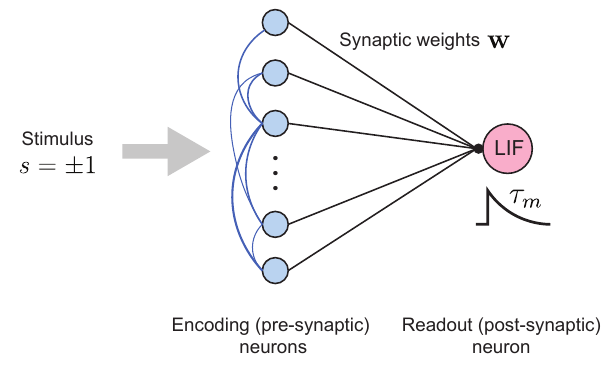}%
\caption{{\it Illustration of the Neural Encoding-Readout Model (NERM).} A binary stimulus $s$ is encoded in the time-dependent  activity of a population of $\N$ encoding pre-synaptic neurons, whose firing rates ${\x}^{(s)}(t)$ are the solution of an Ornstein-Uhlenbeck process, that depends on $s$ through the average firing rate vector ${\bm\mu}^{(s)}$. Given the firing rate vector $\x^{(s)}(t)$, the encoding neural activity consists of a set of $\N$ spike trains in a time window $[0,\Tc]$, with spiking times drawn from a multivariate inhomogeneous Poisson process with underlying firing rate $\x^{(s)}(t)$. The encoding spike trains, weighted by the vector of synaptic weights $\bf w$, constitute the afferent input current of the readout LIF neuron. The  stimulus-specific response of the readout is characterized in terms of the mean and variance ($\rro^{(s)},\vro^{(s)}$) of the readout neuron firing rate. \label{fig:NN_valente}}
\end{figure}
\begin{figure}[h!]
\centering
\includegraphics[width=0.8\textwidth]{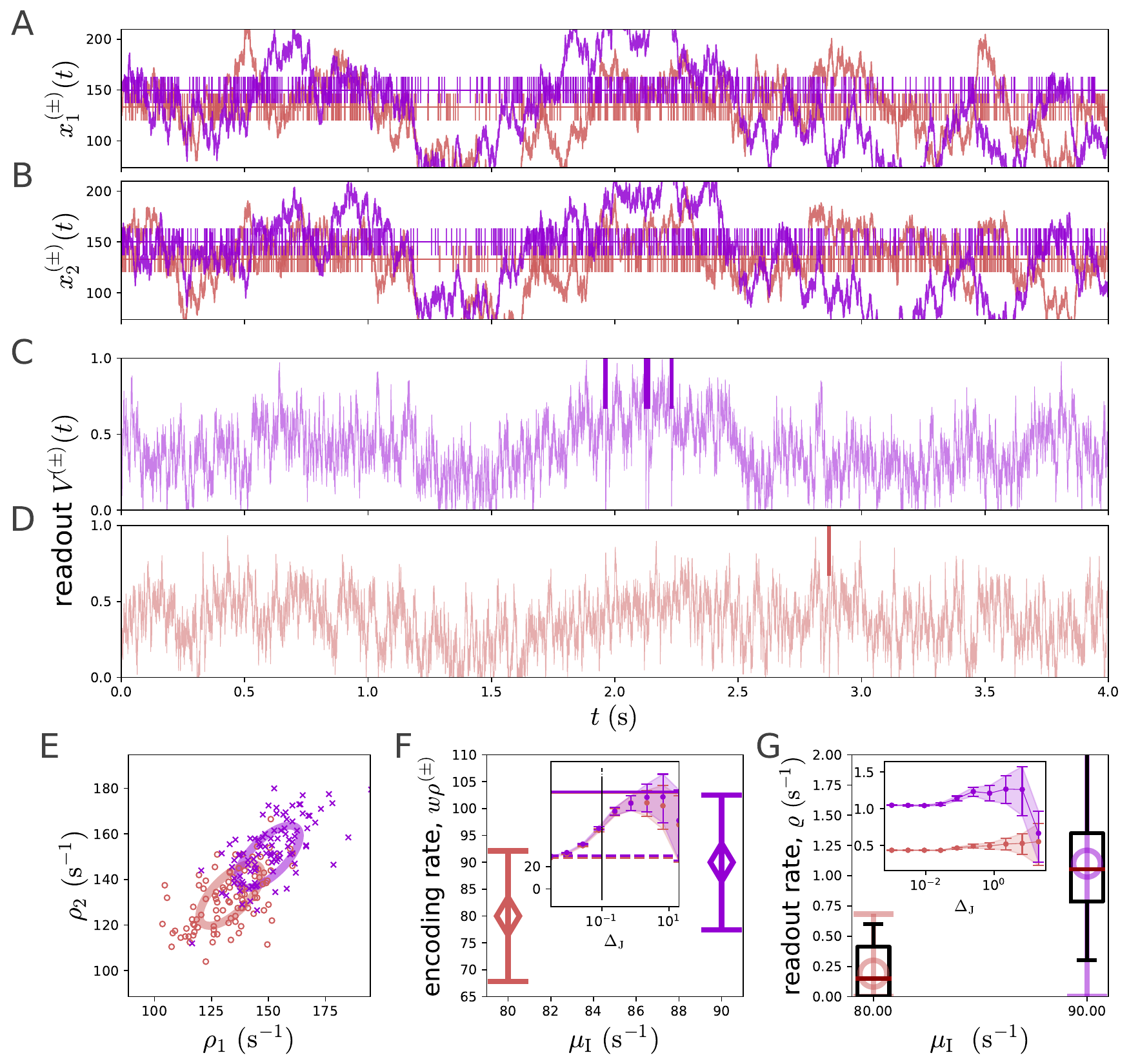}%
\caption{\textit{Neural activity in the Neural Encoding and Readout Model with $\N=6$ encoding neurons and a strong degree of noise correlation alignment $\alphaV=0.9$}. (A-B) Encoding firing rates $x^{(s)}_{1,2}(t)$ (continuous curves in A,B) and  spike trains sampled from an inhomogeneous Poisson process with rate $x^{(s)}_{1,2}(t)$  are shown in a time window $[0,T=4{\rm s}]$ for stimuli $s=\pm 1$ (orange and violet, respectively). (C-D) LIF membrane potential $V(t)$ (lines) and spike times (ticks) for $s=-,+$ (respectively panels C and D). (E) Scatter plot of encoding spike densities $(\rho_1,\rho_2)$, with $\rho_j=n_j(\Delta)/\Delta$. Each point corresponds to a different time window of size $\Delta_{\rm e}=0.05$, for a simulation that lasts up to $\Tc=200$. The ellipse's largest axis is oriented along the $(1,1)$ axis, and the larger and lower axes of the ellipse are set to $\nu$ and $(1-\alphaV^2)^{1/2}\nu$ respectively, where $\nu=\sigmaV+\bar\mu^{1/2}$ is the theoretical noise amplitude in the diffusion approximation. (F) Average (diamonds) and standard deviation (error-bars) of the encoding rate density. While the averages are simply $w\sum_i\mu_i^{(\pm)}$, the standard deviations are their theoretical value, $[(\sigmai^2+\sigmac^2)/w]^{1/2}$. The inset shows the theoretical value (continuous lines) of the rate standard deviation, along with its numerical estimation by means of the Jack-Knife method \cite{amit2005}, using an increasing sliding window $\Delta_{\rm J}$ (in abscissa). (G) Box-plots of the histogram of readout neuron firing rates, across ${\rm int}(\Tc/\Delta_{\rm r})$ different time intervals of length $\Delta_{\rm r}=0.5$. The difference in the averages divided by the (stimulus-averaged) error is the SNR. The rate standard deviation is estimated by means of the Jackknife error (inset ordinates) for increasing sliding window (inset abscissa). See details on this simulation in Sec. \ref{sec:figs23} in Appendix \ref{sec:validity}. The simulation parameters are: $\N = 6$, $\alphaV = 0.9$, $\sigmaV = 8.9442$, $\tauc = 0.1$, $\taum = 0.05$, $\mui^{(-)} = 80$, $\mui^{(+)} = 90$, $w = 0.1$.
See \cite{repository} and Appendix \ref{sec:figs23} for further details.  \label{fig:valente_rates}}
\end{figure}
\begin{figure}[h!]
\centering
\includegraphics[width=0.8\textwidth]{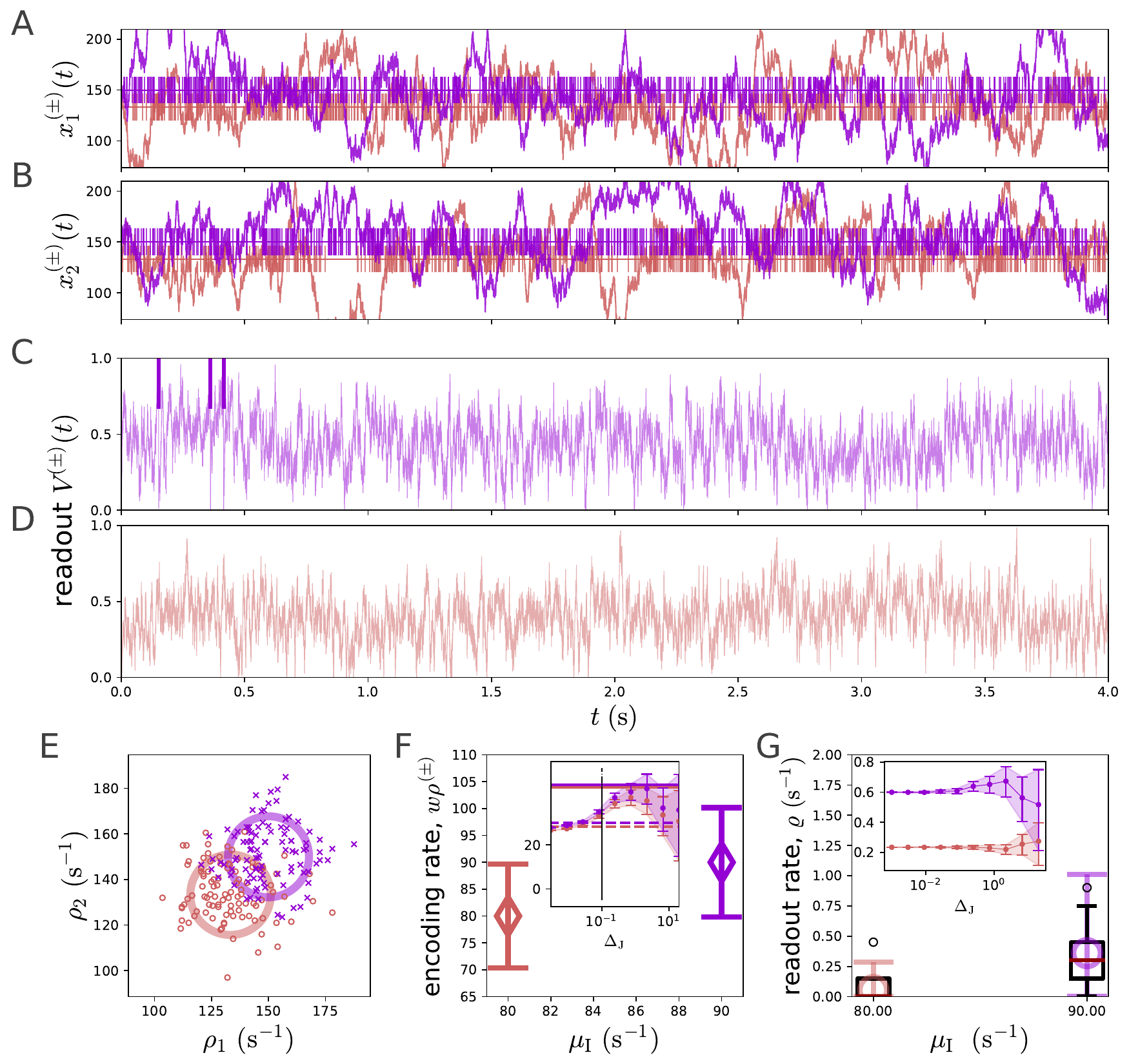}%
\caption{{\it Neural activity in the neural encoding and readout model}, as in Fig. \ref{fig:valente_rates}, but in absence of noise correlations $\alphaV=0$. Note that, even in absence of off-diagonal noise correlations $\alphaV=0$, the total firing rate $x^{(s)}(t)$ oscillates around its mean (because the diagonal of the noise correlation matrix is nonzero: $\sigmac\ne 0$ even if $\alphaV=0$ in Eq. (\ref{eq:dictionary_sigmac})). We can see that the {\it encoding SNR} increases, with respect to the prevalence of noise correlations in Fig. \ref{fig:valente_rates}  (since $\mu^{(+)}-\mu^{-}$ is constant, while $\sigmac$ has decreased). The rest of the parameters are:  $N = 6$, $\alphaV = 0$, $\sigmaV = 8.9442$, $\tauc = 0.1$, $\taum = 0.05$, $\mui^{(-)} = 80$, $\mui^{(+)} = 90$, $w = 0.1$. \label{fig:valente_rates_nocorrelations}}
\end{figure}

Before defining our work, for self-consistence of this study we briefly summarize (and schematize in a summary cartoon in Fig.\ref{fig:cartoon_valente}) the  biological experimental and modeling work that leads to it. Experimental work has shown that noise correlations in local brain circuits are predominantly information-limiting \cite{moreno2014, zohary1994, Rumyantsev2020,bartolo2020,valente2021}, as simultaneously-recorded activity (which contains correlations between the activity of different neurons) encodes less information that when this activity is randomly shuffled across trials to form pseudo-population shuffled responses that have the same single-neuron  probabilities as the real data but no correlations across neurons (Fig.\ref{fig:cartoon_valente}A). However, the same data  also shows stronger across-neuron correlations in trials in which animals make behaviorally correct judgment about the stimulus than in error trials in which animals make incorrect judgments \cite{runyan2017,Francis2022,valente2021, Safaai2023, BalaguerBallester2020, Shahidi2019}. This suggests that these correlations may enhance the readout of correlated activity and aid task performance, even when hindering encoded information (Fig.\ref{fig:cartoon_valente}B). Previous work reconciled these observations using a biophysical Neural Encoding-Readout Model (NERM), where the readout integrated correlated activity from two input units with a short membrane time constant \cite{valente2021} (see Fig.\ref{fig:cartoon_valente}C). To study the impact of noise correlations at the encoding (input, pre-synaptic) stage on a post-synaptic readout neuron, we extend the NERM to a population of $\N$ neurons  encoding a binary stimulus $s=\pm 1$, and a single readout neuron that integrates the neural activity of the encoding population (see Figs. \ref{fig:NN_valente},\ref{fig:valente_rates},\ref{fig:valente_rates_nocorrelations}). 

The pre-synaptic encoding neurons are modeled as $\N$ inhomogeneous Poisson processes with stochastic coupled firing rates. For stimulus $s$, the firing rates of the encoding population $\x^{(s)}(t)$ satisfy the multivariate Ornstein-Uhlenbeck process given by:

\begin{equation}\label{eq:MOU}
    \tauc\, \dot \x(t) = ({\bm \mu^{(s)}}-\x(t)) + \sqrt{2\tauc C}\cdot {\bm \eta}(t),
\end{equation}
where ${\bm \eta}(t)$ is the vector of across-time and across-neuron uncorrelated white noises, ${\bm\mu}^{(s)}$ the stimulus-specific mean activity (or tuning curve), $C$ the stationary across-neuron covariance matrix and $\tauc$ the correlation timescale (see Table \ref{table:symbols} in Appendix \ref{sec:symbols}). Thus, the resulting firing rates are correlated across neurons with covariance $C$ and across time with timescale $\tauc$, with  first and second moments given by 
$\langle x_j^{(s)}(t)\rangle_{\bm\eta} = \mu_j^{(s)}$ and  $\langle(x_i^{(s)}(t')-\mu_i^{(s)})(x^{(s)}_j(t'+t)-\mu_j^{(s)})\rangle_{\bm\eta} = C_{ij} e^{-|t|/\tauc}$,
where the expectation values $\<\cdot\>_{\bm\eta}$ are computed across white noise realizations \cite{risken1996,livi2017}. We  define the  matrix of {\it noise correlations}, $\bar C$, as $\bar C_{ij}:=C_{ij}/(\vV_i\vV_j)^{1/2}$, where $\vV_j:=C_{jj}$ is the \emph{stationary variance} of the $j$-th neuron rate. The corresponding \emph{spike count variance} is $\sigmaV_j^2=2\tauc\vV_j$ (see Appendix \ref{sec:symbols} for a discussion about the stationary vs spike count variance terminology). The noise correlation matrix is defined to exhibit equal off-diagonal elements, i.e. $\bar C_{ij}= \alphaV^2$ for $i\ne j$, where the  fraction of shared noise $\alphaV \in [0,1]$ modulates the shared over private noise ratio, so that the limiting cases $\alphaV=0,1$ represent  completely uncorrelated and correlated encoding firing rates respectively. To ensure that the noise correlations are information-limiting, we set the noise amplitude $\sigmaV_i$ to be proportional to the difference of tuning curves $\sigmaV_i =\nsrV\,{\Delta\mu}_i$, where ${\Delta \mu}_i:=\mu_i^{(+)}-\mu_i^{(-)} > 0$ with $\nsrV>0$ (see \cite{valente2021} and equation (\ref{eq:snrencoding}) below).

Given stimulus-specific firing rates $\x^{(s)}(t)$, the spike train of the $i$-th encoding neuron $(t_{m_i}^{(i)})_{m_i}$ in the time interval $[0,\Tc]$ is  sampled from an inhomogeneous Poisson process with underlying firing rate $x_i^{(s)}(t)$ (see Fig. \ref{fig:valente_rates} and Appendix \ref{sec:diffusion}), so that the encoding spike trains are generated through a doubly stochastic process. Given the encoding spike trains, the response of the readout neuron is deterministic. 

The postsynaptic readout neuron is modeled as a leaky integrate-and-fire (LIF) neuron \cite{burkitt2006,burkitt2006II} that receives  the spike trains of the encoding neurons as inputs, weighted by the vector synaptic strengths $\bf w$. The dynamics of the readout neuron membrane potential $V(t)$ is given by: 

\begin{align}
\dot V(t) &= -V(t)/\taum + I(t) \label{eq:vLIF}\\
I(t) &= \sum_{i=1}^\N w_i \sum_{m_i>0} \delta(t-t_{m_i}^{(i)}) \label{eq:afferentcurrent},
\end{align}
with firing threshold $\Omega=1$, and where $(t^{(i)}_{m_i})_{m_i}$ is the spike train of the $i$-th encoding neuron. Note that, following previous work \cite{siegert1951,moreno2002,valente2021},  the above equation assumes that the synapses are instantaneous (without a rise time or a synaptic decay time constant, which would have been more realistic). This implies that all the temporal post-synaptic integration properties of the readout neuron depend on its membrane time constant $\taum$.  The readout neuron spiking times will be referred to as ${\vec t}=(t_1,\ldots,t_{\nro(\Tc)})$, where $\nro(\Tc)$ is the number of readout spikes in the time window of length $\Tc$ in a single realization of the encoding spike trains. 

Examples of the dynamics of the input spike trains at the encoding stage and of the membrane and spiking dynamics of the readout neuron are given in Fig \ref{fig:valente_rates} for the case of correlated inputs ($\alpha_V=0.9$) and in Fig. \ref{fig:valente_rates_nocorrelations}  for the case when correlations in the input are set to zero ($\alpha_V=0$). Note that the input activity of the model with $\alpha_V=0.9$ exhibits clearly visible covariances over the 100 ms time-scale set by the Ornstein-Uhlenbeck process (Fig \ref{fig:valente_rates}-A,B) compared to the $\alpha_V=0$ case. The output of the readout neuron exhibits higher rate separation $\Delta\rro$ (Fig. \ref{fig:valente_rates}-G), and higher stimulus-discriminative signal between the rates elicited by the two stimuli, because correlations in inputs increase the frequency of events with near-coincident firing in the input compared to the uncorrelated-input case (Fig. \ref{fig:valente_rates_nocorrelations}). However, input correlations also increase the variability, or noise, of the readout neuron output rate, which makes it difficult to understand whether input correlations increase or decrease the output information without a careful quantification of both signal and noise as function of the parameters of the model. 

To examine how well the stimulus can be discriminated from  the readout spike trains,  we consider the (square root of the) {\it Signal-to-Noise Ratio} (SNR) \cite{valente2021}, defined as (mind that $\Delta\mui>0$ and hence $\rrom<\rrop$):

\begin{equation}
\label{eq:SNRsqrt}
\text{SNR}(\Tc) = \frac{\rro^{(+)}-\rro^{(-)}}{\left[(\vro^{(+)}+\vro^{(-)})/2\right]^{1/2}},
\end{equation}
where $\rro^{(s)}$ and $\vro^{(s)}$ are the stimulus-specific average and variance of the  readout rate $\nro(\Tc)/\Tc$ over the distribution of readout spike trains $\vec t$, i.e. $\text{prob}(\vec{t}\,|s)$ (see Table \ref{table:symbols}). From now on, we will define the SNR {\it in the unit time interval} $\text{SNR}:=\text{SNR}(\Tc)\times \Tc^{-1/2}$, since the rate variance decreases as $\Tc^{-1}$ (Eq. (\ref{eq:variancecorrection})) and hence $\text{SNR}(\Tc)$ increases as $\Tc^{1/2}$. 

Although Figs. \ref{fig:valente_rates},\ref{fig:valente_rates_nocorrelations} have an  illustrative scope, the  estimation of the NERM SNR with parameters used in these figures, shows (see Sec. \ref{sec:figs23} in Appendix \ref{sec:validity}) that the introduction of noise correlations in Fig. \ref{fig:valente_rates} significantly enhances the SNR with respect to the case with no noise correlations, and the same value of the rest of the parameters (Fig. \ref{fig:valente_rates_nocorrelations}).

\section{Analytical results on the Readout LIF model with colored-noise input current\label{sec:auxiliary}}
\subsection{The diffusion approximation for the encoding signal\label{sec:diffapp}}

To build an interpretable theory of the NERM, we will first adopt the {\it diffusion approximation} for the readout neuron afferent current $I(t)$ in Eq. (\ref{eq:afferentcurrent}) \cite{burkitt2006,burkitt2006II}. This approximation is  based on the fact that, for high enough input spike count $x_i(t) \delta$ in a time window $\delta$, the Poisson statistics of the count $n_i(t+\delta)-n_i(t)$ approaches a normal distribution with mean and variance equal to $x_i(t) \delta$. In the diffusion approximation, the input current $I(t)$ and the readout neuron membrane potential $V(t)$ in Eq. (\ref{eq:vLIF}) satisfy (see Appendix \ref{sec:diffusion}):

\begin{subequations}
\label{eq:VI_diffusion}
\begin{align}
\dot V(t) &= -V(t)/\taum + I(t) \\
I(t) &= \mui + \sigmai\, \eta(t) + \sigmac\, \xi(t) \label{eq:I_diffusion} \\
\dot \xi(t) &= -\xi(t)/\tauc + \frac{1}{\tauc} \eta'(t)
\end{align}
\end{subequations}
where $\eta,\eta'$ are two different white noises, and $\xi(t)$ is an Ornstein-Uhlenbeck process with timescale $\tauc$. In the diffusion approximation the density of encoding spikes $I(t)$ in Eq. (\ref{eq:afferentcurrent}) can be treated as a scalar stochastic current with {\it mean} $\mui$, {\it white noise amplitude} $\sigmai$, and {\it colored noise amplitude} $\sigmac$ given by:

\begin{subequations}
\label{eq:dictionary}
\begin{align}
\mui^{(s)} &= w N \overline\mu^{(s)} \label{eq:dictionary_mui} \\
{\sigma_I^{(s)}}^2 &= w\mui^{(s)} \label{eq:dictionary_sigmai} \\
\sigmac^2 &=  w^2 N \sigmaV^2 \left( 1 + (N-1)\alphaV^2 \right)  \label{eq:dictionary_sigmac}
\end{align}
\end{subequations}
where $\overline\mu^{(s)}$ and $\sigmaV^2$ are the average of current means and variances $\mu^{(s)}_i$ and $\sigmaV_i^2$ over encoding neurons $i$, respectively. For simplicity, in Eqs. (\ref{eq:dictionary}) we consider a homogeneous gain vector $w_i=w$ $\forall i$ (see Appendix \ref{sec:diffusion} for a generalization to arbitrary correlation matrices $\bar C$ and gain vectors $\bf w$). The time-shifted covariance of the current $I(t)$ takes the form (from Eq. (\ref{eq:I_diffusion})):

\begin{subequations}
\label{Icorrelations1RON}
\begin{align}
\< (I(t) - {\mu_I})(I(t') - {\mu_I}) \> &= {{\sigma_I}^2} \delta(t-t') + \vc\,e^{-|t-t'|/\tauc} \\
\vc&:=  \frac{{\sigmac}^2}{2\tauc} =  w^2 N \vV \left( 1 + (N-1)\alphaV^2 \right)  \label{eq:dictionary_vc}
.
\end{align}
\end{subequations}
where $\vc$ is the {\it stationary variance of the colored noise}. These results are consistent with those in reference \cite{moreno2002}.

In sum, while the encoding is in principle characterized by the $3\N+2$ parameters ${\bm\mu}^{(\pm)},{\bf \vV},\alphaV,\tauc$, in the diffusion approximation it is described by the $4$ parameters ${\mui}^{(\pm)},\vc(\alphaV),\tauc$. The firing statistics of the readout depend on these encoding parameters and on the readout parameters $w,\taum$ (note that $\sigmai^{(\pm)}$ is already determined by $\mui^{(\pm)}$, given $w$, see Eq. (\ref{eq:dictionary})). 

The NERM in the diffusion approximation defined by Eqs. (\ref{eq:VI_diffusion}) will be, from now on, our working hypothesis model, to which our analytical results refer, and to which the numerical simulations refer, unless stated differently. For an assessment of the validity of the diffusion approximation, and a study of the behavior of the SNR in the NERM model with spiking encoding current in Eqs. (\ref{eq:MOU},\ref{eq:vLIF},\ref{eq:afferentcurrent}), we refer to the Appendix \ref{sec:validity}.

{\bf Encoding signal-to-noise ratio.} From Eq. (\ref{eq:dictionary_sigmac}) it follows that the fraction of shared noise $\alphaV$ increases the amplitude $\sigmac$ of colored correlations. In fact, in the unit time interval, the SNR {\it of the input current} $I(t)$ is given by (see Eqs. (\ref{Icorrelations1RON},\ref{eq:variancecorrection})): 

\begin{align}
\label{eq:snrencoding}
\text{input SNR} = \frac{{\mu_I}^{(+)}-{\mu_I}^{(-)}}{\left(\sigmac^2+({\sigmai^{(+)}}^2 + {\sigmai^{(-)}}^2 )/2 \right)^{1/2} }
\end{align}
which monotonically decreases with $\alphaV$ (see Eq. (\ref{eq:dictionary_sigmac})). 


\subsection{The quenched-noise approximation for the LIF neuron rate mean \label{sec:quenchedmean}}

In the following, we derive expressions for the mean and variance (hence, for the SNR, see Eq. (\ref{eq:SNRsqrt})) of the readout neuron defined by Eq. (\ref{eq:VI_diffusion}) as a function of the parameters $\mui^{(\pm)},\vc,\tauc,w,\taum$.

How to compute analytically  the mean and variance of the firing rate of a LIF model with a coloured noise input has been extensively investigated. Approximate expressions for the mean $\rro$ have been derived in the $\tauc\ll\taum$ and $\tauc\gg\taum$ limits \cite{brunel1998,fourcaud2002,moreno2002,morenobote2004,burkitt2006,burkitt2006II}. In particular, Moreno et al \cite{moreno2002} derive an analytical approximation to the mean firing rate $\rro$ of the colored-input noise LIF neuron in the $\tauc\gg\taum$ limit (at first order in $\taum/\tauc$), and  under the further approximations of small $\alpha_{\rm M} := \sigmac^2/\sigmai^2$ and small $\mui/(\sigmai^2+\sigmac^2)^{1/2}$.

Here we present an approximated expression for $\rro$ in the $\tauc\gg\taum$ limit, in the {\it quenched-noise approximation}. The quenched-noise approximation generalizes the Moreno solution \cite{moreno2002} to larger values of $\alpha_{\rm M}$. The approximation stands on a simple argument: whenever $\tauc\gg \taum$, {\it in time windows of length $\ell$, $\taum\ll\ell\ll \tauc$}, the coloured noise $\xi(t)$ is approximately constant and, consequently, {\it the rate of the LIF neuron behaves as that of a white-noise LIF neuron} with input current $I(t)=\tilde\mu + \sigmai\,\eta(t)$, where $\tilde \mu=\mui+\sigmac\,\xi$ is an effectively constant drift. Under this assumption, the firing rate $\rro$ in time intervals of size $\Tc\gg\tauc$ can be approximated as {\it the average of $\rw(\tilde\mu)$ over the stationary distribution of values of $\tilde\mu \sim {\cal N}(\tilde\mu;\mui,\vc)$} with $\vc=\sigmac^2/(2\tauc)$ (see (Eq. \ref{eq:I_diffusion})), where ${\cal N}(\cdot;m,v)$ denotes the normal distribution with mean $m$ and variance $v$. Thus, the readout firing rate in the quenched-noise approximation $\rro(\mui) \simeq \rq(\mui)$ is given by:

\begin{align}
\label{eq:rquenched}
\rq(\mui) &:= \<\rw(\tilde\mu)\>_{{\cal N}(\tilde\mu;\mu_I,\vc)} \qquad \text{(quenched-noise approximation for $\tauc\gg\taum$)}
\end{align}
where $\rw(\mu)$ is the Siegert solution for the mean rate of a \emph{white noise LIF neuron} (see Eq. (\ref{eq:rwhiteLIF}) in Appendix \ref{sec:quenched_mean}) \cite{siegert1951,burkitt2006}. In the saddle-point approximation for low enough $\vc=\sigmac^2/(2\tauc)$, it can be proved (see Appendix \ref{sec:quenched_mean}) that Eq. (\ref{eq:rquenched}) takes the analytic form:

\begin{align}
\label{eq:rquenched_analytical}
\rq(\mui) &= \rw(\mui) \left[\frac{1}{1-\vc g'_1(\mui)}\right]^{1/2} \exp\left( \frac{\vc}{2} \frac{ g_1^2(\mui)}{1-\vc g'_1(\mui)} \right), \qquad \text{$\tauc\gg\taum$, low $\vc$}
\end{align}
where the function $g_1:=\rw'/\rw$ and its derivative $g'_1$ with respect to $\mui$ depend on the white LIF neuron parameters only. Notice that, consistently with the quenched-noise hypothesis in Eq. (\ref{eq:rquenched}), the emerging correction to the white noise case depends on the properties of the colored noise only through $\vc$.

\begin{figure}[h!]
\includegraphics[width=1\textwidth]{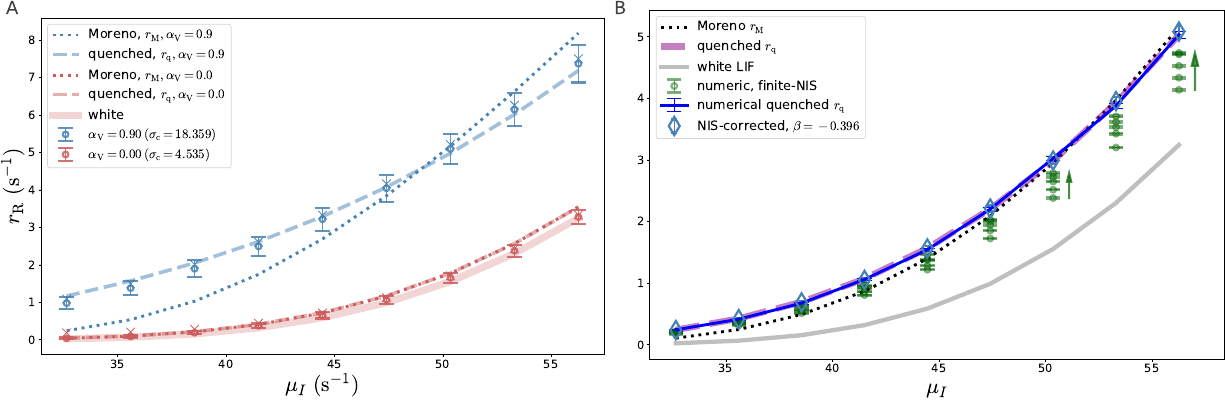}%
\caption{ {\it Readout firing rate $\rro$ vs input current $\mui$: numerical estimations and analytical approximations.} (A) $\rro$ vs $\mui$ for two values of the noise correlation alignment: $\alphaV=0$ (no noise correlations, red curve), and $\alphaV=0.9$ (strong information-limiting noise correlations, blue curve --see the value of $\sigmac$ in the legend). The numerical estimation (points) correspond to a simulation with the highest NIS, $\nis=2^{17}$ (see Appendix \ref{sec:numerical}). The error-bars are the rate standard deviation in the unit time window, $(\tilde\vro\times\Tc)^{1/2}$, estimated as the standard deviation of the spike count $\nro(\Tc)$ in $[0,\Tc]$ across the $\ntrials=10^3$ simulation trials. The analytical activation functions correspond to the white-noise LIF $\rw(\mui)$ in Eq. (\ref{eq:rwhiteLIF}) (red line), the Moreno solution $r_{\rm M}(\mui)$ for $\tauc\gg\taum$ in Eq. (\ref{eq:r_moreno}) \cite{moreno2002} (dotted red and blue lines), and the quenched-noise approximation $\rq$ in Eq. (\ref{eq:rquenched}) (dashed red and blue lines). The simulation parameters are: $\tauc = 0.1$, $\taum = 0.005 $, $\sigmaV = 1.778$, $\N=20$, $w = 0.450 $. (B) Numerical estimations of $\rro$ vs $\mui$ for $\tauc\simeq 0.720$, and all the values of the NIS, $\nis=2^m$ for $m=12,\ldots,17$ (different green points at a common abscissa indicate different values of the NIS). We report as well the finite-NIS corrected estimation of $\tilde\rro(\mui)$ (blue diamonds) and their error in the regression (error-bars), as well as the analytical approximations $r_{\rm M}$ and $\rq$. The finite-NIS corrected estimation is compatible with the quenched-noise approximation only. See Appendix \ref{sec:numerical} for details, and more on the comparison between the raw numerical, NIS-corrected numerical, Moreno, quenched-noise in Appendix \ref{sec:alternativefigures}. Finally, we present a comparison of the analytical expression for the quenched-noise approximation for the mean rate, Eq. (\ref{eq:rquenched_analytical}) with the general expression, Eq. (\ref{eq:rquenched}). The simulation parameters are: $\tauc = 0.720$, $\sigmac = 30.593$, $\taum = 0.005 $, $w = 0.45$. \label{fig:r_quenched}} 
\end{figure}

Interestingly, if we now expand  $[1-\vc g'_1(\mui)]^{-1/2}$ in (Eq. \ref{eq:rquenched_analytical}) to the first order in $\nu_{\rm W}=(\sigmac^2\taum)/(\sigma_I^2\tauc)$ (W stands for Walter et al, who use the same perturbative parameter in reference \cite{walter2021}), and the exponential in (Eq. \ref{eq:rquenched_analytical}) to the first order in $\taum/\tauc$, we recover the Moreno et al expression for $\rro$ at first order in $\taum/\tauc$ in Eq. (\ref{eq:r_moreno}). We conclude that Eq. (\ref{eq:rquenched_analytical}) is therefore a generalization of the Moreno  solution for $\rro$ in the $\tauc\gg\taum$ limit \cite{moreno2002,morenobote2004}, expected to hold for larger values of $\alpha_{\rm M}$ (Figs. \ref{fig:r_quenched},\ref{fig:numSNRsqrt}-A and Appendix \ref{sec:alternativefigures}).

Comparing the Moreno and quenched-noise approximations with the numerical data in the regime $\tauc\gg\taum$ reveals that: (\textit{i}) for low $\sigmac$ both approximations $r_{\rm M}$ and $\rq$ converge (see Eq. (\ref{eq:rquenched_analytical})) to the white-noise LIF activation function $\rw$ and for low enough $\sigmac$ and hence $\alpha_{\rm M}$, both solutions $r_{\rm M}$ and $\rq$ are consistent with each other (Fig. \ref{fig:r_quenched}-A red lines, Fig. \ref{fig:numSNRsqrt}-A and Appendix \ref{sec:alternativefigures}); (\textit{ii}) for larger values of $\sigmac$ the solutions differ, and the numerical estimation of $\rro$ (i.e. $\tilde\rro$; see Appendix \ref{sec:numerical}), is \emph{consistent only with the quenched-noise approximation} (Fig. \ref{fig:r_quenched}-B blue lines); (\textit{iii}) to find full agreement between numerical simulations and $\rq$, we need to apply the finite-Number of Integration Steps (NIS) correction (Fig. \ref{fig:r_quenched}-B, see Appendix \ref{sec:numerical}). 

We stress that Eq. (\ref{eq:rquenched_analytical}) is an analytical approximation of Eq. (\ref{eq:rquenched}). As such, its validity limits are those of Eq. (\ref{eq:rquenched}) (i.e., $\tauc\gg\taum$), plus the further condition that $\vc$ is low enough  (see Appendix \ref{sec:quenched_mean}). Throughout the article, with the exception of Appendix \ref{sec:varyingsigmav}, we always use our analytic expression for the rate mean, Eq. (\ref{eq:rquenched_analytical}), and never the numerical evaluation of the integral in Eq. (\ref{eq:rquenched}). As a matter of fact, our analytical expression for the rate mean is a good approximation, since the conditions for its derivation are satisfied in the range of parameters that we probe in our simulations. This is shown in Fig. \ref{fig:r_quenched}-B and in Appendix \ref{sec:quenched_mean}. The same holds for the expressions of the rate variance in the quenched-noise approximation, that we show below in Sec. \ref{sec:quenchedvar}. Finally, we probe, in a designated simulation in Appendix \ref{sec:varyingsigmav} for constant $\tauc$ and increasing $\sigmac$ (hence, increasing $\vc$), the quenched noise approximation beyond the validity limits of Eq. (\ref{eq:rquenched_analytical}). Since, in this simulation, it is $\tauc\gg\taum$, we are within the validity limits of the quenched-noise approximation and, indeed, the readout firing rate is still well explained, net of the finite NIS effect described below, by Eq. (\ref{eq:rquenched}), although it is no longer well described by Eq. (\ref{eq:rquenched_analytical}).

\subsection{Intuitive interpretation of the enhanced firing rate effect in the quenched-noise approximation \label{sec:intuitive}}

\begin{figure}[h!]
\includegraphics[width=0.7\textwidth]{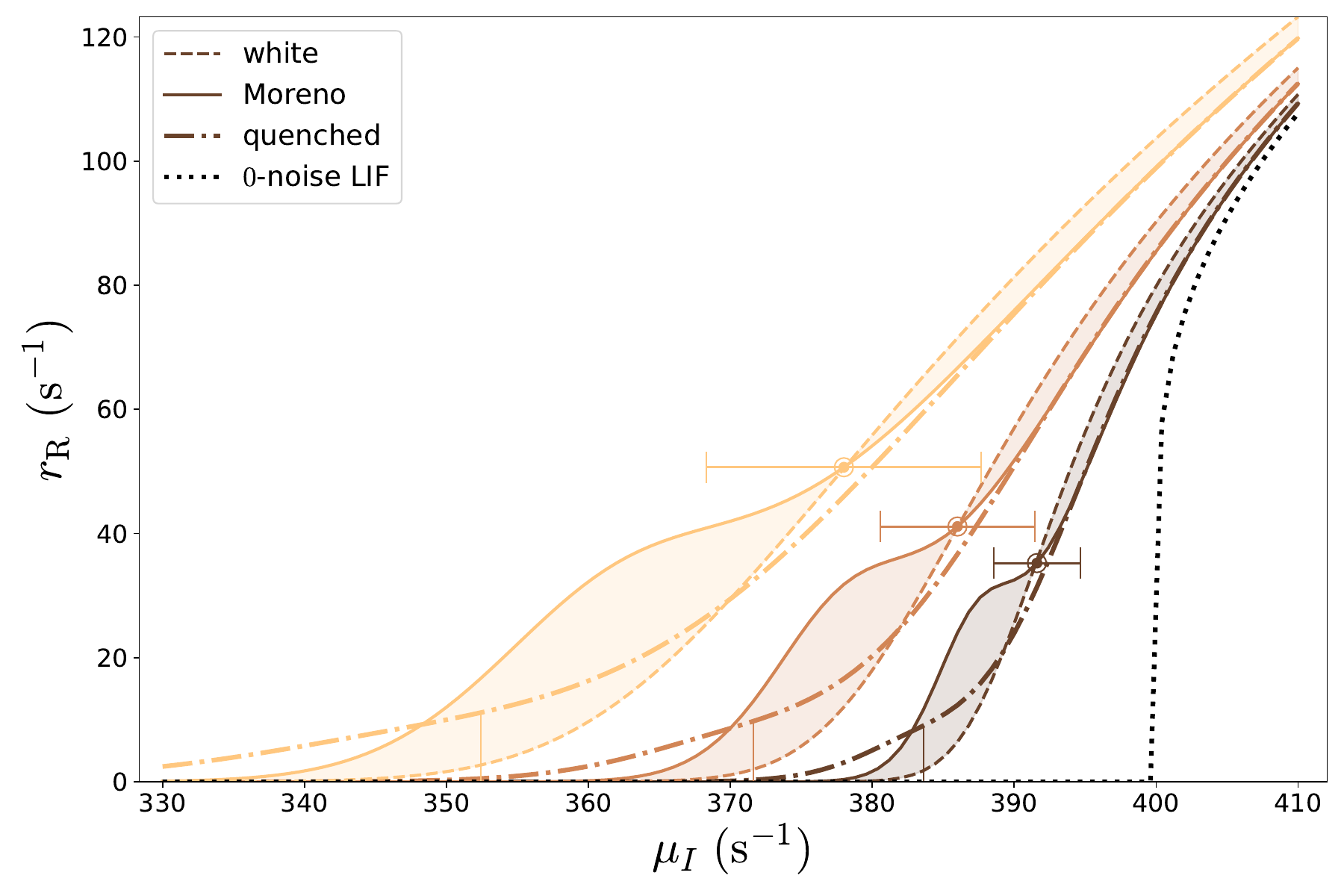}%
\caption{{\it Activation function of the LIF neuron with white and colored input noise}. The dashed lines denote the Siegert expression for the white-noise LIF $\rw(\mui)$; the continuous curves are the Moreno solution (\ref{eq:r_moreno}) for $\tauc\gg\taum$; the dashed-dotted lines are the quenched-noise approximation in (\ref{eq:rquenched}); the black dotted line is the LIF activation function in the absence of noise, $r_0(\mu) = [-\taum \ln(1-(\mu\taum)^{-1})]^{-1}$. Different colors correspond to different values of the gain $w=(0.00031,0.001, 0.00316)$ (from darker to lighter). The rest of the parameters are: $\taum=2.5\cdot 10^{-3}$, $\tauc=10^{-1}$, $\sigmai^2=w\mui$, $\sigmac=8\bar\sigmai$, with $\bar\sigmai^2=w\bar\mui$ and $\bar\mui=370$. The circles over the curves show the inflection point of the white-noise LIF neuron, and error-bars indicate $\vc^{1/2}$ in every direction.  \label{fig:r_interpretation}}
\end{figure}

The quenched-noise approximation provides an intuitive interpretation of the enhancement of the firing rate in the presence of colored noise. Since $\rq(\mui)\simeq \<\rw({\tilde\mu})\>_{\tilde \mu}$ (Eq. (\ref{eq:rquenched})), we expect the introduction of a colored noise to increase the LIF firing rate, $\rro(\mui)  > \rw(\mui)$, for those values of $\mui$ in which $\rw$ is convex, and vice-versa -- since $\<f(\mu)\>_\mu > f(\<\mu\>_\mu)$ is the definition of convex function. Precisely, since $\vc^{1/2}$ is the standard deviation of ${\tilde\mu}$, we expect $\rq(\mui)  > \rw(\mui)$ to hold for values of $\mui$ and $\vc$ such that $\mui$ is lower, in units of $\vc^{1/2}$, than the inflection point $\mu^*$ of the white-noise LIF neuron (satisfying $\d^2\rw(\mu)/\d\mu^2|_{\mu^*}=0$). Fig. \ref{fig:r_interpretation} shows that this is the case for both the quenched-noise and the Moreno approximations. 

The enhancement in firing rate therefore has a simple explanation in the large-correlation timescale regime. The input current signal $I(t)$ fluctuates within a timescale $\tauc$. Although those fluctuations are symmetric around the mean $\mui$, since the activation function of the white LIF model is non-linear, {\it and convex for low enough current $\mui$}, {\it the gain in firing rate for positive fluctuations, $\rw(\mui+\delta)-\rw(\mui)$ is larger than the loss for negative fluctuations $-\rw(\mui-\delta)+\rw(\mui)$}. Thus, for the firing rate to be enhanced by colored noise, {\it the LIF readout neuron must operate in a sufficiently low input current regime} (so that $\rw(\mui)$ is convex, given the relevant parameters $\sigmai,\taum$, see Sec. \ref{sec:discussion}). This effect is known as \emph{Jensen's force} \cite{buendia2019}.

\subsection{The quenched-noise approximation for the LIF neuron rate variance\label{sec:quenchedvar}}

The rate variance $\vro$ of a LIF neuron with white+coloured noise input has been analytically computed only in the $\tauc\to 0$ limit \cite{feng2000}, in the $\tauc\to\infty$ limit with no white noise \cite{schwalger2008}, for a perfect integrator \cite{lindner2004,middleton2003}, for conductance synapses \cite{salinas2000}, for binary-valued inputs \cite{salinas2002} or for oscillating, periodic input \cite{brunel2001}. To our knowledge, there are no closed analytical expressions for it (see also \cite{lindner2009}) \footnote{\cite{walter2021,walter2022} obtained approximated expression for the first-passage time distribution of an active Brownian particle (a non-Markovian random walker) for small $\nu_{\rm W}=\sigmac^2\taum/(\sigmai^2\tauc)$}. Here we derive an expression for the rate variance within the quenched-noise approximation. 

We first provide an expression for the {\it temporal correlation function} $\C(t) := \<\rhoro(t')\rhoro(t'+t)\>-\<\rhoro(t')\>\<\rhoro(t'+t)\>$ of a LIF neuron with white+colored input noise (Eq. (\ref{eq:VI_diffusion})) in the quenched-noise approximation (where $\rhoro(t):=\sum_j \delta(t-t_j)$ is the LIF spike density in a single realization, and $\<\cdot\>$ is the average over realizations of the noises $\xi,\eta$). The rate variance ${\rm Var}({\rm rate})=\Tc^{-2}{\rm Var}\left(\nro(\Tc)\right)$ relates to $\C$ as:

\begin{align}
\label{eq:variancecorrection}
{\rm Var}({\rm rate}) := \vro = \frac{\rro}{\Tc} + \frac{2}{\Tc^2} \int_0^{\Tc}\d t'(T-t')\C_{-}(t') 
\end{align}
where $\C_{-}(t)$ is the regular part of $\C(t) = \rro\delta(t)+\C_-(t)$. We define as well the {\it temporal distribution function} $g(t) := \rro^{-1}\<\rhoro(t')\rhoro(t'+t)\>$, i.e. the spike density at a distance $t$ from each spike, so that $\C(t)=\rro g(t)-\rro^2$.  

We compute $\gq(|t-t'|)$ in the quenched-noise approximation (Appendix \ref{sec:quenched_variance}) by using the facts that ({\it i}) the effective encoding current $\tilde\mu_t:=\mui+\sigmac\xi(t)$ obeys an Ornstein-Uhlenbeck process (Eq. (\ref{eq:I_diffusion})) with known stationary distribution $p(\tilde\mu_t)={\cal N}(\tilde\mu_t;\mui,\vc)$ and Green function $p(\tilde\mu_{t'}|\tilde\mu_t)$; ({\it ii}) in the quenched-noise approximation we have that $\rq\gq(t-t')\simeq \langle\langle\rw(\tilde\mu_{t})\rw(\tilde\mu_{t'})\rangle_{p(\tilde\mu_{t'}|\tilde\mu_{t})}\rangle_{p(\tilde\mu_t)}$. Thus, our derivation addresses the impact of pre-synaptic current fluctuations in the post-synaptic {\it temporal correlations}, beyond their impact in the post-synaptic mean firing rate. The equation $\rq\gq(t-t')\simeq \langle\langle\rw(\tilde\mu_{t})\rw(\tilde\mu_{t'})\rangle_{p(\tilde\mu_{t'}|\tilde\mu_{t})}\rangle_{p(\tilde\mu_t)}$ is the quenched-noise approximation to $\gq$, expected to be valid for resolution timescales $\ell$ satisfying $\taum\ll \ell\ll\tauc\ll\Tc$. In Appendix \ref{sec:quenched_variance} we show that, under the further conditions of small $\vc$ and low $\omega:=\exp(-|t-t'|/\tauc)$, one can derive an analytic expression for $\gq$, given by (see Appendix \ref{sec:quenched_variance}):

\begin{align} \label{eq:g_quenched_loww}
\rro \gq^{(\text{lo-}\omega)}(t-t')  &\simeq \rq^2 \left( 1 + e^{-|t-t'|/\tauc}\, Q\right) \qquad \text{low $\vc$, large $|t-t'|/\tauc$} 
\end{align}
where $Q$ is a constant that depends on the white LIF properties $\mui,\sigmai,\taum$, and on $\vc$ (see Eq. (\ref{eq:Q}) in Appendix \ref{sec:quenched_variance}).

\begin{figure}[h!]
\includegraphics[width=0.7\textwidth]{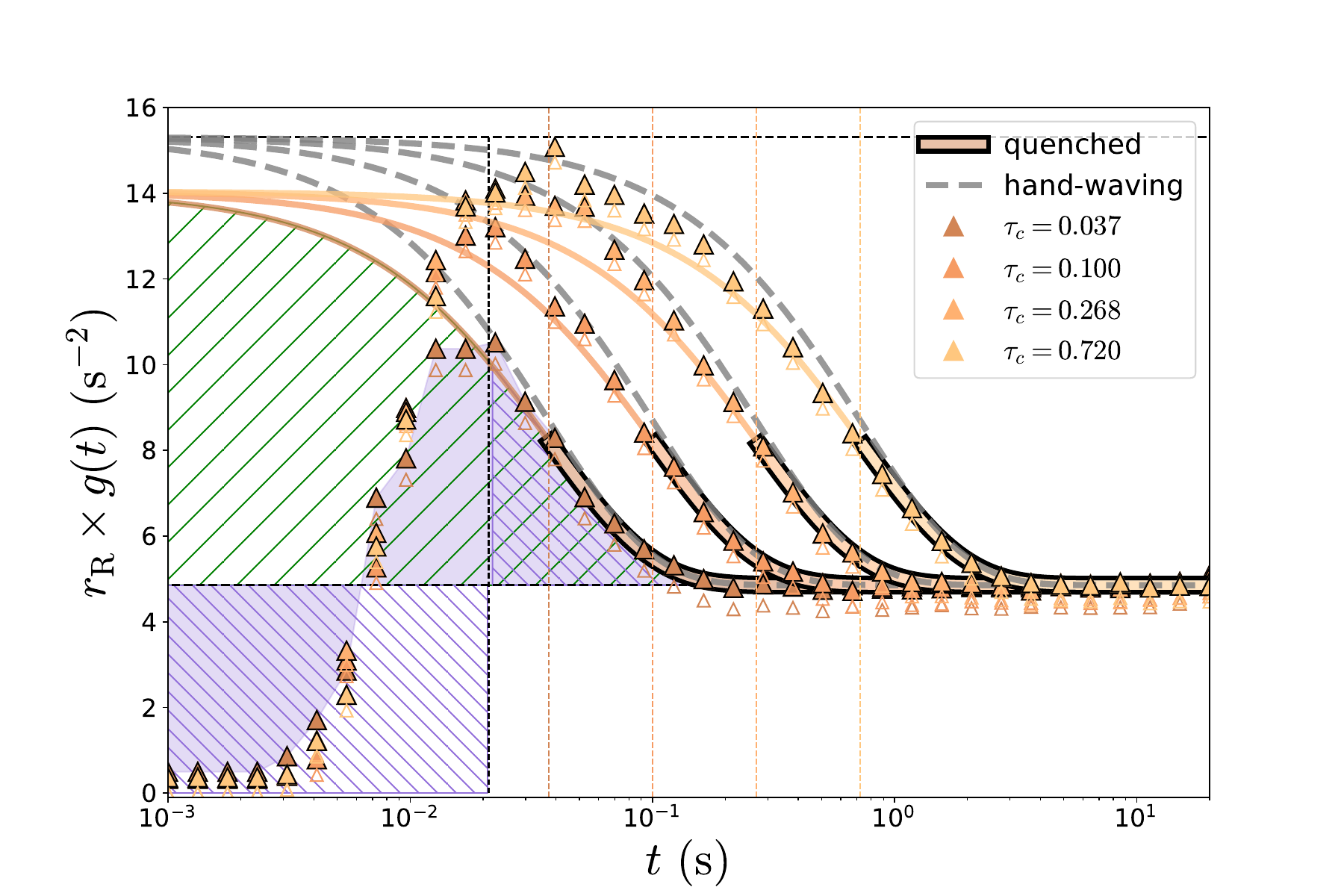}%
\caption{{\it Temporal distribution function $g(t)$ of readout spikes.} Different curves/colors correspond to different values of $\tauc$. The points correspond to the numerical estimation from the simulated spike trains (see Appendix \ref{sec:numerical}); the colored continuous curves correspond to the quenched approximation $\gq$ in (\ref{eq:g_quenched_loww}) (please, note that these curves become bordered by a black line for $t>\tauc$, in the regime in which they have been derived). Importantly, in order to correct the finite-NIS effect, we have shifted the numerical $g$ points by a quantity, $\tauc$-dependent but constant in $\mui$, which is $-\tilde\rro+\rq$, where $-\tilde\rro$ is the numeric estimation of the rate (which is underestimated, due to the finite-NIS effect). To show the effect of such a shift, we report as well the raw, non-shifted numerical estimations (small, void triangles). The gray dashed lines are the hand-waving expression for $g$, (\ref{eq:gansatz}) in Appendix. \ref{sec:handwaving}. The dashed and continuous horizontal lines mark, respectively, the values $\vq+\rq^2$ and $\rq^2$. The black vertical line signals $\mui^{-1}$, while the vertical colored lines mark the corresponding values of $\tauc$. It is interesting to note that the hand-waving approximations for $g$ seem to work slightly worst than $\gq$, as expected. The parameters of the simulation are: $\mui =  47.4$, $w=0.45$, $\sigmac =  30.593$, $\taum=0.005$, $\tauc=0.7196$. \label{fig:g_quenched}}
\end{figure}

The comparison with the numerical estimation of $g$ (Appendix \ref{sec:numerical}) reveals that Eq. (\ref{eq:g_quenched_loww}) captures the behavior of the distribution function for large $t$ (Fig. \ref{fig:g_quenched}). Although the quenched-noise approximation is expected to be valid for resolution timescales $\ell\gg\tauc$, $\gq$ reproduces well the numerical $g$'s even at small values of $t$ of the order of $\tauc$ (colored vertical lines in Fig. \ref{fig:g_quenched}) and even as small as the zero-noise LIF ISI $\mui^{-1}$ (black vertical line in Fig. \ref{fig:g_quenched}). 

\begin{figure}[h!]
\includegraphics[width=1.\textwidth]{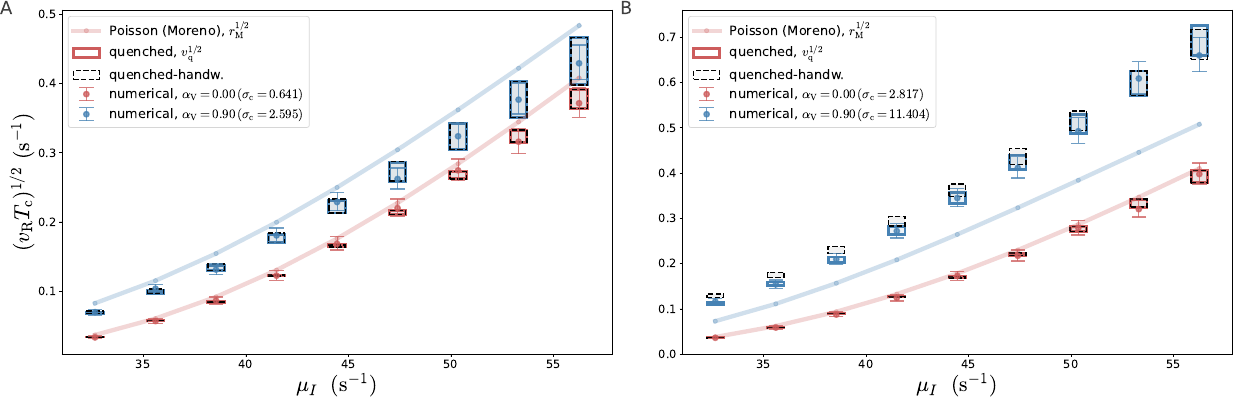}%
\caption{{\it Standard deviation of the firing rate in the unit time interval, ${\rm Std}({\rm rate})\,\Tc^{1/2} = (\vro\Tc)^{1/2}$, vs $\mui$.} Different panels correspond to two simulations with different values of $\tauc$ ($\tauc = 0.0052$ and $0.1$ for panels A and B respectively). Different curves in each panel correspond to different values of $\alphaV=0,0.9$: absence of noise correlations, and highly aligned noise correlations, respectively (see the value of $\sigmac$ in the legend). The points are the numerical estimations, and the error-bars are the confidence interval of the corresponding $\chi^2$ variance, with a probability $\alpha_\chi=0.9$ (see Appendix \ref{sec:numerical}). Continuous lines are the Poisson-quenched hypothesis $\vro=\rq/\Tc$; the filled, continuous-line boxes are our quenched-noise upper and lower bounds for the variance (\ref{eq:varrate_quenched}) (upper and lower limits of the box); the void, dashed-line boxes correspond to the upper and lower bounds in the hand-waving approximation (see Appendix \ref{sec:handwaving}). We observe that in absence of noise correlations when $\alphaV=0$, the Poisson approximation is relatively good, specially for low $\mui$. Only large values of $\sigmac$ induce positive corrections of the Poisson variance. The quenched-noise approximation captures both negative and positive deviations from Poissonian variance. The rest of the parameters are: $\taum = 0.005 $, $w = 0.45$, $N=20$, $\sigmaV = 1.104$.
\label{fig:rstd_quenched}}
\end{figure}

We exploit our expression for $\gq^{(\text{lo-}\omega)}$ to draw two approximations for the rate variance, depending on the value of a temporal cutoff $t_0$. 

The first approximation is an upper bound, computed from Eq. (\ref{eq:variancecorrection}) but taking $g(t)$ {\it to be equal to our low-$\omega$ expression for $\gq$ in Eq. (\ref{eq:g_quenched_loww}) for all times $0<t<\Tc$}. The area below this upper bound is the green, slash-hatched area in Fig. \ref{fig:g_quenched}, which is an upper bound for the integral of $g$ (shadowed violet area below the $g$ curve in the figure). This upper bound is given by $\vq \simeq \rro\Tc^{-1} + 2\Tc^{-2} \int_{0}^{\Tc}\d t'(T-t')\C_{-}^{({\rm low- }\omega)}(t')$.

The second approximation is a lower bound, computed from Eq. (\ref{eq:variancecorrection}) but taking $g(t)$ {\it to be equal to the low-$\omega$ expression for $\gq$ in Eq. (\ref{eq:g_quenched_loww}), only for a given temporal range $t>t_0$, while being equal to $0$ for $t\le t_0$}. The area below this lower bound is the blue, backslash-hatched area in Fig. \ref{fig:g_quenched}, which is a lower bound for the integral of $g$. This lower bound is given by $\vq \simeq \frac{\rro}\Tc^{-1} + 2\Tc^{-2} \int_{t_0}^{\Tc}\d t'(T-t')\C_{-}^{({\rm low- }\omega)}(t') -\rro^2\,\left(\frac{t_0}{\Tc}-\frac{t_0^2}{2\Tc^2}\right)$.

As seen in  Fig. \ref{fig:g_quenched}, in the upper bound approximation we overestimate the refractory period of $g$ (in which $g<\rq<\gq^{(\text{lo-}\omega)}$) for low enough times $t<t_0$. In fact, our approximation is not expected to capture the behavior of $g$ in timescales of the order of $\tauc$ or lower. Vice-versa, the lower bound approximation underestimates the low-$t$ behavior of $g$, since $g$ is set to its minimum possible value, $g(t)=0$ for $t<t_0$. For the lower bound to work, the cutoff time $t_0$ must be close to the maximum of the function $g$. As a natural guess, we choose it to be $t_0=\mui^{-1}$, the zero-noise LIF ISI (see Fig. \ref{fig:g_quenched}). 

In short, taking $\C_{-}^{({\rm low- }\omega)}(t'):=\rq\gq^{(\text{lo-}w)}-\rq^2$, with $\gq^{(\text{lo-}w)}$ in Eq. (\ref{eq:g_quenched_loww}), our upper- and lower-bound expressions for the rate variance take the form:

\begin{subequations}
\label{eq:varrate_quenched}
\begin{align} 
\vq   &\simeq \frac{\rro}{\Tc}+2\rro^2\, Q \frac{\tauc}{\Tc} \qquad &\text{upper bound} \label{eq:varrate_quenched_upper} \\
\vq  &\simeq \frac{\rro}{\Tc}+2\rro^2\, Q \frac{\tauc}{\Tc}\left( e^{-\frac{t_0}{\tauc}}-e^{-\frac{\Tc}{\tauc}}\right) -\rro Q \frac{\tauc^2}{\Tc^2} \left[ e^{-\frac{t_0}{\tauc}}\left(\frac{t_0}{\tauc}+1\right) -  e^{-\frac{\Tc}{\tauc}}\left(\frac{\Tc}{\tauc}+1\right)\right)+  &\nonumber \\
	&-\rro^2\left(\frac{t_0}{\Tc}-\frac{t_0^2}{2\Tc^2}\right) \qquad &\text{lower bound} 
.
\end{align}
\end{subequations}
These approximations capture the rate variances in a broad range of $\mui$ and $\tauc$ parameters (see Fig. \ref{fig:rstd_quenched} and Appendix \ref{sec:alternativefigures}). In particular, the quenched-noise approximation to the variance in Eq. (\ref{eq:varrate_quenched}) captures both the negative deviations from Poissonian noise $\text{Var}(\text{rate})=\rro/\Tc$ (that originate from the refractory period for $t<t_0$, $\C(t)<0$), and the over-Poissonian deviations (originating from the high-$t$ regime, in which $\C(t)>0$).  

Importantly, in Appendix \ref{sec:handwaving}, we develop a further approximation to the firing rate variance, that we will refer to as \emph{hand-waving approximation}, and which admits a straightforward interpretation of the over-Poissonian enhancement. In the {hand-waving approximation} approximation, the expressions for the temporal distribution function and for the rate variance are as in the above equations, for a different value of the constant $Q$, $Q^{({\rm hand})}:=\uq/\rq$, where $\uq$ is, in analogy with Eq. (\ref{eq:rquenched_analytical}), the variance of the white-noise LIF activation function over the stationary distribution of $\mui$'s, $\uq:= \<\rw^2(\tilde\mu)\>_{{\cal N}(\tilde\mu;\mu_I,\vc)}-\rq^2$. Despite the simplicity of its demonstration, the hand-waving approximation for the rate variance is almost as accurate as its more accurate variant derived in Appendix \ref{sec:quenched_variance} (Fig. \ref{fig:g_quenched}). As in the previous subsection, the above analytic expressions for $\gq,\vq$ require, besides the validity condition of the quenched-noise approximation $\tauc\gg\taum$, the further condition of small $\vc$ (see details in \ref{sec:quenched_variance}).

\section{Noise correlations and SNR enhancement in the encoding-readout model \label{sec:valentemodelsolution}}


In this section, we study the behavior of the SNR, Eq. (\ref{eq:SNRsqrt}) as a function of the noise correlation stationary variance $\vc$ (parametrised by $\alphaV$) and timescale $\tauc$. Crucially, for the comparison between numerical estimations and theoretical approximations for different values of $\tauc$, we keep {\it the noise stationary variance $\vc$ constant} (where, throughout this section, by {\it constant} we mean {\it independent of $\tauc$}), which amounts to consider {\it a $\tauc$-dependent count variance} $\sigmac^2=2\tauc\vc$ (see \cite{schwalger2008}). Taking $\vc$ constant is the suitable scenario to explore the $\tauc\gg\taum$ limit without loss of generality (see Appendix \ref{sec:constantvc}). 

The fact that $\sigmac$ depends on $\tauc$ for constant $\vc$ is an arbitrary choice of the correlated noise amplitude that we probe in our simulations for each value of $\tauc$. Since $\tauc,\sigmac$ do not vary independently (in the so called {\it first simulation set} analyzed in this section, see Appendix \ref{sec:numerical}), we decided to set the spike count variance $\sigmac$ to depend on the independent variable $\tauc$ as $\sigmac=(2\tauc\vc)^{1/2}$ (being $\vc$ a $\tauc$-independent proportionality constant), rather than taking the alternative option (i.e., $\vc$ depending on $\tauc$ as $\vc=\sigmac^2/2\tauc$, being $\sigmac$ a $\tauc$-independent proportionality constant) because this scenario makes more sense when studying the large-$\tauc$ limit (see Appendix \ref{sec:constantvc}).


We provide an alternative analysis of the SNR in a complementary probing variables scheme, in which we fix $\tauc$ and vary $\sigmac$ (instead of fixing $\vc$ and varying $\tauc$), in Appendix \ref{sec:varyingsigmav}.


\subsection{Use of the analytical approximations \label{sec:analyticalapps}}

Below in subsection \ref{sec:SNRvstauc}, we present a study of the behavior of the SNR as a function of $\tauc$ by comparing the numerical simulations of $\rro^{(s)}$, $\vro^{(s)}$ and of the SNR, with the following analytical approximations. 

In the region $\tauc\ll\taum$, the only analytical solution that we need in order to explain the numerical estimations of the SNR, is the Moreno et al solution \cite{moreno2002} for the rate, $r_{\rm M}$, in the $\tauc\ll\taum$ limit (Eq. (10) of \cite{moreno2002}). Two different conditions make possible to approximate the $\rro$, $\vro$ and SNR, disposing only of this analytical expression. First, because in the limit $\tauc\ll\taum$, and for fixed $\vc$, $\sigmac$ is small enough for our simulations to be in the validity regime of the Moreno solution (low $\alpha_{\rm M}=\sigmac^2/\sigmai^2$) and, hence, the Moreno et al solution for $\tauc\ll\taum$ reproduces correctly the numerical simulations for $\rro$ (see Fig. \ref{fig:numSNRsqrt}-A). Second, because, for low enough $\tauc$, hence low enough $\sigmac$, we can neglect the over-Poissonian corrections to the variance $\vro$ in Eq. (\ref{eq:variancecorrection}) (see the $\alphaV=0$ case in Fig. \ref{fig:rstd_quenched}). In sum, we simply take $\vro\Tc\simeq\rro \simeq r_{\rm M}$, where $r_{\rm M}$ is the Moreno et al approximation (we will refer to this approximation for the readout rate variance as the {\it Poisson-Moreno approximation}). We construct our analytical expression for the SNR with these expressions for $\rro$ and $\vro$.

For $\tauc\gg\taum$, none of these conditions hold any more. Although in principle we dispose of the alternative Moreno et al solution for $r_{\rm M}$, valid for $\tauc\gg\taum$ (that we report in Eq. \ref{eq:r_moreno}), it is no longer possible to exploit this solution to explain the behavior of the SNR for large $\tauc$ and constant $\vc$. Indeed, for large enough $\tauc$, hence $\sigmac$, the simulation parameters lie outside the validity limit (low $\alpha_{\rm M}$) of the Moreno solution and, consequently, $r_{\rm M}$ differs from the numerical estimation $\tilde\rro$. Second, and more importantly, the over-Poissonian corrections of the rate variance cannot be neglected any longer. Therefore, we cannot approximate $\rro$ with $r_{\rm M}$, and we cannot use any longer the relation $\vro\Tc\simeq\rro \simeq r_{\rm M}$. The quenched-noise approximation for $\tauc\gg\taum$ provides us with a solution for both problems. First, $\rq$ captures well the behavior of $\rro$ also in this regime. Second, the quenched-noise approximation provides us with an equation for the rate variance as well (Eq. (\ref{eq:varrate_quenched}), or the simpler hand-waving quenched-noise approximations Eqs. (\ref{eq:uq},\ref{eq:ratevariance_quenched})). We illustrate these facts in the next subsection, and particularly in Fig. \ref{fig:numSNRsqrt}, in which we compare the Poisson-Moreno approximation with the quenched-noise approximation. 



\subsection{Signal enhancement, noise enhancement and SNR enhancement \label{sec:SNRvstauc}}

\subsubsection{Signal enhancement}

We now present the numerical results for the readout mean rate enhancement $\Delta\rro:=\rro^{(+)}-\rro^{(-)}$ versus $\tauc$, for two values of the fraction of shared noise $\alphaV$. We refer to Appendix \ref{sec:numerical} for the numerical range of probed parameters, and to Appendix \ref{sec:alternativefigures} for different choices of such parameters. 

\begin{figure}[h!]
\includegraphics[width=0.7\textwidth]{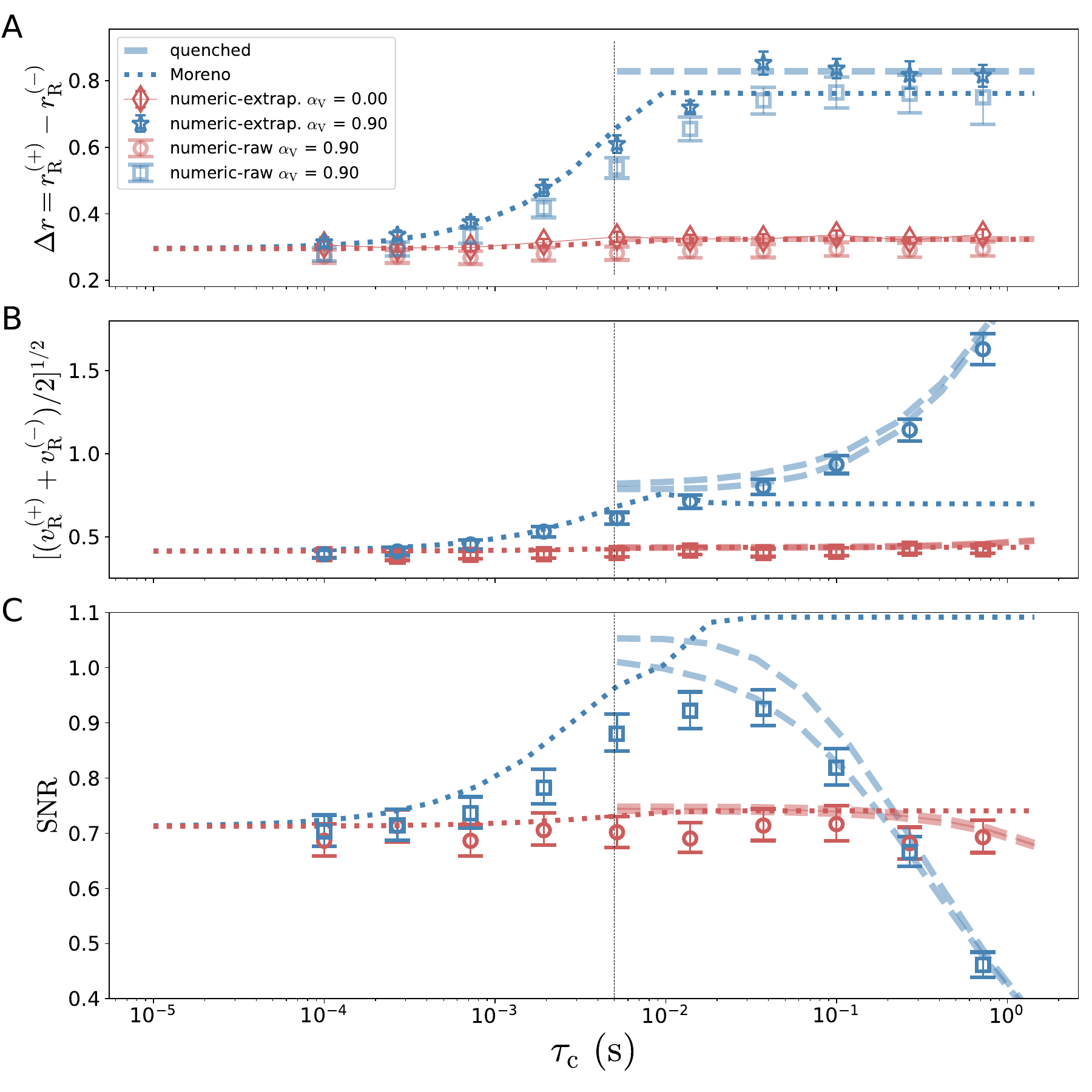}%
\caption{{\it Mean rate enhancement, rate standard deviation and SNR, as a function of $\tauc$.} (A) Readout mean rate increment $\Delta\rro=\rro^{(+)}-\rro^{(-)}$ vs $\tauc$. Different curves/colors in each panel correspond to both values of $\alphaV=0,0.9$. Stars correspond to the infinite-NIS extrapolated procedure (see Appendix \ref{sec:numerical}) (and the error-bars indicate the uncertainty of the regression), while the semi-transparent squares correspond to the raw numerical estimations $\tilde\rro$ for the highest NIS (and the error-bars are $(\tilde\vro\Tc)^{1/2}$). The dotted lines are the interpolated Moreno perturbative solution in \cite{moreno2002}, using an interpolating time $t_{\rm in}=2.5\taum$. The dashed line (shown for $\tauc>\taum$ only) is the quenched-noise analytical expression, Eq. (\ref{eq:rquenched_analytical}). The vertical dotted line signals the value of $\taum$. (B): Standard deviation of the stimulus-averaged readout firing rate in the unit time interval, $[(\vro^{(+)}+\vro^{(-)})/2]^{1/2}\Tc^{1/2}$ vs $\tauc$. Here, only the raw estimations of the variance are shown. The error-bars are the $\chi^2$ confidence interval with a probability $\alpha_\chi=0.95$. The pairs of dashed lines (shown for $\tauc>\taum$ only) correspond to the lower and upper bounds for the variance in the quenched-noise approximation, Eq. (\ref{eq:varrate_quenched}). We attribute the slight disagreement between the numerical estimations and $\vq$ to a finite-NIS underestimation of the numerical variance. (C): Readout SNR in the unit time interval (of which panels A and B are, respectively, the numerator and the denonminator), vs $\tauc$. The numerical points correspond to the raw estimators for both the numerator and the denominator. We remind that $\sigmac$ varies with $\tauc$ according to Eq. (\ref{eq:dictionary_sigmac}). The simulation parameters are: $(\mu_I^{(+)},\mu_I^{(-)}) = (32.65,41.5)$, $w = 0.45$, $\N=20$, $\vV =  9.8$, $\taum=0.005$.  \label{fig:numSNRsqrt}}
\end{figure}

First, we observe that introducing noise correlations enhances the firing rate difference for all the considered values of $\tauc$ (Fig. \ref{fig:numSNRsqrt}-A). This effect is not simply equivalent to {\it the LIF neuron firing rate enhancement in the presence of noise correlations} (see Fig. \ref{fig:r_quenched}), but it corresponds to an increase of the mean firing rate {\it difference} $\Delta\rro$, in the presence of noise correlations. 

For low enough $\tauc$ (and constant $\vc=\sigmac^2/(2\tauc)$), the colored-noise amplitude $\sigmac$ vanishes, so that the numerical $\Delta\rro$ is consistent with that of the white noise LIF $\Delta\rw$. For larger but low enough values of $\tauc$, the Moreno perturbative approximation to $\rro$ for low $\tauc/\taum$ holds as expected. 
For large values of $\tauc\gg\taum$, we observe a disagreement between the numerical estimation and the Moreno  approximation, and an agreement between the numerical estimation and the quenched-noise approximation, as expected (see Sec. \ref{sec:quenchedmean} and Appendix \ref{sec:alternativefigures}) \footnote{Notice that both the Moreno solution and the quenched-noise solution exhibit a constant behavior in $\tauc$ for large $\tauc$ (and constant $\vc$), as expected from Eqs. (\ref{eq:rquenched},\ref{eq:rquenched_analytical}).}.
For intermediate values of $\tauc\lesssim\taum$, although outside the validity limits of both approximations, we observe a nice agreement between the numerical $\Delta\rro$ and the interpolation of both Moreno solutions \cite{moreno2002} (although this agreement is sensitive to the interpolating time, here set to $t_{\rm in}=2.5\taum$). 

Thus, for fixed values of the parameters $\mui^{(\pm)},\taum,w$, increasing noise correlations leads to an increase in the readout neuron firing rate difference $\Delta \rro$, quantitatively understood within the Moreno et al solution for low $\tauc$ and within the quenched-noise approximation for large enough $\tauc$.

As a methodological note, the numerical simulations match the analytical predictions only when applying the finite-Number of Integration Steps (NIS) correction to the numerical solution of the stochastic differential equation, Eq. \ref{eq:VI_diffusion}) (see Appendix \ref{sec:numerical}). 

\subsubsection{Rate variance enhancement}

We now consider the denominator of the SNR, i.e. the stimulus-averaged variance of the readout firing rate, $(\vro^{(+)}+\vro^{(-)})/2$. First, for low enough $\tauc$ we observe an agreement between $\tilde\vro$ and the Poisson-Moreno approximation (Fig. \ref{fig:numSNRsqrt}-B). We attribute the small deviations to a numerical underestimation of the variance due to the finite-NIS effect (see Appendix \ref{sec:varunderestimation}). We  notice that in this case we do not employ the NIS-corrected estimation for the rate variance, since it exhibits larger statistical errors. 
Second, the over-Poissonian increment of $\tilde\vro$ for large values of $\tauc$ in the presence of noise correlations can only be explained, within the quenched-noise approximation, as the result of the integration of the temporal correlation function $\gq$ (see Eq. (\ref{eq:varrate_quenched}), and Appendix \ref{sec:handwaving} for a further, intuitive explanation of the enhancement of $\tilde\vro$ for large $\tauc$). 

Overall, both approximations, the Poisson-Moreno and the quenched-noise approximations describe the numerical behavior of $\tilde\vro$ in the $\tauc\ll\taum$ and $\tauc\gg\taum$ regimes, respectively. For large $\tauc$, the variance $\vro$ increases with $\tauc$ (see the second term in Eq. (\ref{eq:variancecorrection}) and Eq. (\ref{eq:varrate_quenched})), while the mean rate $\rro$ stays constant with $\tauc$ (as expected, see Eq. (\ref{eq:rquenched_analytical})).

\subsubsection{SNR enhancement}

From the results of the previous subsections, it follows that the SNR exhibits a non-monotonic behavior with $\tauc$ (Fig. \ref{fig:numSNRsqrt}-C). 
For large enough $\tauc$, noise correlations lower the readout SNR below its value in absence of noise correlations. For high values of $\tauc$, the increment of the SNR numerator, $\Delta\rro$, depends only on $\vc$, so that it stays constant in $\tauc$ (see Fig. \ref{fig:numSNRsqrt}-A and Sec. \ref{sec:quenchedmean}). Instead, the rate variance increases with $\tauc$ due to the over-Poissonian correction (which increases with \emph{both $\tauc$ and $\vc$}, see Eq. (\ref{eq:varrate_quenched_upper}) and Appendix \ref{sec:handwaving}). Beyond a certain value of $\tauc$, the increment in the denominator becomes prevalent, thus leading to a decrement of the SNR. 
For sufficiently small $\tauc$, when the enhancement in the rate variance $(\vro^{(+)}+\vro^{(-)})/2$ is not enough to overcome that of $\Delta\rro$ in the SNR numerator, we observe that the noise correlations induce an enhancement of the SNR. This constitutes the apparently paradoxical effect according to which  encoding correlations that lower the encoding-SNR,  enhance the readout SNR. 

We have also compared the readout SNR with the \emph{encoding SNR} (see Eq. (\ref{eq:snrencoding})) (see Fig. \ref{fig:numSNRsqrt_alt} in Appendix \ref{sec:alternativefigures} and Fig. \ref{fig:numSNRsqrt_semicorrected} in Appendix \ref{sec:varunderestimation}). By construction, the encoding SNR decreases with the noise correlation amplitude, and it may actually become lower than the readout SNR (please, see a discussion on this point in Appendix \ref{sec:encodingsnr}). 

For details regarding the small theoretical-numerical disagreement for low $\tauc$ in Fig. \ref{fig:numSNRsqrt}-C, we refer to the Appendix \ref{sec:varunderestimation}, where we attribute it to the finite-NIS effect.


In sum, the SNR presents a non-monotonic behavior with $\tauc$, that is quantitatively explained by the quenched-noise approximation. 

\subsection{The conditions for enhanced-by-consistency SNR to occur \label{sec:conditions}}

\begin{figure}[h!]
\includegraphics[width=0.7\textwidth]{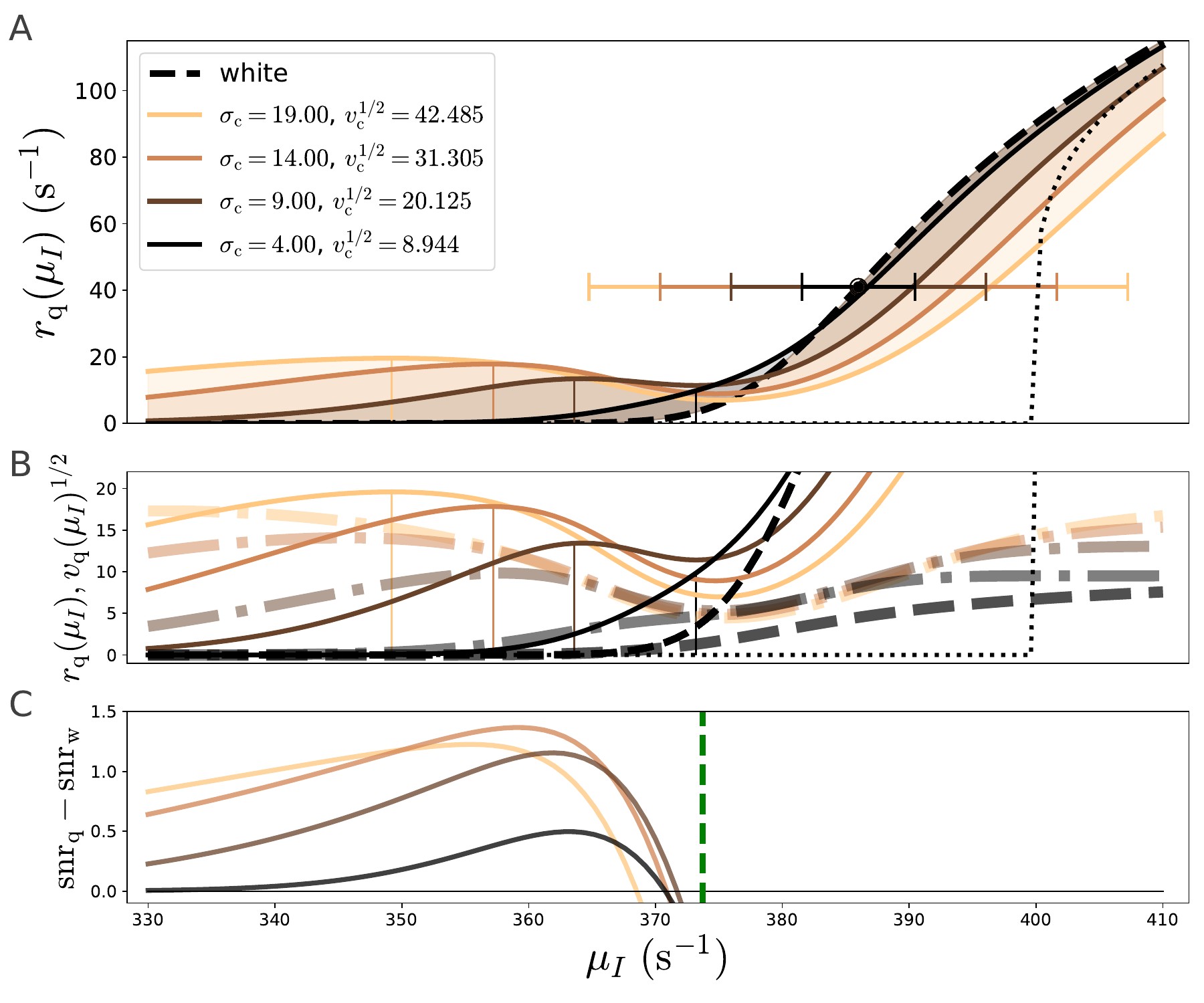}%
\caption{{\it Readout rate mean, variance and excess of SNR for negligible values of $\rro^{(-)}$.} (A) Mean firing rate $\rro(\mui)$ of the white-noise LIF (dashed line), and the quenched-noise colored-noise LIF $\rq(\mui)$ (continuous lines) for various values of $\sigmac$ (or $\vc=\sigmac^2/(\tauc)$, different curves). For each $\mui$, the value of $\sigmai$ is set to $\sigmai=(w\mui)^{1/2}$, with $w=10^{-3}$. As in Fig. \ref{fig:r_interpretation}, the circle and error-bars indicate the inflection point and $\vc^{1/2}$, while the black dotted line indicate the activation function of the deterministic LIF $r_0$. (B) A zoom of the panel A, in which we report as well the quenched-noise rate standard deviation, $\vq^{1/2}$ (dashed-dotted lines). The vertical lines indicate the maximum of each $\rq(\mui)$ curve. (C) The difference between the SNR $[\rq(\mui)-\rq(\mui^{(-)})]/[(\vq(\mui)+\vq(\mui^{(-)})/2]^{1/2}$ (taking a low $\mui^{(-)}=260$, actually equivalent to neglect $\rq^{(-)}$) in the presence of colored noise, minus the white-noise SNR, $=[\rw(\mui)-\rw(\mui^{(-)})]/[(\vw(\mui)+\vw(\mui^{(-)}))/2]^{1/2}$, assuming a Poissonian rate variance, $\vw=\rw$. The vertical dashed line corresponds to our upper bound $\hat\mui$. Rates and variances are calculated with a $\sigmai^2=w\mui$ that varies with $\mui$ for fixed $w$. The rest of the model parameters are: $\tauc = 0.1$, $\taum = 0.0025 $, $w = 0.001 $, $\mui^{(-)} = 260$. \label{fig:snr_vs_muI}}
\end{figure}

In this section we address \emph{for which general values} of the encoding and decoding parameters we expect the SNR enhancement to occur in the presence of noise correlations.
The quenched-noise approximation for the rate mean and variance allows us to draw necessary conditions on the NERM parameters, at least for sufficiently large $\tauc$. Our conclusions are drawn from the behavior of $\rq,\vq$ as a function of $\mui$ and $\sigmac$ (Fig. \ref{fig:snr_vs_muI}). In this section, we use the quenched-noise approximation and, in particular, the hand-waving expression for the rate variance, Eq. (\ref{eq:ratevariance_quenched}) in Appendix \ref{sec:quenched_variance}, which is already a good approximation for the behavior of the rate variance (Fig. \ref{fig:rstd_quenched}). 

{\bf Conditions on $\mui^{(\pm)}$ and dependence on $\sigmac$.} In Sec. \ref{sec:intuitive} we said that, for the colored noise to enhance the firing rate $\rro$, the value of $\mui$ should be safely lower than the inflection point of the LIF (circle in Fig. \ref{fig:snr_vs_muI}-A). Given the parameters of the white LIF $\taum,\sigmai$, the inflection point occurs at the value of $\mui$ for which $g_1^2+g_1'=0$, where $g_1=\rw'/\rw$. The inflection point $\mui^*$ of $\rw(\mui)$ is also the value of $\mui$ below which $\rq>\rw$ for low enough $\vc$ (indeed, expanding $\rq$ in (\ref{eq:rquenched_analytical}) to $o(\vc)$, we get that $\rq\ge\rw$ for $\mui\le\mui^*$, v. Fig. \ref{fig:snr_vs_muI}). We can proof (see Appendix \ref{sec:necessaryconditions}) that the condition $\mui^{(+)}<\mui^*$ is actually a necessary condition for the SNR to be enhanced in the presence of noise correlations in the NERM model, at least for large enough $\tauc\gtrsim\taum$. In essence, for the SNR to increase in the presence of noise correlations, it is necessary that the numerator $\Delta\rro$ increases (which is not sufficient, since the denominator could increase too much). Furthermore, a necessary condition for $\Delta\rro$ to increase is, in its turn, that $\rrop$ increases. Therefore, for the SNR to increase with $\alphaV>0$, $\rrop$ must increase, which implies $\mui<\mui^*$.

\begin{figure}[h!]
\includegraphics[width=1.\textwidth]{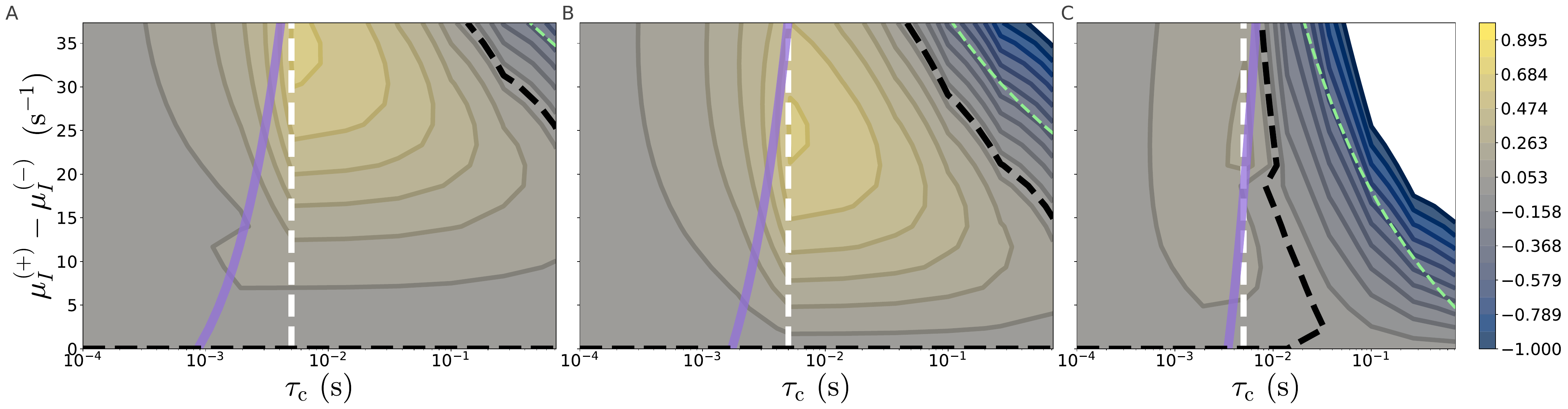}%
\caption{{\it Contour plot of the excess of SNR ($\delta_{\rm SNR}$) with respect to the absence of noise correlations, versus $\tauc$ and $\Delta\mui$.} Different panels correspond to different values of $\mui^{(-)}=10,20,40$ (A, B, C, respectively). The theoretical SNR has been calculated according to the Poisson-Moreno (low $\alpha_{\rm M}$) and quenched-noise (large $\alpha_{\rm M}$) approximations. The thick purple line signals the threshold $\alpha_{\rm M}=0.75$ that we have used to select the approximation: above and at the left of the purple line, we use Poisson-Moreno, while below and at the right, we use the quenched-noise approximation. The dashed green line indicates the the parameter values for which $\mui^{(+)}<\hat\mui$ (a necessary condition for the SNR to increase, assuming a negligible $\mui^{(-)}$). The white dashed line indicates $\taum$, while the dashed black line indicates the $\delta_{\rm SNR}=0$ contour. We remind that $\sigmac$ varies with $\tauc$ according to Eq. (\ref{eq:dictionary_sigmac}). The rest of the model parameters are: $\taum= 0.05$, $\alphaV= 0.9$, $\N=20$, $\sigmaV= 1.104$, $w= 0.45$. \label{fig:theoreticalSNR_vs_muI}}
\end{figure}

In Appendix \ref{sec:necessaryconditions} we also show that this upper bound to $\muip<\mui^*$ can be improved by taking into account the increase in the readout rate variance, i.e., by imposing that, beyond the numerator, the SNR itself should increase in the presence of noise correlations. With the analytic solution at hand, the condition $\text{SNR}_{\rm q} \ge \text{SNR}_{\rm w}$ implies that (see Appendix \ref{sec:necessaryconditions}) $\mui^{(+)}<\hat\mui<\mui^*$, where $\hat\mui$ is the solution of the transcendental equation $g_1'+g_1^2 = 2\tauc\rw g_1^2$. Although  $\hat\mui$ has been derived for low enough values of $\vc$ and $\muim$ (see Appendix \ref{sec:necessaryconditions}), it actually works as an upper bound for the value of $\mu$ below which we observe the SNR enhancement in all our simulations --see the green vertical line in Fig. \ref{fig:snr_vs_muI}. See also Figs. \ref{fig:theoreticalSNR_vs_muI},\ref{fig:snr_vs_muImin},\ref{fig:snr_vs_muImin350} and Appendix \ref{sec:othervariables}. 

Again, this necessary condition is not sufficient to guarantee that $\text{SNR}_{\rm q} > \text{SNR}_{\rm w}$. Too high values of $\vc$, or non-negligible values of $\mui^{(-)}$, hence of $\rro^{(-)}$, may make $\text{SNR}_{\rm q} < \text{SNR}_{\rm w}$. Let us, first, focus on the dependence on $\muim$. Mind that, at least for low enough $\sigmac$,  $\vro^{(-)}$ is an increasing function of $\mui^{(-)}$ (see Fig. \ref{fig:snr_vs_muI}) and, therefore, the SNR denominator enhancement increases with  $\muim$. The larger the value of $\sigmac$, the lower the values of the (leftmost) inflection points of $\rro$ and $\vro$ (see Fig. \ref{fig:snr_vs_muI}).  This implies that the larger $\sigmac$, the lower the value that $\mui^{(-)}$ should take for the SNR to increase (as can be seen in Fig. \ref{fig:snr_vs_muImin} in Appendix \ref{sec:SNRvsmuImin}). 

We now focus on the dependence on $\sigmac$. For large enough values of $\sigmac$ (or $\vc$), the SNR denominator increases faster with $\sigmac$ than the numerator $\Delta\rro$, because of the over-Poissonian correction to the variance, cf. the second term in Eq. (\ref{eq:varrate_quenched_upper}), or Eq. (\ref{eq:ratevariance_quenched}) in Appendix \ref{sec:handwaving}. This is the origin of the non-monotonous dependence of the SNR on the noise amplitude (see as well Appendix \ref{sec:varyingsigmav} for an explicit analysis of this). 

In sum, in order for the SNR to increase in the presence of noise correlations and for large enough $\tauc\gtrsim\taum$, it is necessary that: (c1) $\mui^{(+)}<\hat\mui$ is low enough, and (c2) both $\sigmac$ and $\mui^{(-)}<\hat\mui^{(-)}$ are low enough, with the upper bound $\hat\mui^{(-)}$ being a decreasing function of $\sigmac$. 

{\bf Dependence on other variables.} These considerations are enough to understand the qualitative dependence of the SNR with respect to the rest of the LIF parameters. We exemplify such dependencies in Fig. \ref{fig:theoreticalSNR_vs_muI} and in Figs. \ref{fig:theoreticalSNR_vs_taum},\ref{fig:theoreticalSNR_vs_wgain},\ref{fig:theoreticalSNR_vs_alphaV} in Appendix \ref{sec:othervariables}. In general, we observe a non-monotonous behavior in both $\tauc$ and $\Delta\mui$. The first one is the behavior that we described in Sec. \ref{sec:SNRvstauc}. The non-monotonous behavior in $\Delta\mui$ occurs since, for low input rate increment, the readout rate increment is negligible, while for too large increment $\mui^{(+)}$ approaches $\hat\mui$, and we loose the necessary condition for the SNR to increase. 

In Fig. \ref{fig:theoreticalSNR_vs_muI} we see how the SNR contour plots change with the value of $\mui^{(-)}$. For low enough $\mui^{(-)}$, the signal increment is essentially given by $\rro^{(+)}$, and the picture is as in Fig. \ref{fig:snr_vs_muI}: increasing $\mui^{(-)}$ and $\Delta\mui$ by the same quantity, leads to the same value of the SNR. The consequence is that increasing $\mui^{(-)}$ has the same effect of a shift in the y-axis (Figs. \ref{fig:theoreticalSNR_vs_muI}-A,B). However, beyond a certain value of $\mui^{(-)}$, $\rro(\mui^{(-)})>\rw(\mui^{(-)})$ is no longer negligible, and increasing $\mui^{(-)}$ amounts to reduce $\Delta\rro$, to increase $\vro^{(-)}$ and, consequently, to a reduction of the SNR (Fig. \ref{fig:theoreticalSNR_vs_muI}-C, and Fig. \ref{fig:snr_vs_muImin} in Appendix \ref{sec:SNRvsmuImin}).

With similar arguments one can explain the qualitative dependence of the SNR on $\taum$, $w$ and $\sigmac$ (see Appendix \ref{sec:othervariables}).

\subsection{Further analyses of the SNR in the NERM \label{sec:further}}

{\bf Dependence on $\N$.} For completeness, we have carried out two further analyses regarding the behavior of the SNR in the NERM in the diffusion approximation. The first one, in Appendix \ref{sec:Nscaling} addresses the behavior for large $\N$, whenever the single encoding neuron parameters $\mu_i,\sigmaV_i$ are naturally rescaled with $\N$ as $\mu_i=\mu_i^{(\N_0)} \N_0/\N$, $\sigmaV=\sigmaV^{(\N_0)} \N_0/\N$. In this scenario, the NERM in Eq. (\ref{eq:VI_diffusion}) is expected to become independent of $N$, as the only relevant parameters $\mui,\sigmai,\sigmac$ governing the encoding signal in the NERM take the form (large $N$, see Eq. \ref{eq:dictionary}): $\mui= w \mu_i^{(\N_0)}$, $\sigmai^2=w^2\mu_i^{(\N_0)}$, $\sigmac\simeq w\sigmaV^{(\N_0)} \alphaV$. The numerical simulations in Appendix \ref{sec:Nscaling} are consistent with this picture. 

{\bf Varying $\sigmac$ at constant $\tauc$.} The second complementary analysis (see Appendix \ref{sec:varyingsigmav}) regards the behavior of the SNR in the NERM for fixed $\tauc$ and varying $\sigmaV$ (varying, equivalently, $\sigmac=w\sigmaV N(1+(N-1)\alphaV^2)^{1/2}$). Besides the numerical underestimation of the rate variance induced by the finite-NIS effect, the theory provides us with a correct picture of the non monotonous behavior of the SNR with $\sigmac$, as it did in the above analysis for constant $\vc$ and varying $\tauc$.

\section{Discussion\label{sec:discussion}}

Here we used an Neural Encoding and Readout Model (NERM) of neural activity that combines information encoding in neural populations with a biophysically plausible model of neural readout, to investigate the conditions under which correlations between the temporal activity of input neurons can enhance the transmission of neural information downstream, even if they decrease the amount of information encoded in the input. Addressing this problem is highly relevant in neuroscience because it establishes theoretical grounds for the empirical and seemingly paradoxical discovery that higher correlations in neural activity aid performance at the behavioral level, even when they decrease the information at the encoding stage. 

We formally analyzed  a mechanism of enhanced-by-consistency readout\cite{valente2021}, in which the readout is implemented as a single neuron with an output coincidence-detection nonlinearity, and studied it as a function of the interplay between the encoding and readout timescales. Previous work \cite{valente2021} analyzed this NERM by numerical integration, with $\N=2$ encoding neurons. Here, we extend this work by providing an approximated analytic solution of a more general version of the model, which helps assess the conditions under which the apparently paradoxical effect of lower encoding but higher transmission of neural information occurs. 
Previous works have addressed, with analytical approximations, the effect of temporal input correlations on the \emph{mean} output rate of a LIF neuron \cite{brunel1998,fourcaud2002,moreno2002}. 
Here, we provided alternative, approximated analytical expressions not only for the mean output firing rates (that actually generalize those in \cite{moreno2002}), but also for the SNR of information transmission. This constitutes a key extension to evaluate the implications of correlated activity on neural information processing. 

We found that the non-linearity of the LIF activation function induces asymmetric fluctuations of the output firing rate, following symmetric \emph{input} fluctuations. The quenched-noise approximation developed here, valid for large or moderate values of $\tauc \gtrsim \taum$ and for low stationary noise variance $\vc$, provides us with the conditions under which these asymmetric fluctuations allow information-limiting noise correlations to enhance the readout SNR. In particular, the following necessary conditions for the mean input current must be satisfied (see Sec. \ref{sec:conditions}): the readout neuron must operate in a fluctuation-driven regime for low enough $\mui^{(+)}<\hat\mui$, in agreement with the results of \cite{valente2021}, with $\hat\mui(\vc,\taum)$ being a decreasing function of $\vc$ and $\taum$; the stimulus-specific difference in {\it encoding} firing rates $\mui^{(+)}-\mui^{(-)}$ is high enough (with $\mui^{(-)}<\hat\mui^{(-)}(\vc,\mui)$, being $\hat\mui^{(-)}(\vc,\taum$) a decreasing function of $\vc$ and $\taum$). These conditions on the input firing rates confirm  previous conclusions that input correlations enhance readout information from a postsynaptic neuron  when the readout integration time constant is short enough so that the average excitatory inputs received during an integration window are much smaller than the gap between the spiking threshold and resting potential of the readout neuron, so that output firing is driven by input fluctuations \cite{valente2021}.

Beyond the conditions on the encoding firing rates $\mui^{(\pm)}$, our analytical solutions reveal  necessary conditions on other variables, which were not considered in  previous work \cite{valente2021}. For instance, it reveals conditions on the amplitude of noise correlations, $\sigmac$: this quantity must be large enough for noise correlations to have an impact on the SNR numerator $\Delta\rro$, but not too large for the SNR denominator $\<\vro^{(s)}\>_{s=\pm}$ to overcome the enhancement in the numerator (Fig. \ref{fig:snr_vs_muI} and Sec. \ref{sec:conditions}).
The non-monotonic dependence of the readout SNR on $\sigmac^2=2\vc\tauc$ and, hence, \emph{on $\tauc$ for fixed $\vc$} (Fig. \ref{fig:numSNRsqrt}) can be clearly understood within the quenched-noise approximation. On the one hand, the larger $\tauc$, the larger the minimum value of the amplitude $\sigmac$ of noise correlations that can elicit a rate enhancement $\Delta\rro$ (which depends on $\sigmac$ only through $\vc=\sigmac^2/(2\tauc)$). On the other hand, while $\Delta\rro$ depends on the {stationary variance} $\vc$ only, the readout variance $\vro$ increases with \emph{both $\vc$ and $\tauc$} (and \emph{faster with $\vc$}, cf. Eq. (\ref{eq:varrate_quenched_upper})). The dependence of the mean readout rate $\rro$ on the stationary variance $\vc$ follows from the quenched-noise approximation {\it ansatz} itself (Eq. \ref{eq:rquenched}). In contrast, the readout rate variance $\vro$ increases with $\vc$ and $\tauc$, because it is the integration, Eq. (\ref{eq:variancecorrection}), of the temporal correlation function (which, in the quenched-noise approximation, decays as an exponential with timescale $\tauc$ and amplitude $Q(\vc)$, see Eqs. (\ref{eq:g_quenched_loww}) and (\ref{eq:ratevariance_quenched}) in Appendix \ref{sec:handwaving}). As a result, the enhancement of the readout SNR can occur over a wide range of timescales $\tauc$ (or $\sigmac$ for fixed $\vc$), while it decreases at sufficiently large noise amplitude $\sigmac$.

In sum, for the noise correlations to enhance the readout SNR, $\mui^{(\pm)}$ should be lower than the inflection point of the white-noise LIF activation function, their difference $\mui^{(+)}-\mui^{(-)}$ should be large enough, and  $\sigmac$ should be large enough, but not too large.

It is of interest to consider the neuroscientific implications of the conditions and parameter regions we found to be able to support an enhanced-by-consistency information transmission regime. The first consideration regards how the numerical range of parameters in which this is possible compares with real biological parameters. Figure \ref{fig:theoreticalSNR_vs_muI} explores this regime as a function of two parameters that have been extensively measured in the brain:  the time constant of correlations $\tau_c$  and the firing rate differences between the different stimuli to be detected. The correlation time constant increases systematically when moving from sensory to higher-order areas \cite{murray2014hierarchy}, and it ranges from one or few milliseconds for the retina or sub-cortical structures \cite{bruno2006cortex, meytlis2012determining, totah2018locus}, to few tens of milliseconds for cortical sensory areas, and  hundreds of milliseconds 
for association or decision-making areas \cite{runyan2017, morcos2016history, akrami2018posterior}. The firing rate differences between stimuli to be discriminated can vary from fractions of one Hz (spikes/s) to few tens of Hz \cite{rolls1997information, panzeri2001role, iurilli2017population, georgopoulos1986neuronal, mountcastle1957response, quiroga2005invariant}.   Our results (Fig. \ref{fig:theoreticalSNR_vs_muI}) show an enhanced-by-consistency regime for rate differences  spanning few tens of Hz and correlation time constants r $\tau_c$ larger than the membrane time constant $\tau_m$ (which is typically in the range of 5--20 ms) and reaching several tens to few hundreds of millisecond, well within the range of sensory and association areas, suggesting that the enhanced-by-consistency regime can be sustained in these brain areas. 
The second consideration is that the additional necessary conditions derived here have potential implications for the neural biological  functions.
We have found that information-limiting correlations are useful to enhance information transmission only for intermediate values of the correlation strength $\sigmac$.
Across experiments and brain areas, correlation between neurons are reported consistently to be non-zero, but always of small or moderate value \cite{kohn2016,panzeri2022}. Our result suggests that non-null but moderate values may be useful to design neural codes that trade off effectively the information-limiting and information-enhancing effect of correlations and achieve good levels of information transmission. 
We have also found that the SNR enhancement may occur for a wide range of time constants of input correlations $\tauc$, even at values of $\tauc$ much larger than the readout membrane time constant $\taum$. 
This implies that neural circuits may be able to enjoy the benefits of information-encoding-limiting yet information-transmission-enriching correlations even when encoding information with activity correlated over long times. This is important because long encoding timescales are useful to combine temporally separate inputs and  accumulate consistently information over time, or to implement or compute a goal-oriented navigation plan \cite{runyan2017}. 

From the methodological point of view, we have proposed a (to our knowledge) novel analytical approximation for the rate mean and variance of the LIF model with white+colored input noise or, equivalently, for the first passage time of a random walker with non-Markovian noise (the {\it active Browinan particle}, in the terminology of \cite{walter2021,walter2022}). Despite its conceptual and methodological simplicity, this appears as one of the few existing results regarding the LIF model with temporally and spatially correlated input \cite{lindner2009,walter2021}. 
The quenched-noise approximation reveals that the colored-noise LIF neuron exhibits a higher-curvature activation function with respect to the white-noise LIF activation function $\rw(\mui)$. Moreover, beyond the effect on the average firing rate, the quenched-noise approximation addresses as well how temporal auto-correlations in the input are transmitted into auto-correlations in the colored-LIF output firing rate and, finally, how such auto-correlations shape the rate variance below or above Poissonian fluctuations $\vro=\rro/\Tc$.

Our work has considered a single readout neuron. While this simplification is useful to understand in detail the role of neural non-linearities in the transmission of information from temporally correlated inputs, information transmission in the brain happens from input to output populations. Other work \cite{zylberberg2017} has considered the conditions for effective transmission of information between neural populations, although considering only time-averaged spike count responses and neglecting the temporal aspects of correlated activity and neural post-synaptic integration that we considered here. Extending this and previous works will be important to merge the principles of population readout with those of time-correlation and time-integration, to understand how the spatio-temporal structure of neural activity shapes information transmission.

The function of populations of sensory neurons includes encoding the time course of time-varying stimuli, beyond the encoding of the identity of static stimuli as considered in our study. It would be important in future studies to extend our work to consider how noise correlations and readout nonlinearities impact the encoding and readout of information about the time course of time-varying stimuli. Indeed, an empirical study \cite{mazurek2002limits} suggested that noise correlations limit the temporal precision of the information that the time course of the average firing in a neural population can convey about the time course of a stimulus. Other studies have proposed that readout nonlinearities can enhance the detection of very rapid stimulus variations in a way that is robust to the detrimental encoding effects of noise correlations \cite{panzeri2010information}. While studying the time-varying encoding and readout is beyond the scope of the present article, we speculate that extending our formalism to this problem will give insights about the precise conditions in which the temporal parameters of stimulus dynamics, noise correlations and readout nonlinearities allow for a beneficial effect of time-dependent noise correlations on the transmission of information about dynamic stimuli.

A simplification of our work, shared by other previous attempts to characterize analytically how input correlations change the output firing characteristics of a LIF neuron \cite{siegert1951,moreno2002}, is that all steps of post-synaptic time integration have been summarized through the membrane potential time constant. We considered synapses having instantaneous dynamics, neglecting both their finite rise time and decay time constants, and we neglected the timescales of dendritic integration \cite{Ariav2003, London2005}. Although our simple model still allows to study effectively some of the consequences of lengthening or shortening of integration time constants by non-membrane contributions (for example, the effect of slower synaptic decay time constants could be recapitulated by increasing the membrane time constant), these simplifications prevent us to study the effect of the interplay between membrane, synaptic and dendritic integration in shaping the input-output information of a neuron \cite{koch1996brief}. Studying these issues is beyond the scope of the present study and the reach of analytical calculations, and is left for further work.


\section*{Acknowledgments}

We acknowledge Sandro Azaele for pointing out references \cite{constable2013,pavliotis2014,bucher2015,chow2015}. M. I. is supported by: the European Union under the scheme HORIZON-INFRA-2021-DEV-02-01 – Preparatory phase of new ESFRI research infrastructure projects, Grant Agreement n.101079043, “SoBigData RI PPP: SoBigData RI Preparatory Phase Project”; by the project ``Reconstruction, Resilience and Recovery of Socio-Economic Networks'' RECON-NET EP\_FAIR\_005 - PE0000013 ``FAIR'' - PNRR M4C2 Investment 1.3, financed by the European Union – NextGenerationEU; by the European Union - Horizon 2020 Program under the scheme `INFRAIA-01-2018-2019 - Integrating Activities for Advanced Communities', Grant Agreement n.871042, `SoBigData++: European Integrated Infrastructure for Social Mining and Big Data Analytics' (\url{http://www.sobigdata.eu}); Grant No. PID2023-149174NB-I00 financed by MICIU/AEI/10.13039/501100011033 and EDRF/EU funds. M. I. also benefits from the IMT computing resources funded by the DM, Italy n. 289, 25-03-2021, CUP D67G22000130001, through the project “Scientific computing for natural sciences, social sciences, and applications: methodological and technological development”. S. P. is supported by the the German Federal Ministry of
 Education and Research, grant  BMBF 01GQ2404.

\clearpage
\appendix
\section{Symbols and acronyms and terms used in the article\label{sec:symbols}}

\subsection{Stationary and count variances}

We here explain the  {\it count} ($\sigma^2$) vs {\it stationary ($v$) variance} terminology that we employ in the article. Let $y(t)$ be an Ornstein-Uhlenbeck process, obeying the Langevin equation: $\tau \d y = -\d t\, y  + \sigma\,\d W$, where $\d W$ is a Wiener process. Then \cite{risken1996,livi2017}, the stationary distribution of $P(y)$ has variance equal to  $\<y^2\>-\<y\>^2=v=\sigma^2/(2\tau)$, and temporal correlation function ${\sf c}(t):=\langle y(t')y(t+t')\rangle = v \exp(-|t|/\tau)$, where $\<\cdot\>$ is the expectation value with respect to realizations of $\d W$. If $y(t)$ represents the rate of an inhomogeneous Poisson process, the variance of the number of spikes $\text{Var}(\text{rate})$ in a time interval of length $T$ is $\text{Var}(\text{rate})=\frac{2}{T^2} \int_0^{T}\d t'(T-t'){\sf c}(t')$. The rate variance in an interval of length $T\gg\tau$ is, hence, $\text{Var}(\text{rate}) = 2\tau v/T = \sigma^2/T$. Hence, $v$ is the {\it stationary variance of the rate}, while $\sigma^2$ is the {\it variance of the spike count in the unit time interval}.

\subsection{Table of symbols and acronyms}

See the acronyms and main symbols used in the article in Tables \ref{table:acronyms} and \ref{table:symbols}, respectively.

\begin{table}[!ht]
\caption{\bf Acronyms used in the article.\label{table:acronyms}}
\begin{tabular}{l|l} 
	acronym & definition   \\
\hline
	NERM	& neural encoding-readout model			\\
	SNR	& signal-to-noise ratio				\\
	NIS	& number of integration steps		\\
	LIF	& leaky integrate-and-fire 			\\
	ISI	& inter-spike interval 			
\end{tabular}
\end{table}

\begin{table}[!ht]
\caption{\bf Symbols used in the article.\label{table:symbols}}
\begin{tabular}{l|c|l} 
	symbol & definition & description  \\
\hline
	${\bf x}^{(s)}(t)$ & 	$\dot \x(t) = ({\bm \mu^{(s)}}-\x(t))/\tauc + (2C/\tauc)^{1/2}\cdot {\bm \eta}(t)$ & vector of instantaneous firing rates of encoding neurons\\
	$\<\cdot\>_{{\bm\eta}}$ & & average with respect to the distribution of white noises \\
	${\bm\mu}^{(s)}$ & 	$=\<{\bf x}^{(s)}(t)\>_{\bm\eta}$	& vector of encoding average firing rates (tuning curves)\\
	$\Delta{\bm\mu}$ & 	${\bm\mu}^{(+)}-{\bm\mu}^{(-)}$	& \\
	$C_{ij}$ & 	$=\text{Cov}\left(x_i(t),x_j(t)\right)$	& covariance matrix (element) between encoding rates \\
	$\tauc$ & 	$\text{Cov}\left(x_i(t),x_j(t')\right)=C_{ij}e^{-|t-t'|/\tauc}$	& noise correlation time \\
	$\vV_i$ & 	$=C_{ii}$	& stationary variance of single encoding neurons \\
	$\sigmaV_i$ & 	$\vV_i=\sigmaV_i^2/\tauc$	& count standard deviation of single encoding neurons \\
	$\bar C_{ij}$ & 	$=C_{ij}/(C_{ii}C_{jj})^{1/2}$	& noise correlation matrix (element) \\
	$\alphaV$ & 	$\bar C_{ij} = \delta_{ij} + (1-\delta_{ij})\alphaV$	& Valente angle \\
	$t_k^{(i)}$ 		& 		& $k$-th spike of the $i$-th encoding neuron \\
	$\rho_i(t)$ 		& $=\sum_{k>0}\delta(t-t_k^{(i)})$ 		& density of spikes of encoding neuron $i$ (single realization) \\
	$\rho(t)$	& 	$=\sum_j \rho_j(t)$		& density of encoding spikes in a single realization   \\
	$n(\Tc)$	& 	$\int_0^\Tc\d t \rho(t)$	& number of encoding spikes in a single realization   \\
	$V(t)$ & 		& membrane potential of the readout LIF neuron \\
	$\taum$ & 		& membrane potential timescale of the readout LIF neuron \\
	$I(t)$ 		& 	$\dot V(t)=-V(t)/\taum+I(t)$	& input current (the $s$-dependence is often omitted) \\
	$\w$ 		& 	$I(t)=\w \cdot {\bm\rho}(t)$	& vector of gains \\
	$\sigmai$	& 	$I(t)=\mui+\sigmai\eta(t)+\sigmac\xi(t)$	&  white-noise amplitude of the afferent current (diffusion app.)\\
	$\sigmac$	& 				&  colored-noise (count) amplitude of the afferent current (diffusion app.)\\
	$\vc$		& 	$=\sigmac^2/(2\tauc)$	&  colored-noise stationary variance of the afferent current (diffusion app.)\\
	$\mui$ 		& 	$=\<I(t)\>_{\eta,\xi}$	&  average input current (diffusion app.)\\
	$\eta(t),\xi(t)$& 		& white and colored noises (diffusion app.) \\
	$t_k$ 		& 				& $k$-th spike of the readout neuron \\
	$\rhoro(t)$ 	& 			$=\sum_{k>0}\delta(t-t_{k})$	& density of spikes (firing rate) of the readout neuron (single realization)  \\
	$\rro$ 		& 	$=\<\rhoro(t)\>_{\eta,\xi}$	&   \begin{tabular}{@{}l@{}}mean of the LIF readout neuron firing rate\\ (or mean of a colored-noise LIF neuron firing rate) \end{tabular}  \\
	$\vro$ 		& 	$=\text{Var}_{\eta,\xi}\left(\rhoro(t)\right)$	&\begin{tabular}{@{}l@{}}variance of the readout neuron firing rate\\ (or variance of a colored-noise LIF neuron firing rate) \end{tabular}   \\
	$\Tc$		& 	&  length of the time chunk (in which the readout spikes are registered) \\
	$\nro(\Tc)$	& $\int_0^\Tc\d t \rhoro(t)$	& number of readout spikes in a single realization   \\
	$\rw$		& 	& Siegert solution for the mean rate of a white-noise LIF neuron \\
	$\tw$		& $=\rw^{-1}$	& Siegert solution for the mean ISI of a white-noise LIF neuron \\
	$\alpha_{\rm M}$& $=\sigmac^2/\sigmai^2$ 	& Moreno parameter \\
	$\rq$		&  	& quenched-noise approximation for $\rro$ \\
	$\vq$		&  	& quenched-noise approximation for $\vro$ \\
	$r_{\rm M}$		&  	& Moreno et al approximation for $\rro$ \\
	$g_1$		& $=\rw'/\rw$  	& \\
	$\mui^*$	& 	& inflection point of $\rw$ \\
	readout rate		&  $=\nro(\Tc)/\Tc$ 				& \begin{tabular}{@{}l@{}} number of readout spikes per unit time in a window of length $\Tc$ \\ (or number of spikes per unit time of a general LIF neuron) \end{tabular}  \\
	$\C(t)$		&  $=\<\varrho(t')\varrho(t'+t)\>_{\xi,\eta}-\rro^2$ 	& temporal correlation function of the readout neuron \\
	$\C_{-}(t)$	&  $\C(t)= a\delta(t) + \C_{-}(t)$		 	& regular part of $\C$  \\
	$g(t)$		&  $\C(t)= \rro g(t) - \rro^2$		 	& readout neuron temporal distribution function \\ 
	$\Delta\rro$	&  $=\rro^{(+)}-\rro^{(-)}$		 	&  \\
	$\mui^*$	&  		 	&  inflection point of $\rw$ \\
	$\hat\mui$	&  		 	&  upper bound of $\mui^{(+)}$ for the SNR to increase (see \ref{sec:conditions})\\
	$\hat\mui^{(-)}$&  		 	&  lower bound of $\mui^{(-)}$ for the SNR to increase (see \ref{sec:conditions})\\
	$\nis$	&  		 	&  number of integration steps per unit time (see \ref{sec:numerical})\\
\end{tabular}
\end{table}

\def\N{{N}}

\def\ffunc{{\sf f}}
\def\gfunc{{\sf g}}
\def\hfunc{{\sf h}}

\def\vc{{v_{\rm c}}}
\def\v1{{v_1}}

\def\x{{\bf x}}
\def\w{{\bf w}}
\def\taum{{\tau_{\rm m}}}
\def\tauc{{\tau_{\rm c}}}
\def\sigmac{{\sigma_{\rm c}}}
\def\sigmai{{\sigma_{\rm I}}}
\def\mui{{\mu_{\rm I}}}
\def\alphaV{{\alpha_{\rm V}}}
\def\nsrV{{\nu_{\rm V}}}
\def\sigmaV{{\sigma_{\rm V}}}
\def\vV{{v_{\rm V}}}
\def\xiC{{\xi_{\rm C}}}
\def\fc{{v}}
\def\rro{{r_{\rm R}}}
\def\rq{{r_{\rm q}}}
\def\gq{{g_{\rm q}}}
\def\vq{{v_{\rm q}}}
\def\rw{{r_{\rm w}}}
\def\tw{{t_{\rm w}}}
\def\vro{{v_{\rm R}}}
\def\rhoro{{\varrho}}
\def\nro{{n_{\rm R}}}

\def\sigmacref{{\sigma_{\rm c}^{({\rm ref})}}}
\def\taucref{{\tau_{\rm c}^{({\rm ref})}}}

\def\Tc{{T_{\rm c}}}

\def\nis{{n_{\rm is}}}
\def\ntrials{{n_{\rm tr}}}

\section{The diffusion approximation for the encoding signal\label{sec:diffusion}}

\subsection{Single, inhomogenenous Poisson process with Ornstein-Uhlenbeck rate}

We consider an inhomogeneous Poisson process whose rate is $r(t)=\mu+\chi(t)$. The fluctuations around its average are such that $\<\chi(t)\>_\chi = 0$, and $\<\chi(t)\chi(t')\>_\chi$ is a given function of $|t-t'|$, and where $\<\cdot\>_\chi$ is the average over realizations of the noise $\chi$. Let $\rho(t)=\sum_k \delta(t-t_k)$ be the density of spikes of such process, where $t_k$ is the $k$-th spike. In other words, $\rho(t)=\dot n(t)$ where $n(t)$ is the number of spikes $t_k<t$ in a single realization (conditioned to a fixed realization of $r$), and the dot stands for the temporal derivative. We are interested in the two-time correlation function, or the covariance

\begin{align}
{\rm Cov}\left(\rho(t),\rho(t')\right) = \<\rho(t)\rho(t')\> -  \< \rho(t) \> \<\rho(t')\>
\end{align}

Importantly, the expected value $\<\cdot\>:=\<\<\cdot\>_{\rm P}\>_{\chi}$ is the annealed average over both sources of stochasticity (over the Poisson spikes given $r(t)$, $\<\cdot\>_{\rm P}$, and over the realizations of $r(t)$, $\<\cdot\>_{\chi}$). Let us observe that the variable $n(t+\delta)-n(t)$ is, for a fixed realization of $r(t)$, and using the definition of inhomogeneous Poisson process, a stochastic variable with average and variance equal to $R(t+\delta)-R(t)$, where $R$ is the primitive of $r$. To first order in $\delta$, such single realizations of $n(t+\delta)-n(t)$ may be written as: 

\begin{align}
n(t+\delta)-n(t) \simeq \delta\, r(t) + r(t)^{1/2}(W_{t+\delta}-W_t)
\label{eq:diffusionapproximation}
\end{align}
where $W_{t+\delta},W_t$ are random numbers such that the variance of $W_{t+\delta}-W_t$ is $\delta$. Now, for sufficiently large $\delta\, r(t)$, the variable $n(t+\delta)-n(t)$ can be taken as normally distributed, and $W_t$ is a Wiener process, since $W_{t+\delta}-W_t \sim {\cal N}(0,\delta)$. In this case, dividing Eq. (\ref{eq:diffusionapproximation}) by $\delta$ and taking the $\delta\to 0$ limit one gets, formally,

\begin{align}
\rho(t) = r(t) + r(t)^{1/2} \eta(t)
\label{eq:diffusionapproximation2}
\end{align}
where $\<\eta(t)\eta(t')\>_\eta=\delta(t-t')$ is a white noise (and we now write $\<\cdot\>_{\rm P}$ as $\<\cdot\>_{\eta}$). Equation (\ref{eq:diffusionapproximation2}) is the so called {\it diffusion approximation} to the Poisson process. 

In the diffusion approximation (Eq. \ref{eq:diffusionapproximation2}), it is immediate to see, first, that $\<\rho(t)\>_{\eta,\chi}=\<r(t)\>_\chi =\mu$. It is also immediate to compute the temporal covariance, or time-correlation function of $\rho$'s. Using Eq.  (\ref{eq:diffusionapproximation2}) and the correlation functions of the noises $\chi$, $\eta$:

\begin{align}
{\rm Cov}\left(\rho(t),\rho(t')\right) &= \<\< \rho(t)\rho(t') \>_{\eta}\>_{\chi} - \mu^2    =\\
&=\<\chi(t)\chi(t')\>_\chi + \delta(t-t')\<r^{1/2}(t)r^{1/2}(t')\>_\chi =\\
&=\<\chi(t)\chi(t')\>_\chi + \delta(t-t')\<r(t)\>_\chi =\\
&=\<\chi(t)\chi(t')\>_\chi + \mu\,\delta(t-t')
.
\end{align}

\subsection{Many encoding neurons}

We now consider a multivariate Poisson process whose $\N-$dimensional vector of rates $\x(t)={\bm\mu} + {\bm\chi}(t)$ is such that the fluctuations around the mean $\bm\mu$ satisfy the Ornstein-Uhlenbeck process: $\dot {\bm\chi}(t) = -{\bm\chi}(t)/\tauc+(2 C/\tauc)^{1/2}\cdot{\bm\eta}(t)$, where $^{1/2}$ is the matrix square root, and where $C$ is a covariance matrix. Indeed, the stationary distribution for the solutions $\bm\chi$ of this Ornstein-Uhlenbeck process is a multivariate Gaussian with null mean and covariance equal to $C$. We define as well the Pearson correlation matrix $\bar C$ as: $\bar C_{ij}=C_{ij} (v_i v_j)^{-1/2}$, where the $v_i:=C_{ii}$'s are the stationary variances of the $\chi_i(t)$'s. Finally, we define the spike count variance of the $j$-th neuron, $\sigma_j^2:=2\tauc v_j$. Following the same reasoning lines of the previous subsection, we learn that, if $\rho_i(t)$ is the firing rate of the $i$-th Poisson process, it is:

\begin{align}\label{eq:multivariate densities}
\<\rho_i(t)\> & = \mu_i \\
\<\rho_i(t)\rho_j(t')\> &= \mu_i \delta_{ij}\delta(t-t')  + (v_i v_j)^{1/2} \bar C_{ij} e^{-|t-t'|/\tauc} 
.
\end{align}

Now, if $I(t) = {\bf w}^\dag \cdot {\bm\rho}(t)$ is the density of spikes of the $\N$ processes, weighted by a vector of synaptic weights $\bf w$, the mean of $I$ and the covariance between $I$'s, ${\rm Cov}\left( I(t),I(t') \right) = \< I(t) I(t')  \>  - \<I(t)\>\<I(t')\>$, take the form:

\begin{subequations}
\label{eq:currentstatistics}
\begin{align}
\< I(t) \> &= \mui \\
{\rm Cov}\left( I(t),I(t') \right) & = \sigmai^2 \delta(t-t')  + \vc e^{-|t-t'|/\tauc}  
\end{align}
\end{subequations}
where

\begin{subequations}
\label{eq:currentparameters}
\begin{align}
\mui &= {\bf w}^\dag \cdot {\bm\rho} \\
\sigmai &= \sum_{j=1}^\N w_j^2 \mu_j \\
\vc &= \frac{\sigmac^2}{2\tauc} \\
\sigmac^2 &= {\bf w}^\dag\cdot C\cdot {\bf w}
.
\end{align}
\end{subequations}

\subsection{The Valente et al 2021 model \label{sec:1RONinputI}}

The NERM analyzed in this article is actually the generalization, for an arbitrary number of encoding neurons $N$, of the encoding-and-readout model originally defined in section {\it ``A model of enhanced-by-consistency information transmission''} in \cite{valente2021}). In reference \cite{valente2021} the encoding neurons are represented by inhomogeneous Poisson processes whose firing rate behaves, given the binary stimulus $s=\pm 1$ as: 

\begin{align}
r_i^{(s)}(t)=\mu_i^{(s)} + \sigmaV_i\left(\alphaV \xiC(t) + (1-\alphaV^2)^{1/2}\xi_i(t)  \right)
\end{align}
where $\mu_i^{(s)}$ is the tuning curve and $\alphaV$ is a real number in $[0,1]$ modulating the fraction of shared and private noise (in such a way that $\<\xi_i(t)\xi_j(t'+t)\>=\delta_{ij}(2\tauc)^{-1}\exp(-|t|/\tauc)$, $\<\xiC(t)\xiC(t'+t)\>=(2\tauc)^{-1}\exp(-|t|/\tauc)$, $\<\xi_i(t)\xiC(t'+t)\>=0$). In this way, the noise variance of the $i$-th encoding neuron is independent of $\alphaV$ and equal to $\vV_i=\sigmaV_i^2/(2\tauc)$. The spike rate standard deviation $\sigmaV_i$ is set to be proportional to the difference of the tuning curves ${\Delta_\mu}_i=\mu_i^{(+)}-\mu_i^{(-)}$:

\begin{align}
\sigmaV_i&=\nsrV\,{\Delta_\mu}_i \\
{\Delta_\mu}_i&:=\mu_i^{(+)}-\mu_i^{(-)} > 0
\end{align} 
and $\nsrV\ge 0$ is a positive noise-to-signal ratio. Taking the noise amplitude $\sigmaV$ proportional to ${\Delta_\mu}_i$, one ensures that for any $\alphaV>0$ the noise correlations are information-limiting. 

We make notice that Valente et al model is of the form as in the previous subsection, with:

\begin{align}
& \sigma_i = \sigmaV_i^2 = \nsrV^2\,{\Delta_\mu}_i^2\\
& \bar{C}_{ij} = \alpha_V^2\qquad i\ne j 
.
\end{align}
If we now suppose that all encoding neurons have equal tuning curves increments, ${\Delta_\mu}_m={\Delta_\mu}$, and that the vector $\bf w$ has constant components $w_j=w$, we obtain that the current encoding $I(t)$ satisfies Eq. (\ref{eq:currentstatistics}) with (see (\ref{eq:currentparameters})):

\begin{align}
\mu_I^{(s)} &= w N \overline\mu^{(s)} \\
\left(\sigma_I^{(s)}\right)^2 &= N w^2 \overline\mu^{(s)}   \\
\sigmac^2 &= 2\tauc \fc w^2 N \sigmaV^2 \left( 1 + (N-1)\alphaV^2 \right) 
\end{align}
where $\overline\mu^{(s)}$ is the average of the encoding neuron tuning curves. From these last equations we see that the signal/noise ratio of the input current ${\rm snr} = {\Delta_{\mu_I}}/(\sigmac^2+\sigma_I^2)^{1/2}$ monotonically decreases with $\alphaV$. Eq. (\ref{eq:currentstatistics}) has the equivalent form:

\begin{align}
I(t) = \mui + \sigmai\, \eta(t) + \sigmac\, \xi(t) \\
\dot \xi(t) = -\xi(t)/\tauc + \frac{1}{\tauc} \eta'(t)
\end{align}
(we omit the $\mu_I$-and $\sigma_I$-dependence on $s$), where $\eta,\eta'$ are two different continuous time white noises, and $\xi(t)$ is an Ornstein-Uhlenbeck process with timescale $\tauc$.

\subsection{Analogy with the random walk problem} 

In the diffusion approximation, valid for $\mui\gg(\sigmai^2+\sigmac^2)^{1/2}$ (see \cite{burkitt2006,moreno2002}, the stimulus-readout probability distribution $\text{prob}({\vec t}|s)$ (on which the SNR depends, through $\rro,\vro$ in Eq. (\ref{eq:SNRsqrt})), becomes the distribution of sequences of first passage times of a continuous-time random walker $V(t)$ (see Eq. (\ref{eq:I_diffusion})), with damp $1/\taum$, constant drift $\mui$, white noise drift $\sigmai\eta(t)$, colored noise drift $\sigmac\xi(t)$, and the reset mechanism of the LIF neuron (that conditions the stationary distribution of the $\xi$'s \cite{middleton2003}). The readout neuron Inter-Spike Interval (ISI) time is the first passage time of such a random walker (see \cite{walter2021,walter2022}, and references therein).

\subsection{A difference between the present notations and that of Valente et al}

Please, mind the notation difference between \cite{valente2021} and the present article: there, it is assumed that ${\rm Cov}\left(x_i^{(s)}(t), x_j^{(s)}(t) \right) = \sigma^2\tauc \exp(-|t-t'|/\tauc)$, so that their parameter $\sigma^2$ has a different dependence on $\tauc$ than our $\sigmaV_i^2$'s (and than our $\vV_i$'s). Indeed, in their work, the instantaneous Poisson rate $r(u)$ is converted into a time-correlated Poisson rate $r(t;\tauc)$ in the following way (mind the misprint in their low-pass filter, equation 19):

\begin{align}
r(t,\tauc)=\int_0^t \d u\,e^{-(t-u)} r(u)
\end{align}
so that $\<(r(t;\tauc)-\mu)(r(t';\tauc)-\mu)\>=\sigma^2\tauc\exp(-|t-t'|/\tauc)$.

\section{Numerical methods \label{sec:numerical}}

\subsection{Numerical integration of the stochastic differential equation for the readout neuron membrane potential}

In the main text, we have presented a stochastic-processes theory of the NERM defined in section \ref{sec:model}. The diffusion approximation (see \ref{sec:diffapp}) allows us to treat the membrane potential of the readout LIF neuron as the solution of a stochastic differential equation. In the article, we compare the approximated analytical solutions to such equations, with numerical simulations, that help us to asses the approximations' validity limit. We have, in particular, and as in references \cite{moreno2002,morenobote2004}, numerically solved Eqs. (\ref{eq:VI_diffusion}) using the Euler-Maruyama method \cite{mannella2002,kloeden2011}, completed with a numerical protocol (see below) that corrects the bias induced by the finiteness of the Number of Integration Steps (NIS) per unit time, $\nis$.  

In brief, our numerical estimation for the rate mean and variance, $\rro,\vro$ respectively, are performed as follows. First, we employ the Euler-Maruyama method with $\Tc\nis$ integration steps to numerically integrate Eqs. (\ref{eq:VI_diffusion}) with the desired set of parameters $\mui,\sigmai,\sigmac,\tauc,\taum$ in a time interval of length $\Tc$. We perform a number of $\ntrials$ different realizations of such integration, differing in the independent random samples of the white noises $\eta,\eta'$, and in the initial conditions for $\xi(0),v(0)$ (that can either be independently sampled from their stationary distributions, or chosen, as for our numerical results, to coincide with those of the previous realization in the last instant value). For each realization, we store the corresponding sequence of readout neuron spike times $\vec t$. Afterwards, we compute the across-trials average and variance of the rate $\nro(\Tc)/\Tc$, where $\nro(\Tc)=\text{len}(\vec t)$ in single realizations, and where $\Tc=20$. These are, respectively, our finite-$\nis$ estimations for the rate mean and variance, $\tilde\rro,\tilde\vro$ (circles and squares in Figs. \ref{fig:r_quenched},\ref{fig:rstd_quenched},\ref{fig:numSNRsqrt}-A,\ref{fig:numSNRsqrt}-B,\ref{fig:numSNRsqrt}-C). Importantly, and as already specified in the main article text, {\it all the rate variance estimations that we report are rate variances per unit time}: $\vro \Tc$. Indeed, both the singular and the regular part of the temporal correlation function $\sf C$ induce rate variances that decreases as $1/\Tc$: to eliminate this dependency, we report the variance in intervals of unit length. Consequenly, the SNR analised in Secs. \ref{sec:SNRvstauc},\ref{sec:conditions} is the SNR per unit time window, or: $\text{SNR}=\Tc^{-1/2}\times \text{SNR}(\Tc)$. The so defined standard deviation of the rate in the unit time interval, $(\vro\Tc)^{1/2}$ coincides with the value of the rate standard deviation, $\vro^{1/2}$ that we obtain if we divide each of our $\ntrials$ realizations of the readout spike trains $\vec t_{r}$ in $m={\rm int}(\Tc)$ chunks, in such a way that we have a new set of $m\times\ntrials$ shorter realizations, of length $1$ each, and we compute the standard deviation of $\text{len}(\vec t)$ across this novel set of spike trains. The same is true for the SNR per unit time interval. 

\subsection{Error estimation of rate mean, variance, and SNR \label{sec:errorestimation}}

As said in the main text, the statistical uncertainty (error-bars) of our estimation $\tilde\rro$ in Figs. \ref{fig:r_quenched},\ref{fig:numSNRsqrt}-A is computed as the standard deviation of the rate $\nro(\Tc)/\Tc$ across the $\ntrials$ trials (hence, multiplied by $\Tc^{1/2}$). The errors of our estimations {\it of the rate variance} in Figs. \ref{fig:rstd_quenched},\ref{fig:numSNRsqrt}-B are the confidence intervals of the Chi-squared distribution with $\ntrials=10^3$ degrees of freedom and a probability of $\alpha_\chi=0.95$. The errors of the SNR in Fig. \ref{fig:numSNRsqrt} correspond to the standard deviation across bootstrap realizations of the $\ntrials$ realizations of spike times --for each of which we compute the SNR (see \cite{repository}). Finally, the uncertainties of the NIS-corrected rates (and $\beta_r$, see below) are the errors of the regression.

\subsection{Parameter fixing \label{sec:parameterfixing}}


We have performed seven sets of numerical simulations. In Table \ref{table:simulationsoutline} we specify the name, purpose and characteristics of each of such (set of) simulations. Among them, the so called 1st simulation is to assess the dependence of the SNR with $\mui,\tauc,\vc(\alphaV),\nis$. It is the set that we use to perform the NIS-corrected estimations of the SNR as a function of $\tauc$ and $\alphaV$ (in Figs. \ref{fig:r_quenched}-A,\ref{fig:g_quenched},\ref{fig:rstd_quenched},\ref{fig:numSNRsqrt}-A,\ref{fig:numSNRsqrt}-B,\ref{fig:numSNRsqrt}). 
In tables \ref{table:parameters1}--\ref{table:parameters7} we report the values of the parameters of each set of simulations (see \cite{repository} as well). In each set of simulations, given the value of the varying independent parameters, the rest of the parameters are consequently set according to Eqs. (\ref{eq:dictionary},\ref{eq:dictionary_vc}). Analogously, given $w$ and $\mui$, we set $\sigmai=(w\mui)^{1/2}$. In the tables, ``diffusion'' refers to the NERM model in the diffusion approximation, Eq. (\ref{eq:VI_diffusion}) (for this model, the number of encoding neurons $\N$ plays the only role of changing the relation between $\vc$ and $\alphaV$ in Eq. (\ref{eq:dictionary_vc})). ``diffusion-N'' model refers to the NERM in diffusion approximation with the sum of single stochastic differential equations for each encoding neuron as input current (see Sec. \ref{sec:figs23}). Finally, ``spike-enc'' refers to the NERM model with spiking encoding input current (see Sec. \ref{sec:validityreadout}). 

\begin{table}
\caption{Outline of simulations. \label{table:simulationsoutline}}
\begin{ruledtabular}
\begin{tabular}{c|c|c|c|c|c}
simulation name	& purpose 							& model		& varying 	& NIS-extrapolation alg. & figures\\
\hline
1st		& \makecell{SNR estimation in the \\diffusion NERM vs $\tauc$, fixed $\vc$} 		& diffusion	& $\mui,\alphaV,\tauc,\nis$	& yes			& \ref{fig:r_quenched}-B,\ref{fig:g_quenched},\ref{fig:rstd_quenched},\ref{fig:numSNRsqrt},\ref{fig:rRregression},\ref{fig:g_illustration},\ref{fig:rRvstauc},\ref{fig:sRvsmui},\ref{fig:sRvstauc},\ref{fig:numSNRsqrt_alt},\ref{fig:g_quenched_alt},\ref{fig:numSNRsqrt_semicorrected}	\\
\hline
2nd		& \makecell{quenched-noise vs Moreno \\difference assessment} 		& diffusion	&	$\mui,\alphaV$	& no			& \ref{fig:r_quenched}-A	\\
\hline
3rd		& \makecell{SNR estimation in the \\diffusion NERM vs $\sigmaV$, fixed $\tauc$}  	& diffusion	& $\mui,\sigmaV,\nis$	& no			& \ref{fig:SNR_varyingsigmac}	\\
\hline
4th		& \makecell{Figs. \ref{fig:valente_rates},\ref{fig:valente_rates_nocorrelations} SNR \\ estimation}  	& diffusion-N 	& $\mui,\alphaV,N$	& no			& \ref{fig:diffusionN}\\
\hline
5th		& \makecell{Figs. \ref{fig:valente_rates},\ref{fig:valente_rates_nocorrelations} SNR \\ estimation}  	& diffusion 	& $\mui,\alphaV,\sigmaV$	& no			& \ref{fig:diffusionN}\\
\hline
6th		& \makecell{spike-enc. LIF vs diffusion LIF \\ comparison}  	& spike-enc 	& $\mui,w$	& no			& \ref{fig:diffusion_Lifspikes},\ref{fig:diffusion_Lifspikes_noise}\\
\hline
7th		& \makecell{SNR estimation in the \\spike-enc. NERM vs $\sigmaV$}  	& spike-enc	& $\mui,\sigmaV$	& no			& \ref{sec:LIFspikessimulation},\ref{fig:SNR_LIFspikes}	\\
\hline
\end{tabular}
\end{ruledtabular}
\end{table}

\begin{table}
\caption{Parameters of the first simulation set. \label{table:parameters1}}
\begin{ruledtabular}
\begin{tabular}{c|c|c}
variable	& varying & value/range\\
\hline
$\mui$ 			& yes 	& $32+m \delta_\mui$, with $m=(0,\ldots,8)$ and $\delta\mui=2.95$ 	\\
 			&  	& =\sf{32.65,35.6, 38.55, 41.5, 44.45, 47.4, 50.35, 53.3, 56.25}	\\
\hline
$\alphaV$ 		& yes 	& $0.0,0.9$	\\
\hline
$\tauc$ 		& yes 	& $10^{-4+a}$, $a=(0,b,2b,\ldots,9b)$ with $b=3/7$	\\
 			&  	& =\sf{1.000e-4, 2.682e-4, 7.196e-4, 1.930e-3, 5.179e-3, 1.389e-2, 3.727e-2, 1.000e-1, 2.682e-1, 7.196e-1}	\\
\hline
$\sigmac$ 		& yes 	& $\vc/(2)\tauc$	\\
\hline
$\nis$ 			& yes 	& $2^{m}$, with $m=(13,\ldots,18)$ 	\\
 			& 	& $=\sf{8192, 16384, 32768, 65536, 131072}$ 	\\
\hline
\hline
$\taum$ 		& no 	& 	$0.005$						\\
\hline
$N$ 			& no 	& 	$20$						\\
\hline
$\sigmaV$		& no 	& 	$1.104241$						\\
\hline
$w$ 			& no 	& 	$0.45$						\\
\hline
$\ntrials$ 		& no 	& 	$1000$						\\
\hline
$\Tc$ 		& no 	& 	$20$						\\
\hline
\end{tabular}
\end{ruledtabular}
\end{table}

\begin{table}
\caption{Parameters of the second simulation set. \label{table:parameters2}}
\begin{ruledtabular}
\begin{tabular}{c|c|c}
variable	& varying & value/range\\
\hline
$\mui$ 			& yes 	& $32+m \delta_\mui$, with $m=(0,\ldots,8)$ and $\delta\mui=2.95$ 	\\
 			&  	& =\sf{32.65,35.6, 38.55, 41.5, 44.45, 47.4, 50.35, 53.3, 56.25}	\\
\hline
$\alphaV$ 		& yes 	& $0.0,0.9$	\\
\hline
\hline
$\tauc$ 		& no 	& $0.1$	\\
\hline
$\taum$ 		& no 	& 	$0.005$						\\
\hline
$N$ 			& no 	& 	$20$						\\
\hline
$\nis$ 			& no 	& $131072$ 	\\
\hline
$\sigmaV$		& no 	& 	$1.77777$						\\
\hline
$\vc$			& no 	& 	$1.1042$						\\
\hline
$w$ 			& no 	& 	$0.45$						\\
\hline
$\ntrials$ 		& no 	& 	$1000$						\\
\hline
$\Tc$ 			& no 	& 	$20$						\\
\hline
\end{tabular}
\end{ruledtabular}
\end{table}

\begin{table}
\caption{Parameters of the third simulation set. \label{table:parameters3}}
\begin{ruledtabular}
\begin{tabular}{c|c|c}
variable	& varying & value/range\\
\hline
$\mui$ 			& yes 	& $80+10m/11$, with $m=0,\ldots,11$ 	\\
\hline
$\sigmaV$		& yes 	& 	$8.9442\times (m/9)$, with $m=0,\ldots,9$						\\
 			&  	& =\sf{0.      , 1.118025, 2.23605 , 3.354075, 4.4721  , 5.590125,  6.70815 , 7.826175, 8.9442}	\\
\hline
$\nis$ 			& yes 	& $2^m$ for $m=15,\ldots,18$ 	\\
\hline
\hline
$\alphaV$ 		& no 	& $0.9$	\\
\hline
$\tauc$ 		& no 	& $0.1$	\\
\hline
$\taum$ 		& no 	& 	$0.005$						\\
\hline
$N$ 			& no 	& 	$20$						\\
\hline
$w$ 			& no 	& 	$0.1$						\\
\hline
$\ntrials$ 		& no 	& 	$100$						\\
\hline
$\Tc$ 			& no 	& 	$200$						\\
\hline
\end{tabular}
\end{ruledtabular}
\end{table}

\begin{table}
\caption{Parameters of the fourth simulation set. \label{table:parameters4}}
\begin{ruledtabular}
\begin{tabular}{c|c|c}
variable	& varying & value/range\\
\hline
$\mui$ 			& yes 	& $80,90$ 	\\
\hline
$\alphaV$ 		& yes 	& $0,0.9$	\\
\hline
$N$ 			& yes 	& $6,10,20$	\\
\hline
$\sigmaV$		& yes 	& 	$=\sigmaV^{(N=20)}\times 20/N$, with $\sigmaV^{(N=20)}=4.47213$ \\ 
\hline
\hline
$\nis$ 			& no 	& $32768$ 	\\
\hline
$\tauc$ 		& no 	& $0.1$	\\
\hline
$\taum$ 		& no 	& 	$0.005$						\\
\hline
$w$ 			& no 	& 	$0.1$						\\
\hline
$\ntrials$ 		& no 	& 	$200$						\\
\hline
$\Tc$ 			& no 	& 	$200$						\\
\hline
\end{tabular}
\end{ruledtabular}
\end{table}

\begin{table}
\caption{Parameters of the fifth simulation set. \label{table:parameters5}}
\begin{ruledtabular}
\begin{tabular}{c|c|c}
variable	& varying & value/range\\
\hline
$\mui$ 			& yes 	& $80,90$ 	\\
\hline
$\alphaV$ 		& yes 	& $0,0.9$	\\
\hline
$\sigmaV$		& yes 	& $0,4.47213$ \\ 
\hline
\hline
$N$ 			& no 	& $20$	\\
\hline
$\nis$ 			& no 	& $32768$ 	\\
\hline
$\tauc$ 		& no 	& $0.1$	\\
\hline
$\taum$ 		& no 	& 	$0.005$						\\
\hline
$w$ 			& no 	& 	$0.1$						\\
\hline
$\ntrials$ 		& no 	& 	$200$						\\
\hline
$\Tc$ 			& no 	& 	$200$						\\
\hline
\end{tabular}
\end{ruledtabular}
\end{table}

\begin{table}
\caption{Parameters of the sixth simulation set. \label{table:parameters6}}
\begin{ruledtabular}
\begin{tabular}{c|c|c}
variable	& varying & value/range\\
\hline
$\mui$ 			& yes 	& $70+130m/11$ with $m=0,\ldots,11$ 	\\
\hline
$\sigmaV$		& yes 	& $0,4.47213$ \\ 
\hline
$w$ 			& yes 	& $0.05,0.1,0.2$						\\
\hline
\hline
$\alphaV$ 		& no 	& $0.9$	\\
\hline
$N$ 			& no 	& $20$	\\
\hline
$\nis$ 			& no 	& $65536$ 	\\
\hline
$\tauc$ 		& no 	& $0.1$	\\
\hline
$\taum$ 		& no 	& $0.005$						\\
\hline
$\ntrials$ 		& no 	&  $20$							\\
\hline
$\Tc$ 			& no 	&  $200$						\\
\hline
\end{tabular}
\end{ruledtabular}
\end{table}

\begin{table}
\caption{Parameters of the seventh simulation set. \label{table:parameters7}}
\begin{ruledtabular}
\begin{tabular}{c|c|c}
variable	& varying & value/range\\
\hline
$\mui$ 			& yes 	& $40+50m/15$ with $m=0,\ldots,15$ 	\\
\hline
$\sigmaV$		& yes 	& $0,2,4.47213$ \\ 
\hline
\hline
$\alphaV$ 		& no 	& $0.9$	\\
\hline
$N$ 			& no 	& $20$	\\
\hline
$\nis$ 			& no 	& $32768$ 	\\
\hline
$\tauc$ 		& no 	& $0.1$	\\
\hline
$\taum$ 		& no 	& $0.005$						\\
\hline
$w$ 			& yes 	& $0.1$							\\
\hline
$\ntrials$ 		& no 	&  $200$							\\
\hline
$\Tc$ 			& no 	&  $200$						\\
\hline
\end{tabular}
\end{ruledtabular}
\end{table}

\subsection{Algorithm of finite-NIS effect reduction \label{sec:finitenis}}

\begin{figure}
\includegraphics[width=1.\textwidth]{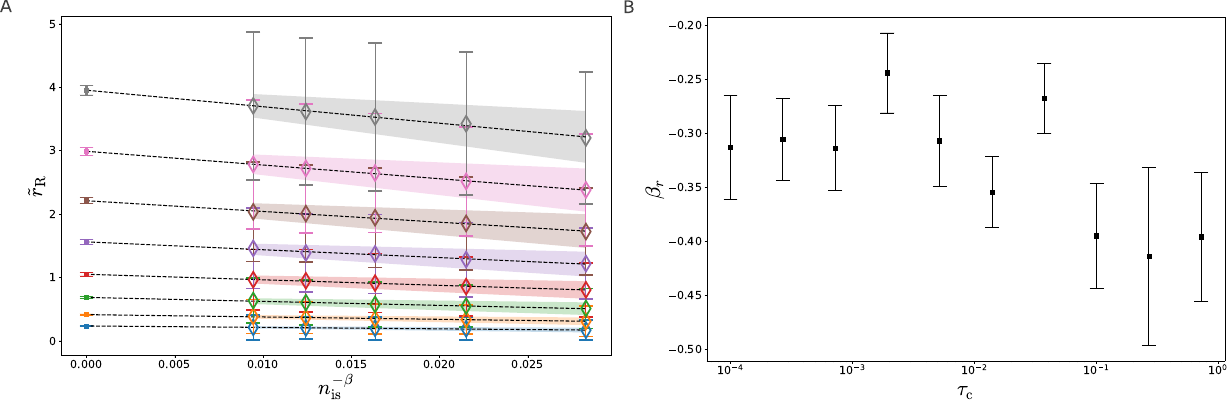}%
\caption{{\it Algorithm of finite-NIS-effect reduction.} (A) Numerical estimation of the mean and standard deviation ($\tilde\rro,\tilde\vro^{1/2}$, diamonds and error-bars, respectively) of the firing rate in single, finite-NIS simulations belonging to the first set (see Sec. \ref{sec:parameterfixing}), for $\tauc=0.7196$ and $\alphaV=0.9$, versus $\nis^{-\beta}$ (see Table \ref{table:parameters1}). Notice that the lowest value of $\nis=2^{14}$ has been omitted, so that $\nis=2^m$ with $m=15,16,17,18$. Different curves correspond to the different values of $\mui$ in the 1st simulation set. See Sec. \ref{sec:errorestimation} for details on the estimation of finite-NIS averages and standard deviation $\tilde\rro,\tilde\vro^{1/2}$. The dotted lines are a linear fit, and the filled area is compressed between the curves $r_\infty(\mui)+(b(\mui)\pm s_b(\mui))\,\nis^{\beta_r}$ (see Sec. \ref{sec:finitenis}), where $s_b(\mui)$ is the standard deviation of the fit result $b(\mui)$ for each $\mui$ (see \cite{repository} and \cite{lmfit}). The points at zero abscissa are the extrapolation of the fit results to $\nis\to\infty$, $r_\infty(\mui)$: these are the NIS-corrected values of $\tilde\rro$ and its error, in the 1st simulation set (and for this value of $\tauc,\alphaV$). (B) The fitted values of $\beta_r$ for each value of $\tauc$. \label{fig:rRregression}}
\end{figure}

Essentially, our finite-NIS correction algorithm consists in the following simple procedure. Given a simulation {for fixed $\tauc$}, we first estimate $\rro$ it for all values of $\mui,\nis$, obtaining $\tilde\rro(\mui,\nis)$ and its statistical uncertainty. Afterwards, we perform a regression of this function as a function of $\nis$, assuming a fitting function $\tilde\rro(\mui,\nis) \simeq r_\infty(\mui) + b(\mui)\, \nis^{\beta}$, but {\it forcing the exponent $\beta$ to assume a common value across the different data $\tilde\rro(\mui,\nis)$}, i.e., performing a unique fit for all the $\mui$ datasets, in which $\beta$ is $\mui$-independent and the rest of the fit parameters $r_\infty,b$ are $\mui$-dependent. The $\nis=\infty$ extrapolated regression parameter $r_\infty(\mui)$ is our estimation for the rate mean (for the given $\tauc$, and the rest of the parameters). Employing this regression protocol, we observe a lower $\chi^2$ variable per degree of freedom, than fixing $\beta$, or fitting it for each $\mui$ separately (please, see the details of the regression protocol in \cite{repository}). We repeat this procedure for each $\tauc$. In Fig. \ref{fig:rRregression}-A, we report the fitted values of $r_\infty$ for a given value of $\tauc$. In Fig. \ref{fig:rRregression}-B, we report the fitted values of $\beta$ vs $\tauc$. Note the low values of $\beta$, that make the raw estimation of the first passage time so slowly dependent on $\nis$ and, hence, so inaccurate --and, consequently, the finite-NIS effect mitigation procedure so necessary.  

Future completions of this work could include a comparison with alternative numerical schemes of integration of stochastic differential equations beyond Euler-Maruyama, as those based on the path integral \cite{bucher2015,chow2015}.

\subsection{Numerical estimation of the temporal distribution function $g$ \label{sec:numericalgt}}

In order for the numerical estimation of the temporal distribution function $g(t)$ to be as clear as in Fig. \ref{fig:g_quenched}, it is necessary to account for the finite-$\Tc$ bias of the estimation. In particular, we account for the finite-$\Tc$ bias, by computing {\it the probability couples of spikes to be at a distance $d=|t-t'|$, when the spikes are sampled with {\it uniform probability} in the $[0,T]$ time interval}. This cumulative probability can be easily seen to be $\text{Prob}\left(d<D\right)=2D/T-D^2/T^2$, and the probability density is: $\text{prob}\left(d\right)=2/T-2d/T^2$. Consequently, if we choose arbitrary bins $b_1<\ldots <b_B<T$, it is $\text{Prob}\left(b_{j-1}<D<b_j \right)=2(b_j-b_{j-1})/T-({b_j^2-b_{j-1}^2})/T^2$. 

Our $g$ estimation algorithm works, consequently, as follows: (i) we initialize to zero an array $\texttt g$ of length $B$. (ii) We put into ${\texttt g[i]}$ the number of couples of spikes $t,t'$ belonging to the same spike train, across all the $\ntrials$ spike trains in the simulation, whose distance $d=|t-t'|$ lies in the $i$-th bin: $b_{i-1}<d\le b_i$. (iii) $\texttt g[i]$ is redefined dividing it by $\text{Prob}\left(b_{i-1}<D<b_i \right)$. We expect that, for large enough $d$, the result is the expected value of the number of couples in an interval $\Tc$ (since $\text{Prob}\left(b_{i-1}<D<b_i \right) = $ (\# couples in the interval $\Tc$ whose distance lies in the $i$-th bin) divided by the (\# couples in the interval $\Tc$)). (iv) We consequently divide by the expected value of the number of couples of (independent) spikes in $\Tc$, $\<(\nro-1)\nro/2\>=(\rro\Tc)^2/2$, using our numerical estimation $\tilde\rro$ at the place of $\rro$; (v) Finally, we multiply this quantity by $\tilde\rro$, to obtain our estimation of $g$. In this way, the assymptotic value is, as required, $\rro$. Summarising:

\begin{itemize}
\item[(i)] initialize $\texttt g[i] = 0$ $\forall i=1,\ldots, B$ 
\item[(ii)] put in $\texttt{g[j]}$ the number of couples of spikes such that $b_{j-1}< |t-t'| \le b_j$, $\forall j$ and $\forall$ trials
\item[(iii)]  $\texttt{g[j]  /= } 2(b_j-b_{j-1})/\Tc-({b_j^2-b_{j-1}^2})/\Tc^2$, $\forall j$
\item[(iv)]  $\texttt{g[j] /= } \ntrials(\tilde\rro\Tc)^2/2$, $\forall j$
\item[(v)]  $\texttt{g[j] *= } \tilde\rro$, $\forall j$
\end{itemize}

\begin{figure}
\includegraphics[width=0.7\textwidth]{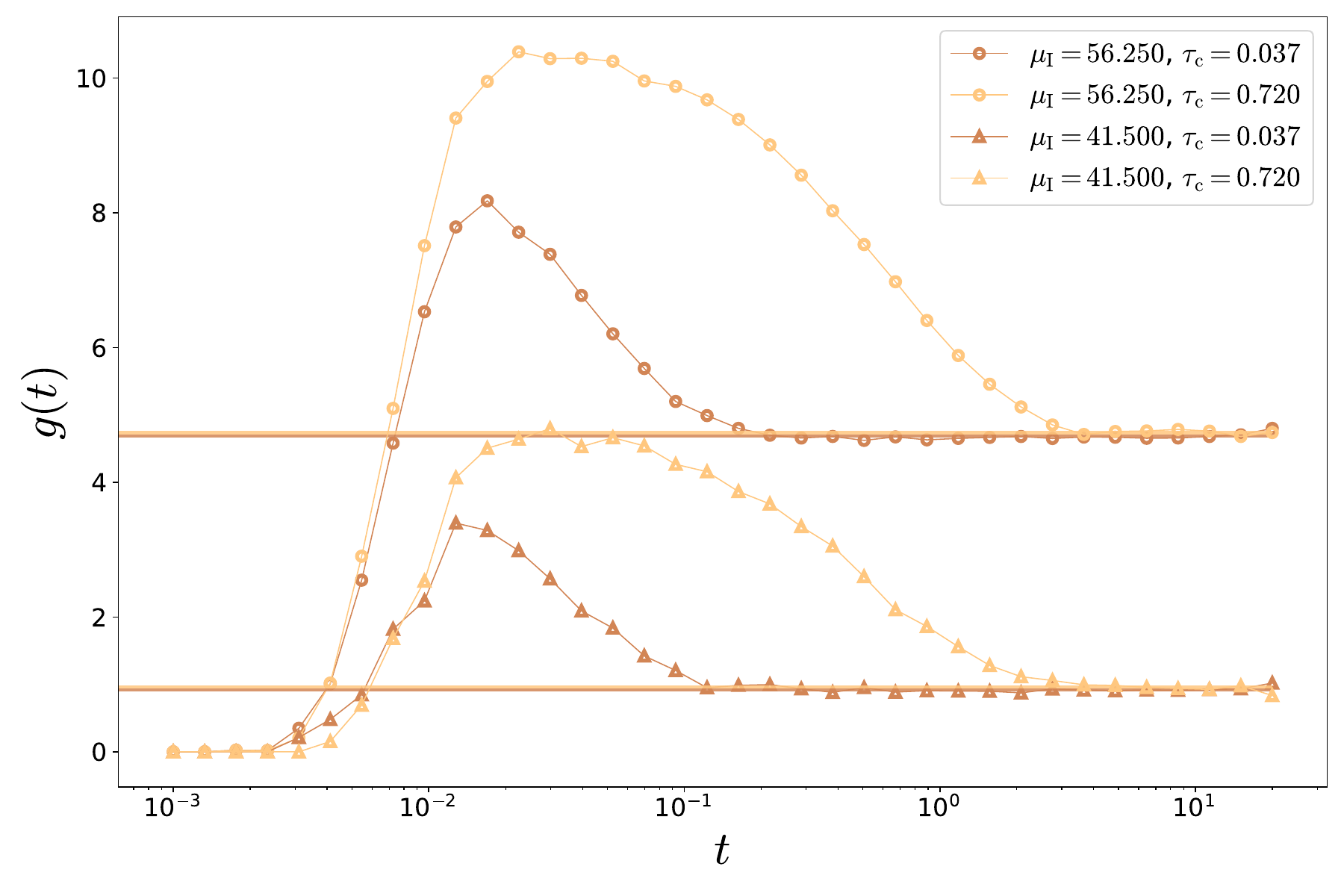}%
\caption{\label{fig:g_illustration} {\it Numerical estimation of the readout neuron temporal distribution function $g(t)$}. Different curves correspond to different values of $\mui$ and of $\tauc$ (see the figure key) in the 1st simulation set, with $\alphaV=0.9$ and $\nis=2^{18}$. The horizontal lines are the NIS-corrected estimations for $\rro$ corresponding to each $\mui,\tauc$. See the rest of the parameter values in Table \ref{table:parameters1}.}
\end{figure}

In Fig. \ref{fig:g_illustration} we illustrate the resulting $g$'s for two values of $\mui$ and of $\tauc$. We can see how, indeed, for large enough values of $t$, $g(t)$ becomes statistically compatible with $\tilde\rro$. This cannot be achieved without correcting for the finite-$\Tc$ bias.

\section{Validity range of the approximations\label{sec:validity}}

\subsection{Validity of the diffusion approximation for the encoding current \label{sec:validityencoding}}

The membrane potential of the LIF readout neuron in the Valente model is described, in the diffusion approximation, by the system of equations (\ref{eq:VI_diffusion}). Such approximation consists in approximating the statistics of the number of encoding spikes in the time window $\Delta$ (sampled from a Poisson distribution with average $\Lambda=\int_\Delta\d t\, {\bf w}\cdot {\bm \rho}(t)$) by the statistics of the normal distribution. For the diffusion approximation to the encoding current to be correct, the average encoding current in the diffusion approximation $\mui$, must be large enough in units of the noise amplitude, $(\sigmai^2+\sigmac^2)^{1/2}$ (the square root of the total spike count variance per unit time). If $\mui$ were not large enough in units of $(\sigmai^2+\sigmac^2)^{1/2}$, negative values of the spike count rate could exhibit non-negligible probability and, consequently, one would overestimate the count variance. We have assessed the validity of the diffusion approximation for the encoding current, which is illustrated for the 1st simulation in the inset of panel F of Fig. \ref{fig:valente_rates}: the numerical estimation of the standard deviation of the spike count variance (performed through the Jack-Knife method for increasingly large time window size $\Delta_{\rm J}$) is, indeed, statistically consistent with its theoretical value $(\sigmai^2+\sigmac^2)^{1/2}$. 

For completeness, in Fig. \ref{fig:approximations} we report the parameters $\alpha_{\rm M},\nu_{\rm W}$ and $\alpha_{\rm d}:=(\sigmac^{2}+\sigmai^2)^{1/2}/\mui$ (see Table \ref{table:symbols}), versus $\tauc$, for two values of $\vV(\alphaV)$, and for {\it the lowest value of $\mui$} in the second simulation. The largest value of $\alpha_{\rm d}$ is at most $\simeq 0.1$ implies that, in the worst case, the mean encoding rate is roughly ten times larger than its standard deviation. This confirms that the parameters of the simulations considered in the article lie in the validity range of the diffusion approximation for the encoding current.

\subsection{Construction of Figs. \ref{fig:valente_rates},\ref{fig:valente_rates_nocorrelations} \label{sec:figs23}}

Figs. \ref{fig:valente_rates},\ref{fig:valente_rates_nocorrelations} are performed in the following way. We first simulate a multivariate $N=6$-dimensional Ornstein-Uhlenbeck process $\x(t)$, solution of Eq. (\ref{eq:MOU}), in a time window $[0,\Tc]$ with $\Tc=200$. More specifically, we solve the unit-variance and zero-mean Ornstein-Uhlenbeck equation $\tauc\dot{\bm\xi}=-{\bm\xi}+2\tauc\,\bar C\cdot{\bm\eta}$, where $\bm\eta$ is a vector of white noises and ${\bar C}_{ij}=\delta_{ij}+(1-\delta_{ij})\alphaV^2$, and we therefore define $x_i(t)=\bar\mu^{(s)}+\sigmaV\,\xi_i(t)$. We show (panels A,B) the firing rates $x_{1,2}(t)$ of two of such $N=6$ neurons, in a shorter time window, $[0,T_{\rm plot}=4]$, and for two different values of the stimulus averages $\mu_{1,2}^{(s)}=\bar\mu^{(s)}$. We also generate a set of spikes sampled from the inhomogeneous Poisson distribution whose firing rate is $x_{1,2}(t)$. As we said in Appendix \ref{sec:numerical}, we do not simulate, for these figures, the LIF neuron with the inhomogeneous-Poisson distributed spikes (Eq. \ref{eq:afferentcurrent}) as afferent current. Rather, and consistently with the simulations analysed in the main article (in which we consider the solution of the LIF model in the diffusion approximation, Eqs. \ref{eq:VI_diffusion}), we solve the stochastic differential equation for $V$ in Eq. \ref{eq:I_diffusion} but using, as input current $I(t)$:

\begin{subequations}
\label{eq:diffusionN}
    \begin{align}
I(t) &= w\left( \sum_{i=1}^N x_i(t) + \bar\mu^{1/2}\sum_{i=1}^N\eta_i'(t) \right) =\\
 &= w\left( N\bar\mu + \sigmaV \sum_{i=1}^N \xi_i(t) + \bar\mu^{1/2}\sum_{i=1}^N\eta_i'(t) \right)
\end{align}
\end{subequations}
where ${\bm\eta}'(t)$ is a different realization of a white noise that simulates, in the diffusion approximation for the single neuron current, and consistently with (\ref{eq:multivariate densities}), the randomness of the inhomogeneous Poisson count statistics of each encoding neuron. Finally, in panel G, we show our estimation for the average $\hat \rro$ and standard deviation $\hat\vro_{\Delta_{\rm J}}^{1/2}$ of readout firing rate corresponding to such an encoding current. Our estimations for the readout spike count standard deviation {\it in such a single realization of the stochastic differential equation}, are performed with the Jack-Knife method, in the whole interval of length $\Tc=200$. In the inset of panel 3, we show the Jack-Knife estimation as a function of the window length $\Delta_{\rm J}$. As a final estimation, we choose the estimation corresponding to the third to the last estimation for $\Delta_{\rm J}\simeq 2$. The black box-plots in panel G correspond to the set of readout firing rates in the simulation, in time windows of length $\Delta_{\rm r}=0.5$. The upper error-bar of the readout rate in panel G has length $=\hat\vro_{\Delta_{\rm J}}^{1/2}$, while the lower error-bar has length $={\rm min}[\hat\vro_{\Delta_{\rm J}}^{1/2},\hat\rro]$. 

We will refer to this model, in which we simulate the LIF RO receiving the sum of $N$ spike currents in the diffusion approximation, defined by equations (\ref{eq:vLIF},\ref{eq:diffusionN}), as {\it the diffusion-N model}. In order to evaluate the SNR of the diffusion-N model with the parameters of Figs. \ref{fig:valente_rates},\ref{fig:valente_rates_nocorrelations}, we have performed the so called {\it 4th simulation}. It consists in $\ntrials=200$ realizations of the solution of Eqs. (\ref{eq:vLIF},\ref{eq:diffusionN}) for two values of $\mui^{(\pm)}=80,90$, for two values of $\alphaV=0,0.9$, for three values of $N=6,10,20$, and for two values of $\sigmaV=0,\sigmaV^{(N=20)}\times 20/N$, where we take the value $\sigmaV^{(N=20)}=4.4721$. 

We present the results for the SNR per unit time (confidence intervals at $95\%$ of probability) in Fig. \ref{fig:diffusionN}. In the figure, the parameters corresponding to those of Figs. \ref{fig:valente_rates},\ref{fig:valente_rates_nocorrelations} are: $\alphaV=0.9$, $\sigmaV=14.907$, $N=6$, and $\alphaV=0$, $\sigmaV=14.907$, $N=6$, respectively. First, we observe that the numerical simulations of the diffusion-N model are consistent with the theoretical predictions for the diffusion model (horizontal filled intervals in Fig. \ref{fig:diffusionN}, taken from Fig. \ref{fig:SNR_varyingsigmac}). Second, only the SNR corresponding to $\alphaV=0.9$ is significantly higher. Indeed, for each value of $N$, we have performed three simulations in the 4th simulation set (see Fig. \ref{fig:diffusionN}): one for $\alphaV=0.9,\sigmaV=\sigmaV^{(N=20)}\times(20/N)$; one for $\alphaV=0,\sigmaV=\sigmaV^{(N=20)}\times(20/N)$; and one for $\alphaV=0,\sigmaV=0$. The first one corresponds to correlated, noisy-rate encoding neurons; the second one, to uncorrelated noisy-rate encoding neurons; the third one, to uncorrelated, deterministic-rate $x_i(t)=\bar\mu$ encoding neurons. As one can see in the figure, and in Eq. (\ref{eq:dictionary_sigmac}), $\sigmac$ is non zero even if $\alphaV=0$, whenever $\sigmaV\ne 0$. This contribution to $\sigmac$ in absence of noise correlations comes from the self-correlation of every neuron, that induces a small self-correlation in $I(t)$ ($N$ times smaller than the contribution induced by the neuron correlations, proportional to $\alphaV$). We make notice that, as we can see in Fig. \ref{fig:diffusionN}, this small value of $\sigmac$ is not enough to significantly enhance the SNR with respect to the zero-noise case $\sigmaV=0$: the SNR enhancement occurs in the presence of across-neuron noise correlations only. 

\subsection{Scaling for large $N$ \label{sec:Nscaling}}

\begin{figure}
\includegraphics[width=0.7\textwidth]{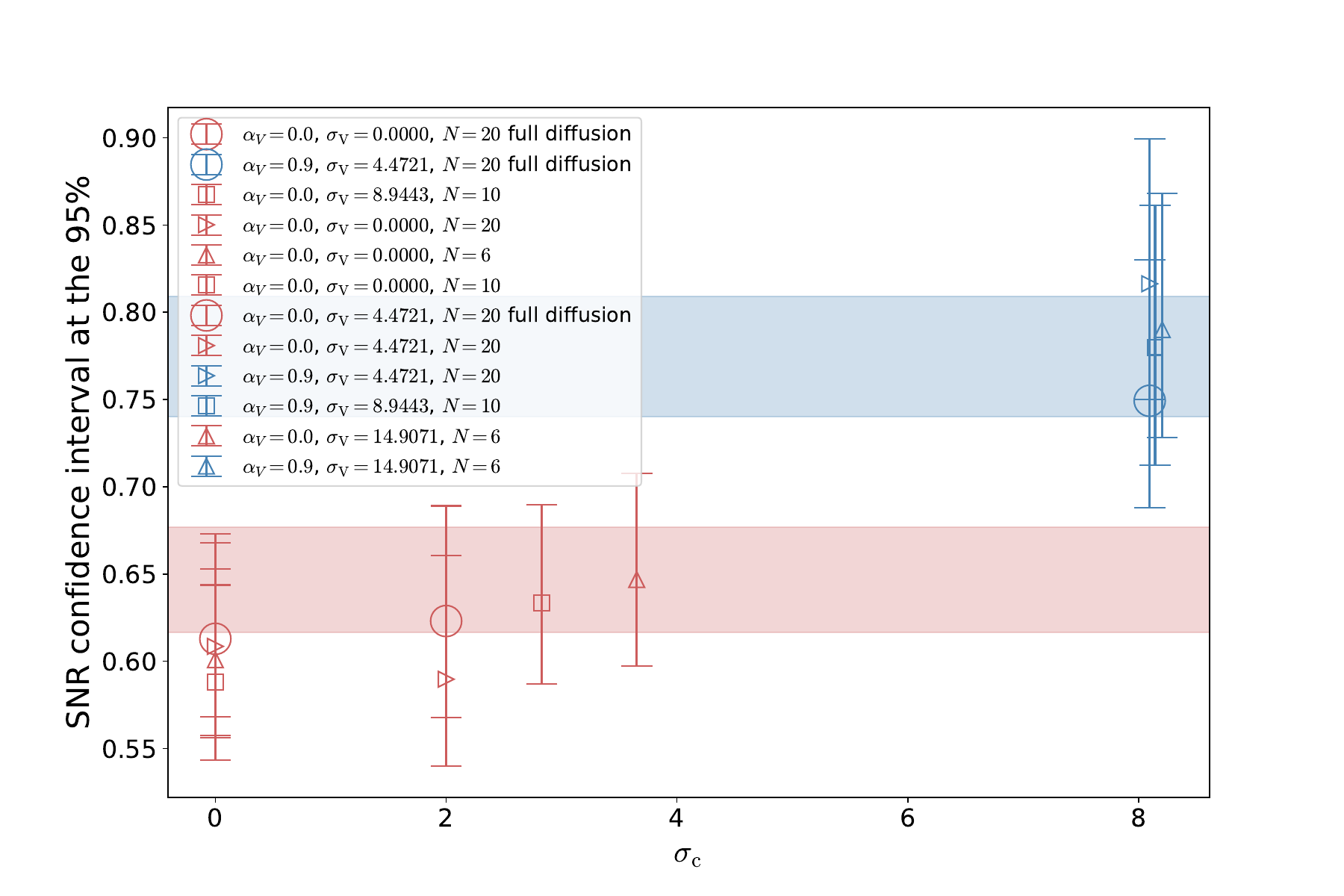}%
\caption{\label{fig:diffusionN} {\it Comparison of the signal-to-noise ratio SNR in the diffusion-N NERM and in the full diffusion NERM}, for several values of $N$, $\alphaV$, and $\sigmaV$, versus $\sigmac$ (see Eq. \ref{eq:sigmacvsN}). The horizontal filled intervals correspond to the theoretical SNR confidence intervals at the 95\% (from Fig. \ref{fig:SNR_varyingsigmac}). See the rest of the parameter values in Table \ref{table:parameters4}.}
\end{figure}

As said above, in Fig. \ref{fig:diffusionN} we present an analysis of the diffusion-N model for varying values of $N=6,10,20$. We remark that, in our simulations of the diffusion-N model, the mean and noise amplitude of the encoding neurons' firing rates $x_i$ are scaled with $N$ using the natural scaling: $\mu_i\propto 1/N$, $\sigmaV\propto 1/N$, which amounts to scale by {\it a single}, $N$-dependent quantity the firing rate $x_i(t)=\bar\mu+\sigmaV\xi(t)$. More precisely:

\begin{subequations}
\label{eq:scaling}
\begin{align}
\bar\mu &= \bar\mu^{(N=N_0)}\times\frac{N_0}{N} \\
\sigmaV &= \sigmaV^{(N=N_0)}\times\frac{N_0}{N} 
\end{align}
\end{subequations}
in such a way that the mean input current $\mui=wN\bar\mu$ is constant, independent of $N$, and that 

\begin{align}\label{eq:sigmacvsN}
\sigmac^2 &=  w^2\sigmaV^2 N^2 \left(\frac{N-1}{N}\alphaV^2+N^{-1}\right) \simeq\\
&\simeq w^2 \sigmaV^{(N=N_0)}\alphaV^2 + O[N^{-1}]
\end{align}
becomes more and more independent of $N$, the larger the value of $N$. In Fig. \ref{fig:diffusionN} we illustrate the behavior of the SNR confidence intervals vs $\sigmac$ in three simulations with increasing values of $N$ (but with a common value of $\mui=wN\bar\mu$ and of $N\sigmaV$, i.e., in the presence of the scaling in (\ref{eq:scaling})). As we have just mentioned, specially for high values of $\alphaV$ (see the blue symbols with $\alphaV=0.9$ in the figure), the value of $\sigmac$ quickly becomes independent of $N$. As we will see, this implies that also the SNR becomes independent of $N$ for large $N$.

Finally, in Fig. \ref{fig:diffusionN} we also present a comparison between the diffusion-N NERM and the {\it full diffusion} NERM defined by Eq. (\ref{eq:VI_diffusion}), that we have analysed in this article, and whose current is (compare with Eq. (\ref{eq:diffusionN})):

\begin{align}
I(t) &= \mui + \sigmai \eta(t) + \sigmac \xi(t)
\end{align}
where the parameters $\mui,\sigmai,\sigmac$ are given by Eqs. (\ref{eq:dictionary}). Now, since the simulations analysed in this article, and the 4th simulation corresponding to Figs. \ref{fig:valente_rates},\ref{fig:valente_rates_nocorrelations} lie in the validity range of the diffusion approximation for the encoding, the SNR properties of the diffusion-N model and of the corresponding full diffusion model are statistically consistent. As far as both models are equivalent, the dependence of the readout properties on the colored noise are given by $\sigmac$ only. For larger and larger values of $N$, and in the presence of the scaling in Eq. (\ref{eq:scaling}), we expect a value of the the full diffusion parameters $\mui,\sigmai,\sigmac$ which is quasi-independent of $N$. As a consequence, also the SNR and the readout properties quickly become independent of $N$, as figure \ref{fig:diffusionN} illustrates. 

\subsection{Validity of the diffusion approximation for the readout firing rate mean and variance \label{sec:validityreadout}}

The analytical results presented in this article exploit the Siegert result for the firing rate of the LIF model with white-colored noise current, Eq. (\ref{eq:rwhiteLIF}), that was derived in the diffusion approximation \cite{burkitt2006}. Consequently, {\it our analytical results refer to the LIF model with coloured input noise in the diffusion approximation, Eq. (\ref{eq:VI_diffusion})}. In this article, we have numerically demonstrated that our analytical approximations explain quantitatively the full diffusion NERM in Eq. (\ref{eq:VI_diffusion}). It is a natural question to ask to what extent the diffusion approximation itself is accurate in the region of parameters that we explore in this article or, in other words, to what extent Eq. (\ref{eq:VI_diffusion}) is a good model of the {\it LIF model with spiking encoding current} in Eqs. (\ref{eq:vLIF},\ref{eq:afferentcurrent}). We will show in this section that, actually, the diffusion approximation is not accurate in the fluctuation-driven regime of the LIF model, in which the SNR enhancement phenomenon occurs. However, the SNR enhancement phenomenon, and its qualitative dependence with $\mui^{(\pm)}$ and $\sigmac$ explained in Sec. \ref{sec:valentemodelsolution} are also present in the LIF model with spiking encoding current, albeit in range of parameters that are shifted with respect to those that apply for the diffusion LIF model in Eq. (\ref{eq:VI_diffusion}). Therefore, the SNR enhancement effect and its dependence on the relevant quantities are not an artifact of the diffusion approximation. 

\begin{figure}
\includegraphics[width=0.7\textwidth]{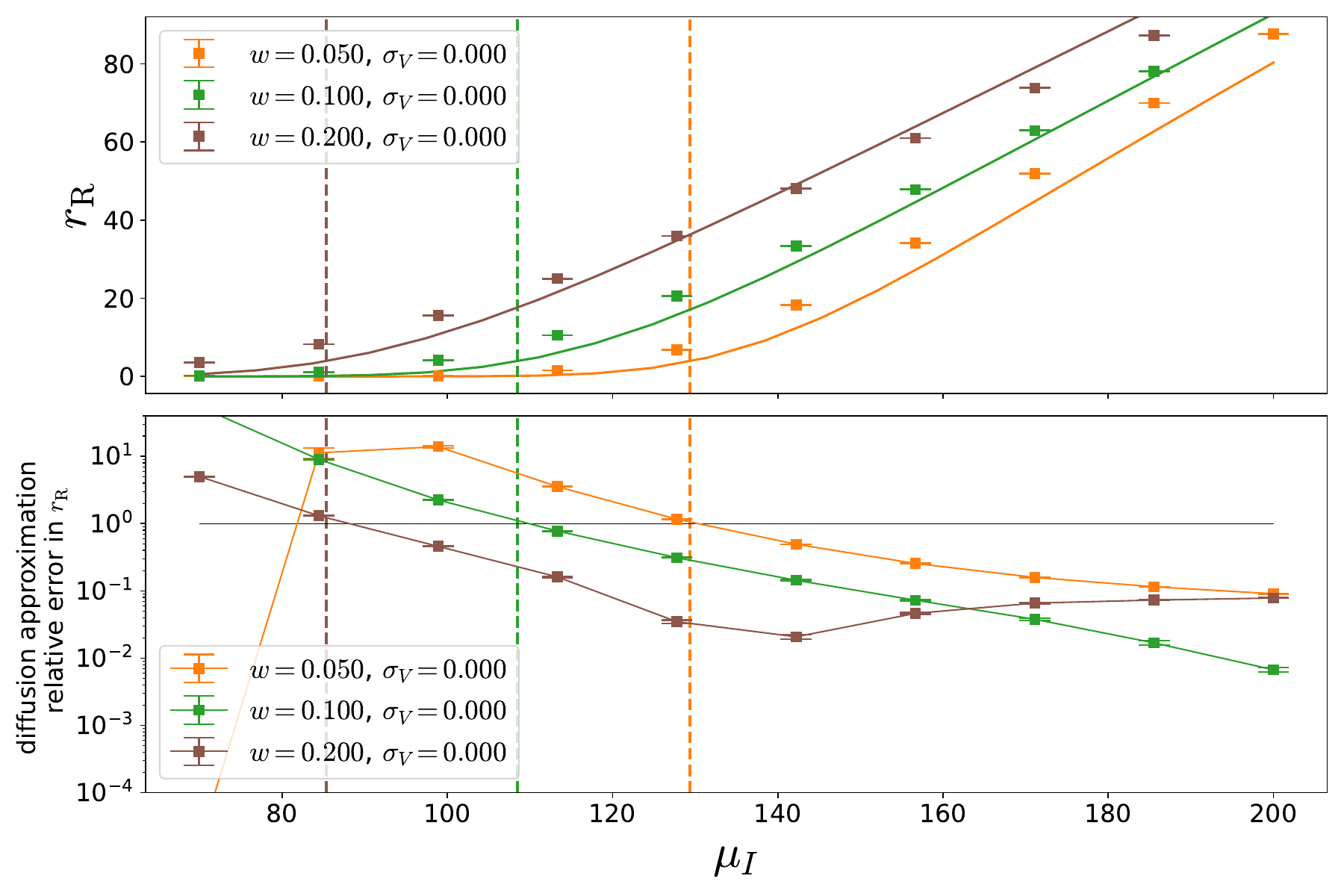}%
\caption{\label{fig:diffusion_Lifspikes} {\it Comparison between the firing rate of the LIF model with spiking encoding current, and the diffusion approximation (absence of encoding noise correlations).} Upper panel: white-noise LIF firing rate in diffusion approximation, $\rw$ (curves), and numerical estimation of the firing rate in the spiking encoding current LIF simulations (points), $\tilde\rro$, versus $\mui$. The points' error bars are smaller than the symbol sizes. Different colors correspond to different values of $w$, in the legend. Lower panel: relative error $|\rro-\rw|/\rw$ versus $\mui$. The dashed vertical lines signal the value of $\hat\mui$ for each value of $w$. See the rest of the parameter values in Table \ref{table:parameters6}.}
\end{figure}

\begin{figure}
\includegraphics[width=0.7\textwidth]{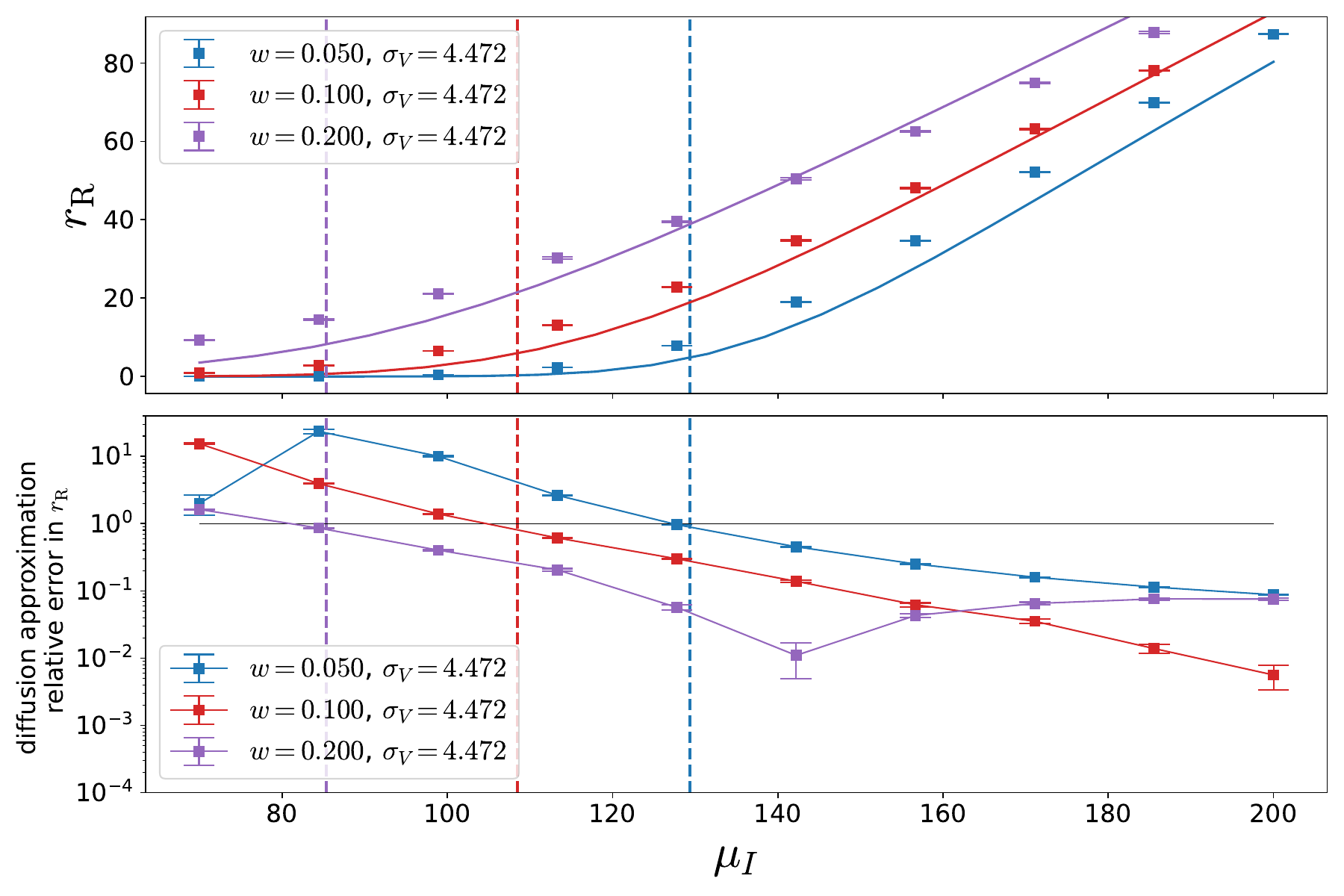}%
\caption{\label{fig:diffusion_Lifspikes_noise} {\it Comparison between the firing rate of the LIF model with spiking encoding current, and the diffusion approximation (with encoding noise correlations).} As Fig. \ref{fig:diffusion_Lifspikes} but for a nonzero colored noise amplitude $\sigmaV=4.4721$. We use the quenched-noise analytical expression for $\rq$ as a theoretical value. See the rest of the parameter values in Table \ref{table:parameters6}.}
\end{figure}

We have simulated (the {\it 6th set of simulations}, see Table \ref{table:parameters6}) the LIF model with spiking encoding current with the same parameters $\tauc,\taum,N,\sigmaV$ of the fourth simulation, in a finer and broader grid of values of $\mui$ and with three values of the gain $w$. The simulations of the LIF model with spiking encoding current are performed through an event-driven exact integration algorithm \cite{brette2007,pittorino2017,repository}. The comparison between the diffusion approximation for the mean readout firing rate $\rw$ (in the case of zero colored noise amplitude $\sigmaV=0$) and the numerical firing rate of the LIF model with spiking encoding current is shown in Fig. \ref{fig:diffusion_Lifspikes}. Figure \ref{fig:diffusion_Lifspikes_noise} is a variant of this analysis for nonzero encoding noise correlations, $\sigmaV=4.4721$.
In the figures, the vertical lines signal the value of $\hat\mui$ for each value of $w$, or the upper bound derived in section \ref{sec:valentemodelsolution} for the value of $\mui^{(+)}$ below which the SNR enhancement can occur. We observe that, despite the absolute error decreases for decreasing $w$, as expected, the relative error (lower panels in Figs. \ref{fig:diffusion_Lifspikes},\ref{fig:diffusion_Lifspikes_noise}) actually decreases with $w$ for given values of $\mui$. In any case, we observe that in the interesting region of $\mui<\hat\mui$ in which the phenomenon that we analyse occurs, the relative error of the diffusion approximation becomes higher than one (i.e., of the same order, or higher than the theoretical value itself) {\it for all the values of $w$}. This picture holds unchanged in the presence of noise correlations, and also for wider ranges of $w$. Although we expect the relative error to decrease for LIF models with lower values of $\taum$, we conclude that, in the regime in which we observe the SNR enhancement, the diffusion approximation is not accurate in general. These results are consistent with previous assessment of the accuracy of the diffusion approximation \cite{helias2010,cupera2014}, that reveal that it is only relatively accurate.


In this circumstance, a natural question is whether the behavior of the SNR that we described in Sec. \ref{sec:valentemodelsolution} is an artifact of the diffusion approximation. We checked whether the NERM with spiking encoding current actually exhibits the behavior predicted by the theory (Sec. \ref{sec:valentemodelsolution}) and by the NERM in diffusion approximation, Eq. (\ref{eq:VI_diffusion})). In this section, we find that this is the case.

\begin{figure}
\includegraphics[width=0.7\textwidth]{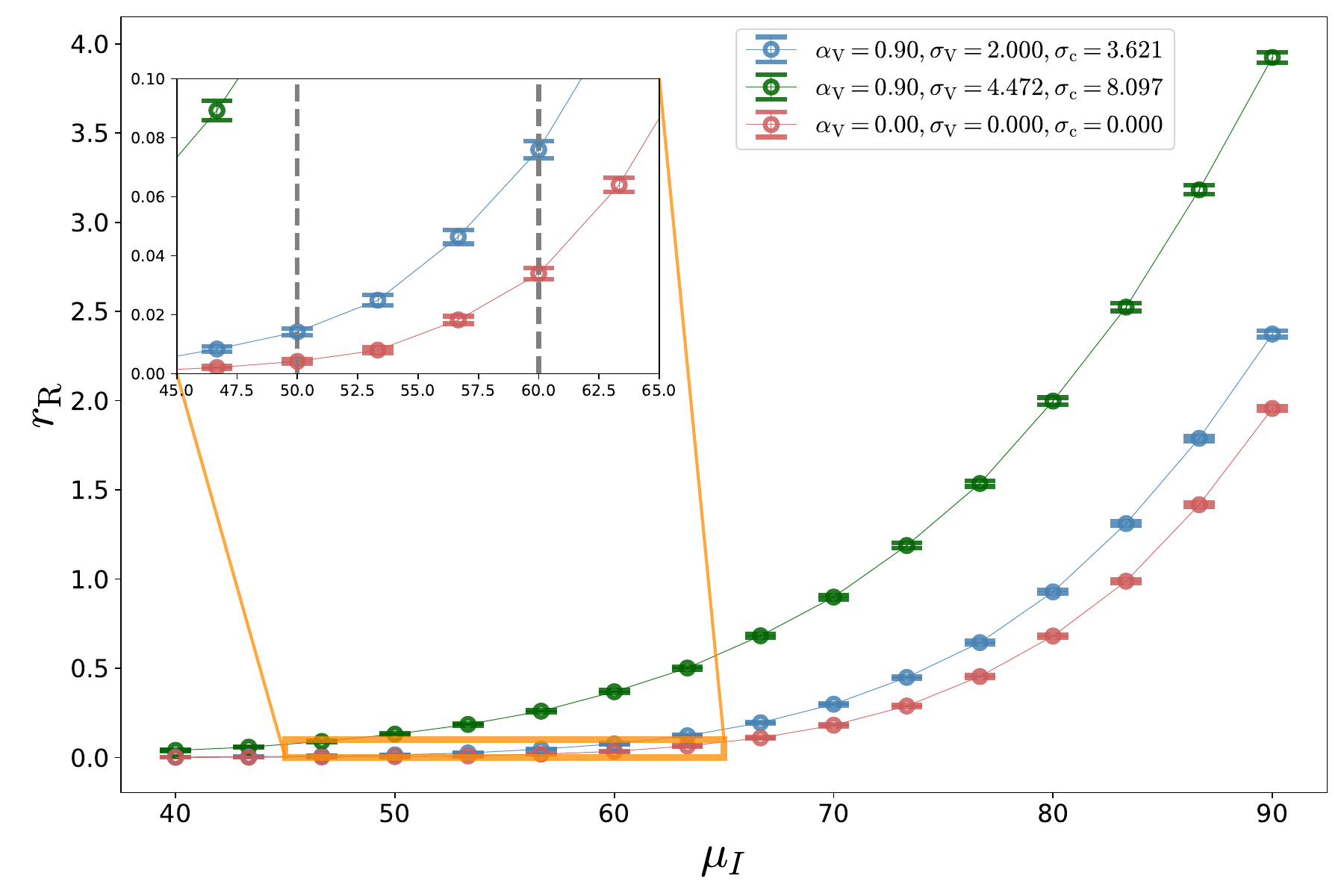}%
\caption{\label{sec:LIFspikessimulation} {\it Summary of the 7th simulation}, or the simulation of the spiking encoding current LIF, that we use to evaluate the SNR as a function of $\mui^{(\pm)}$ and $\sigmaV$. We plot the numerical estimation of $\rro$ versus $\mui$, for three considered values of $\sigmaV$. The lines are guides to the eye. The dashed vertical lines in the inset and in the main figure signal, respectively, the two values of $\mui^{(\pm)}$ used to compute the SNR in panels A,B of figure \ref{fig:SNR_LIFspikes}. See the rest of the parameter values in Table \ref{table:parameters7}. }
\end{figure}

\begin{figure}
\includegraphics[width=0.7\textwidth]{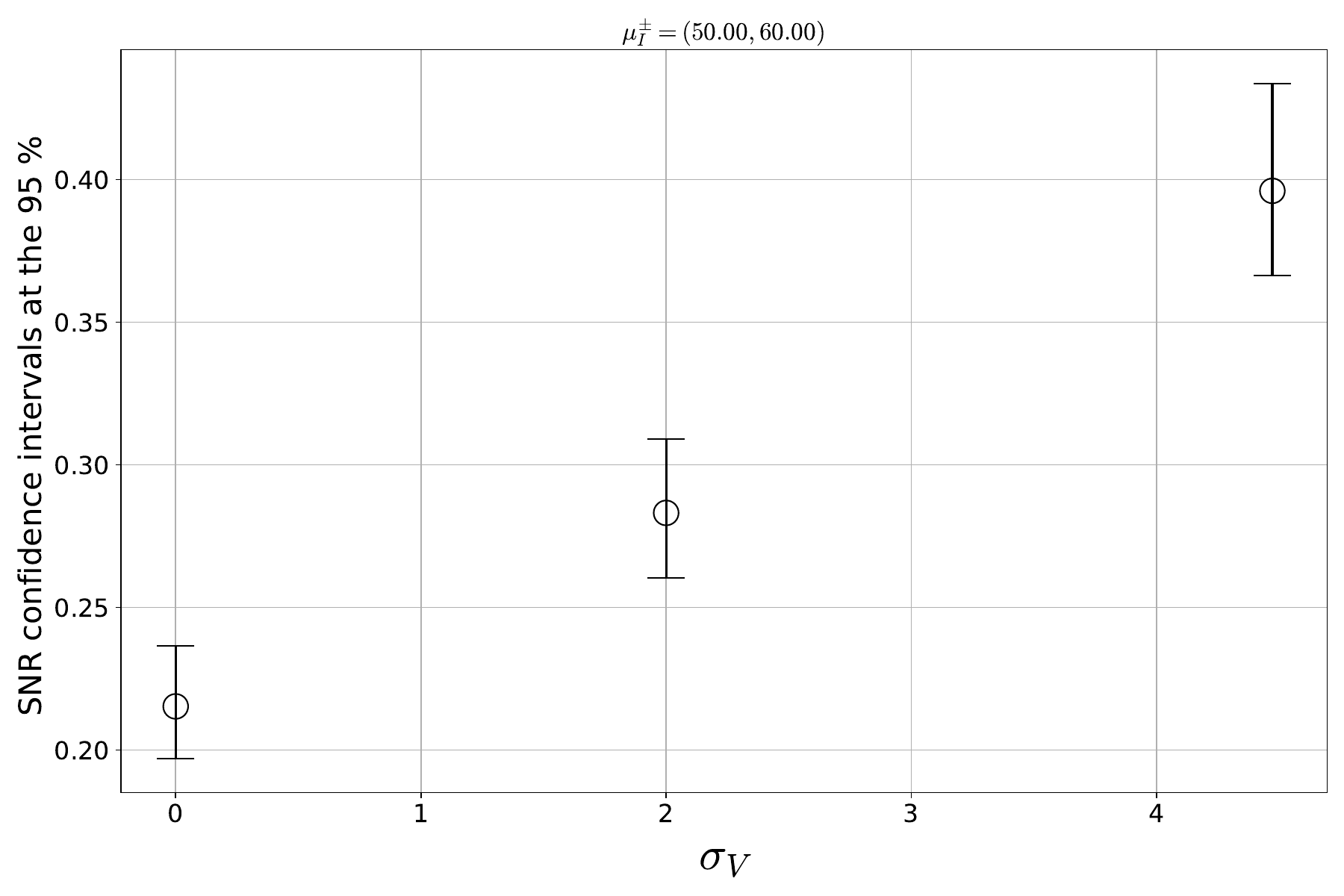}%
\caption{\label{fig:SNR_LIFspikes} {\it SNR of the NERM with spiking encoding current}, versus $\sigmaV$, taking $\mui^{(\pm)}=50,60$. See the rest of the parameter values in Table \ref{table:parameters7}.}
\end{figure}

We have assessed this with the help of the {\it 7th simulation} (see Table \ref{table:parameters7}), whose results are summarized in Fig. \ref{sec:LIFspikessimulation}. The figure serves as a guide for our numerical tests of the SNR of the NERM with spiking encoding current, that we present in Fig. \ref{fig:SNR_LIFspikes}, where we show the SNR for the couple $(\muip,\muim)=(50,60)$. These values, $\mui^{(\pm)}$ are marked with vertical lines in the inset of Fig. \ref{sec:LIFspikessimulation}, to show them in relation to the readout firing rate for three different values of $\sigmaV$. In Fig. \ref{fig:SNR_LIFspikes}, we show how, for the interval $(\muip,\muim)=(50,60)$, the SNR significantly increases with $\sigmaV$, in such a way that noise correlations indeed induce an enhancement of the SNR, as far as both $\mui^{(\pm)}$ are low enough. This result is also consistent with \cite{valente2021}.

\begin{figure}
\includegraphics[width=0.7\textwidth]{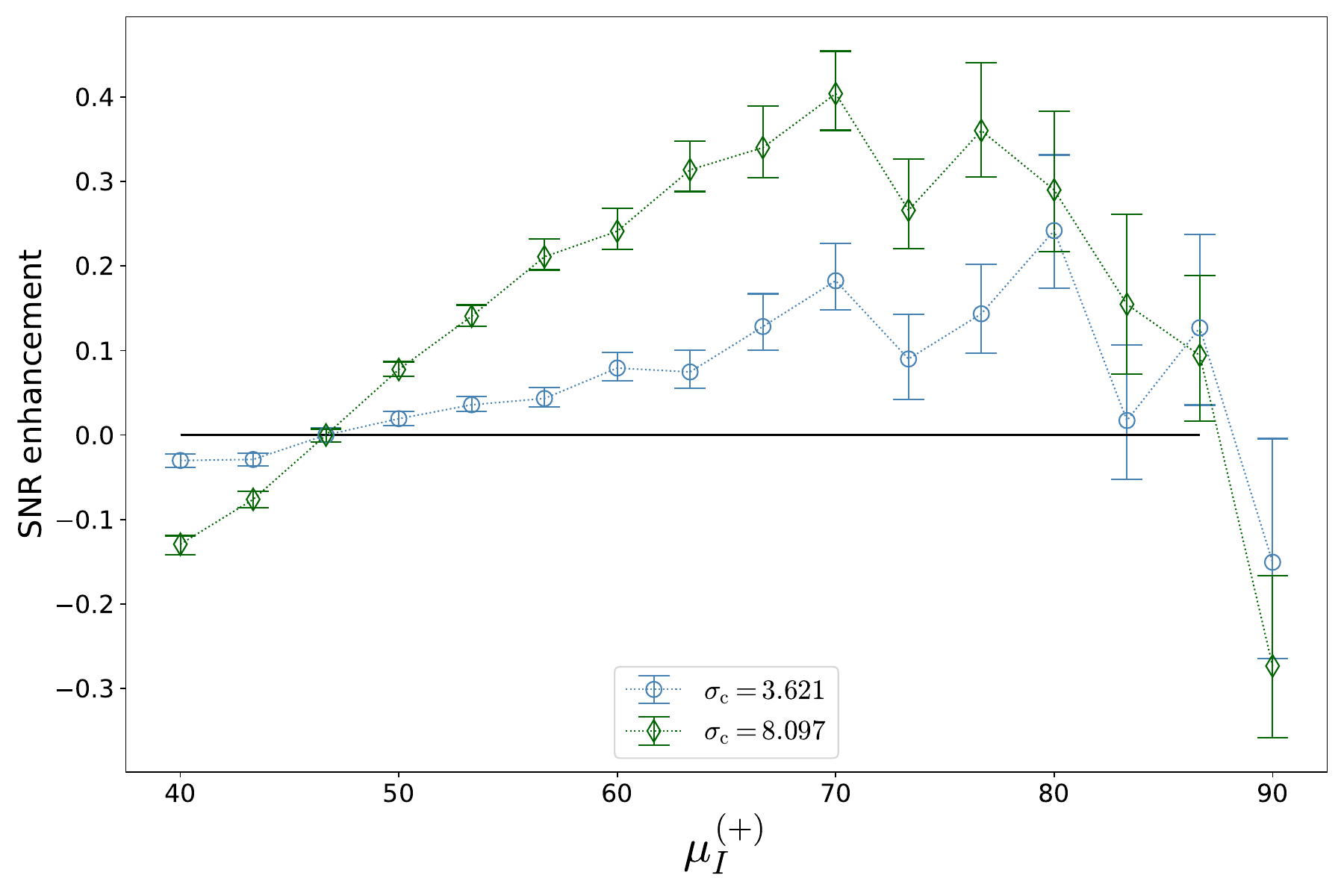}%
\caption{\label{fig:SNR_LIFspikes_muim46} {\it Excess of SNR with respect to the white-noise model for $\sigmac=0$ in the NERM with spiking encoding current}, versus $\muip$ for a fixed value of $\muim=46.667$. Error bars are the confidence intervals at the $75\%$ through bottstrapping across data realizations. Red circles and green diamonds correspond $\sigmac=3.621$ and $8.097$, respectively (or, alternatively, to $\sigmaV=2$ and $4.472$, with $\alpha=0.9,\N=20$). See the rest of the parameter values in Table \ref{table:parameters7}.}
\end{figure}

\begin{figure}
\includegraphics[width=0.7\textwidth]{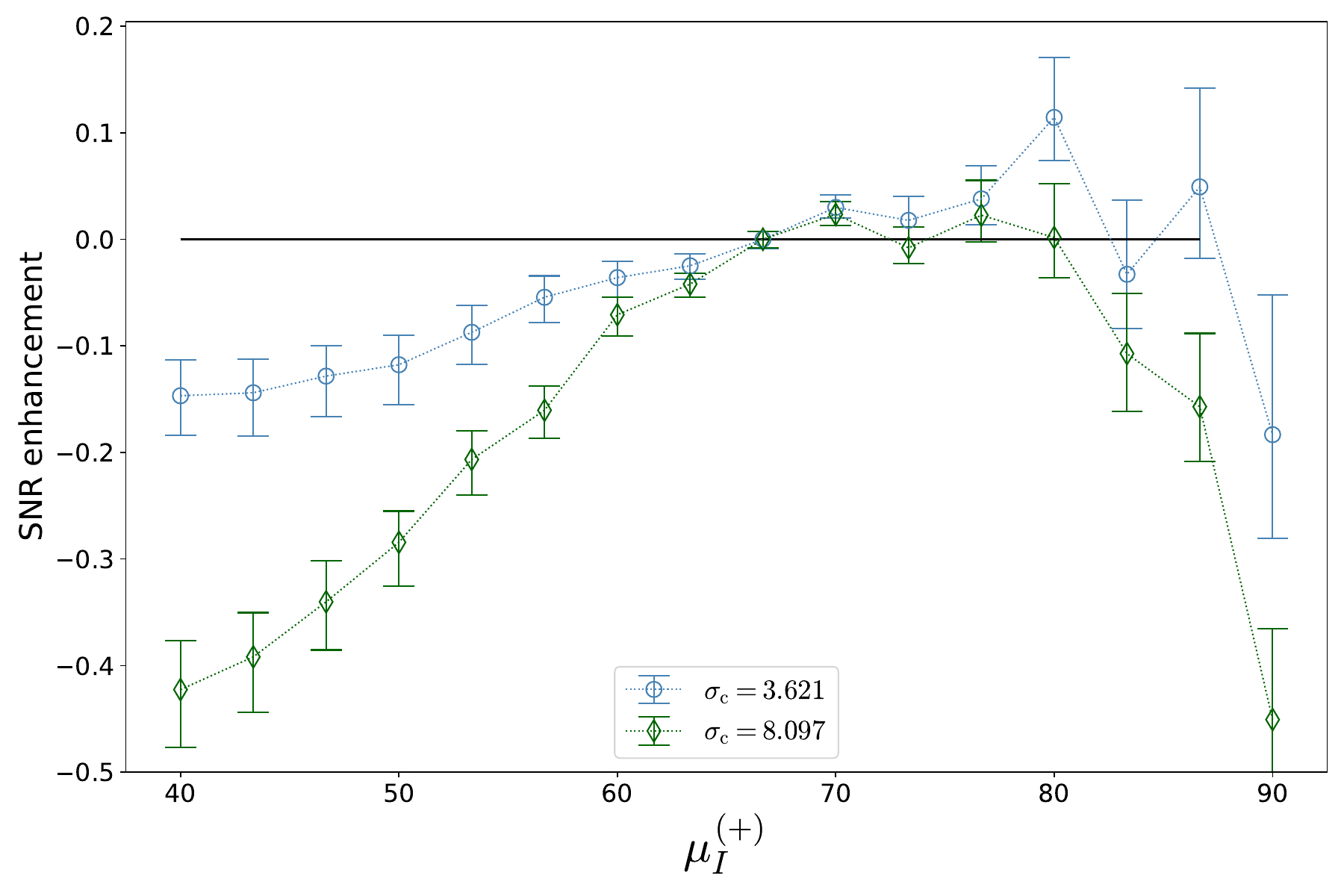}%
\caption{\label{fig:SNR_LIFspikes_muim73}  {\it Excess of SNR with respect to the white-noise model for $\sigmac=0$ in the NERM with spiking encoding current}, versus $\muip$, as in Fig. \ref{fig:SNR_LIFspikes_muim46}, but for $\muim=73.33$. See the rest of the parameter values in Table \ref{table:parameters7}.}
\end{figure}

\begin{figure}
\includegraphics[width=0.7\textwidth]{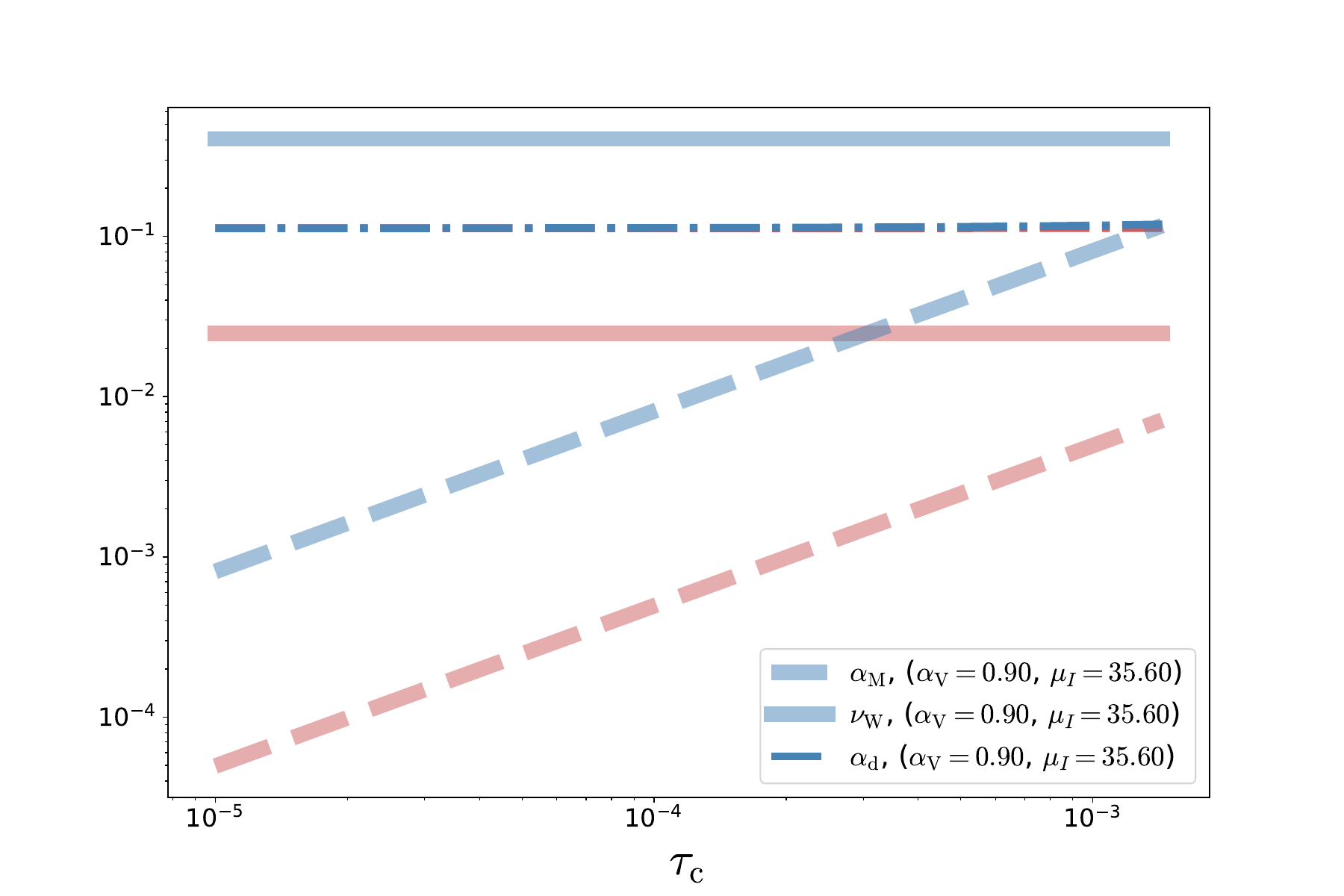}%
\caption{\label{fig:approximations} {\it The approximation parameters $\alpha_{\rm M},\nu_{\rm W}$ and $\alpha_{\rm d}:=(\sigmac^{2}+\sigmai^2)^{1/2}/\mui$, vs $\tauc$}, for $\mui=35.6$ and both values of $\alphaV={0,0.9}$.}
\end{figure}

Furthermore, also the non-monotonic behaviors of the SNR versus $\muip$ and $\sigmaV$ described in Sec. \ref{sec:valentemodelsolution} are observed in the spiking current NERM. We illustrate this in Figs. \ref{fig:SNR_LIFspikes_muim46},\ref{fig:SNR_LIFspikes_muim73}, where we plot the excess of SNR, $\text{SNR}-\text{SNR}_\text{w}$ versus $\muip$, for two values of $\muim=46.67,73.33$ (in Figs. \ref{fig:SNR_LIFspikes_muim46},\ref{fig:SNR_LIFspikes_muim73} respectively), and where of course $\text{SNR}_\text{w}$ is the SNR for the white noise case with $\sigmac=0$ (red points in Fig. \ref{sec:LIFspikessimulation}). First, Fig. \ref{fig:SNR_LIFspikes_muim46} illustrates the non-monotonic dependence of the SNR on $\muip$: when $\muip$ approaches the inflection point $\mui^*$ of the $\rw$ curve, also the SNR numerator $\Delta\rro$ begins decreasing with $\muip$, and so the SNR. Second, Fig. \ref{fig:SNR_LIFspikes_muim73} illustrates that, for high values of $\muim$, the readout SNR is no longer enhanced by encoding noise correlations. Moreover, for large enough $\muim$, the SNR decreases with $\sigmac$. This is consistent with the qualitative dependence on $\sigmac$ that we described in Sec. \ref{sec:conditions}. Given $\mui^{(-)}$, there is a value of $\sigmac$ beyond which the SNR decreases with $\sigmac$. We expect, in this way, that for sufficiently low $\muim$, the SNR increases with $\sigmac$, while for sufficiently high $\muim$, the SNR decreases with $\sigmac$. Vice-versa, the value of $\mui^{(-)}$ below which the SNR enhancement is observed, decreases for increasing $\sigmac$. Figs. \ref{fig:SNR_LIFspikes_muim46},\ref{fig:SNR_LIFspikes_muim73} clearly suggest (compare with Figs. \ref{fig:snr_vs_muI},\ref{fig:snr_vs_muImin}) that the behavior of the NERM with spiking encoding current exhibits the same qualitative behavior than its counterpart in the diffusion approximation.

We conclude that, despite the diffusion approximation does not accurately reproduce the behavior of the NERM with spiking encoding current in Eqs. (\ref{eq:MOU},\ref{eq:vLIF},\ref{eq:afferentcurrent}) in the relevant region of parameters, the theory explains qualitatively well the behavior of this model (besides reproducing quantitatively well the behavior of the NERM in the diffusion approximation, Eqs. (\ref{eq:VI_diffusion})). 

As a remark note, it is somehow natural that the SNR enhancement effect and its dependence on the relevant quantities are not an artifact of the diffusion approximation. This behavior can, indeed, be simply explained in terms of some essential features of the analytic solution, as the faster dependence on $\sigmac$ of the over-Poissonian correction to the rate variance (the SNR denominator), with respect to the rate mean (the SNR numerator, see Sec. \ref{sec:valentemodelsolution}). As far as we expect such general features to apply for the LIF model with spiking encoding current, we also expect the SNR behavior in the NERM to stay qualitatively unchanged in the NERM with spiking encoding current. 

An in-depth analysis of the spiking encoding current NERM (for instance, its dependence on the NIS), and its relation with the theory presented here is out of the scope of the present article.

\section{Quenched-noise approximation for the LIF neuron firing rate mean \label{sec:quenched_mean}}

\subsection{The Moreno \textit{et al.} solution}

As said in the main text, Moreno et al have shown in \cite{moreno2002} that, in the $\tauc\gg\taum$ limit on which we focus, the average firing rate of the coloured-noise LIF model $\rro$ may be approximated, {\it to first order in $\taum/\tauc$}, and {\it under the further approximations} of small $\alpha_{\rm M} = \sigmac^2/\sigmai^2$ and small $\mui/(\sigmai^2+\sigmac^2)^{1/2}$ (needed for the diffusion approximation), as:

\begin{subequations}
\label{eq:r_moreno}
\begin{align}
\rro(\mui) &\simeq \rw(\mui) \left[ 1+\frac{\rw(\mui)\taum^2\alpha_{\rm M}}{\tauc} \left(\taum \rw(\mui)\left[R(\Theta_0)-R(H_0)\right]^2- \frac{\Theta_0 R(\Theta_0)-H_0 R(H_0)}{\sqrt{2}}\right) \right] \\
\alpha_{\rm M} &:={\sigmac^2}/{\sigma_I^2} 
\end{align}
\end{subequations}
where $\rw(\mui)$ is the mean firing rate of a LIF model {\it with white input noise} (or $\sigmac=0$), taking the well-known expression due to Siegert \cite{siegert1951,burkitt2006}:

\begin{subequations}
\label{eq:rwhiteLIF}
\begin{align}
\rw(\mu) &= {\tw(\mu)}^{-1}  \\
\tw(\mu) &=\sqrt{\pi}\taum\,\int_{H(\mu)}^{\Theta(\mu)}\d t\,\phi(t) 
\end{align}
\end{subequations}
and where $\Theta_0:=\Theta(\mui)$, $H_0:=H(\mui)$, and:

\begin{align}
\phi(t)&:=e^{t^2}\,(1+{\rm erf}(t)), \qquad R(x):=(2/\pi)^{1/2}\phi(x) \\
\Theta(\mu) &:= \frac{1-\mu\taum}{\sigma_{\rm I}\sqrt{\taum}}, \qquad H(\mu) := \frac{-\mu\taum}{\sigma_{\rm I}\sqrt{\taum}}
\label{eq:rwhiteLIFfunctions}
.
\end{align}

\subsection{Derivation of the mean firing rate in the quenched-noise approximation \label{sec:demonstrationrq}}

The readout firing rate in the quenched-noise approximation is:

\begin{subequations}
\label{eq:quenchedmeandef}
\begin{align}
\rq(\mui) &:= \<\rw(\tilde\mu)\>_{{\cal N}(\tilde\mu|\mu_I,\vc)} \qquad \text{quenched-noise app.}
\end{align}
\end{subequations}
where $\rw(\mu)$ is the Siegert solution in Eq. (\ref{eq:rwhiteLIF}). More in general, we want an expression for the integral of a real-valued function $r(y)$, weighted by the density of a Gaussian probability distribution ${\cal N}(y;y_0,v)$ with average $y_0>0$ and variance $v$, which will be taken to be very small, but in the positive real axis $\Delta=[0,\infty]$. More specifically, the integral we want to perform is:

\begin{align}  \label{eq:integraltosolve}
I := \<r(y)\>_{{\cal N}(y;y_0,v)} &= \int_\Delta \d y\,{\cal N}(y;y_0,v) r(y) =\\
&= \frac{1}{(2\pi v)^{1/2}} \int_\Delta \d y\, e^{-f(y)} \\
f(y)&:=   \frac{(y-y_0)^2}{2v} -\ln(r(y))   \\
M&:=v^{-1}
.
\end{align}  

Let us Taylor-expand around $y_0$: $f(y)=-\ln(r(y_0)) - (y-y_0)\, g_1(y_0) +(1/2)(y-y_0)^2[M-g'_1(y_0)]+O[(y-y_0^3)]$, where we have defined $g_1(y):=r'(y)/r(y)$ and $M:=1/v$ and where $O[\cdot]$ is the big-o notation. Hence:

\begin{align}  
\label{eq:taylor}
I &= r(y_0) \int_\Delta \d y\, \exp{ \left( -\frac{1}{2}(y-y_0)^2[M-g_1'(y_0)]+(y-y_0)g_1(y_0)+O[(y-y_0)^3] \right) }
.
\end{align}  

From now on, we will consider $M$ to be large. Values of $y$ such that $|y-y_0|\gg M^{-1/2}$ are exponentially suppressed in the integral. If $M$ is large enough, we can consequently neglect the terms of order $(y-y_0)^3$ in front of those of order $(y-y_0)$ and $(y-y_0)^2$ (even those that are not proportional to $M$). Suppose that this is the case, that $O[(y-y_0)^3]$ can be neglected in the exponential of (\ref{eq:taylor}). This amounts to approximate $\ln r$ by its Taylor series up to order two in the relevant interval of length $M^{-1/2}$. If, moreover, $y_0 \gg M^{-1/2}$ (or if $y_0>0$, and $M$ is large enough), we can neglect the error that we make in substituting the integrating interval $[0,\infty]$ by $[-\infty,\infty]$. Summarising:

\begin{align}  \label{eq:almostthere}
I &\simeq r(y_0)\frac{1}{(2\pi v)^{1/2}} \int_{-\infty}^{\infty} \d z\, \exp{ \left( -\frac{1}{2}z^2[M-g_1'(y_0)]+z g_1(y_0)] \right) } \qquad \text{large $M$ and $y_0>0$}
,
\end{align}  
where we have changed the integration variable $z=y-y_0$.  
Under these assumptions, we can use Gaussian formulae in (\ref{eq:almostthere}) and get:

\begin{align}  \label{eq:there} 
I &\simeq r(y_0) \left(\frac{1}{1-M^{-1}g_1'(y_0)}\right)^{1/2} \exp{\left( \frac{1}{2}\frac{M^{-1}g_1^2(y_0)}{1-M^{-1}g_1'(y_0)} \right)}
.
\end{align}  

Notice that, if we take the $M\to\infty$ limit in eq. \ref{eq:there}, we obtain $I\simeq r(y_0)$, which is what we obtain approximating our $I$ in Eq. (\ref{eq:integraltosolve}) using the standard saddle-point equation  (where $\sf f$ is a generic function):

\begin{align}
\int\, r\, \exp(M{{\sf f}})\simeq r(y_0)(2\pi)^{1/2}(M|{\sf f}''(y_0)|)^{-1/2}\exp(M{\sf f}(y_0))
.
\end{align}

Now, substituting $r$ by $\rw$, $y_0$ by $\mui$, and $v$ by $\vc$ in Eq. (\ref{eq:there}), we get the quenched-noise approximation for the mean rate:

\begin{align}\label{eq:rqanalytical}
\rq(\mui) &= \rw(\mui) \left[\frac{1}{1-\vc g'_1(\mui)}\right]^{1/2} \exp\left( \frac{\vc}{2} \frac{ g_1^2(\mui)}{1-\vc g'_1(\mui)} \right)
\end{align}
where the function $g_1(\mu)$ and its derivative $g'_1(\mu)$ depend on the white LIF parameters only:

\begin{align}
g_1 &= \frac{\rw'}{\rw} = -\frac{\tw'}{\tw} \\
g_1' & = -\frac{\tw'' \tw-{\tw'}^2}{\tw^2} \\
\end{align}

We report the explicit expressions for the derivatives of the Siegert mean ISI:

\begin{subequations}
\label{eq:twprimeprime}
\begin{align}
\tw' &= - \frac{\pi^{1/2} \taum^{3/2}}{\sigmai} \left[\phi(\Theta)-\phi(H)\right] \\
\tw'' &= \frac{2\pi^{1/2} \taum^{2}}{\sigmai^2} \left[\Theta\phi(\Theta)-H\phi(H)\right]
\end{align}
\end{subequations}

where $\Theta,H$ denote the functions in Eq. (\ref{eq:rwhiteLIFfunctions}).

\subsection{Relation with asymptotic integral expansions \label{sec:asymptotic}}

Taylor-expanding $r$ around $y_0$ within the integrand in (\ref{eq:integraltosolve}) would lead to the assymptotic integral expansion:

\begin{align}  
I &= \left(\frac{1}{2\pi v^2}\right)^{1/2} \int_\Delta\d y\, \left( \sum_{m=0}^{\infty}  (y-y_0)^m\frac{r^{'m}(y_0)}{m!} \right) {\cal N}(y;y_0,v) = \sum_{m=1}^{\infty}  \frac{r^{' m}(y_0)}{m!} \<(y-y_0)^m\>_{{\cal N}(y;y_0,v)} =\\
&\simeq  \sum_{n=0}^{\infty}  \frac{r^{' 2n}(y_0)}{(2n)!} v^{n} \label{tiziano}
\end{align}  
where $r^{'p}$ means the $p$-th order derivative. The last equation is an approximation, because we have approximated the integrating interval $[0,\infty]$ by $[-\infty,\infty]$. We notice that the standard Laplace approximation for the integral is the $0$-th order of the assymptotic expansion in $v$. Our expression, Eq. (\ref{eq:there}) coincides with the assymptotic expansion {\it up to first order in $v$}. Indeed, it is easy to expand Eq. \ref{eq:there} to the first order in $v$, one finds: 

\begin{align}  
I & = r(y_0) + \frac{1}{2} v r''(y_0) + O[v^3] 
\end{align}  
which trivially coincides with (\ref{tiziano}) up to $O[v]$. This confirms again that our expression, eq. \ref{eq:there} is a refinement of the standard Laplace expression.

\newcommand{\Ilb}[2]{I_{#2}^{{\rm lb}(#1)}}
\newcommand{\Iub}[2]{I_{#2}^{{\rm ub}(#1)}}
\subsection{Formal mathematical proof \label{sec:formalproof}}

Let $g$ and $h$ be two functions, twice continuous and differentiable in an interval $\Delta$ in the positive real line, that can also be unbounded. Now, suppose that $g$ has a single absolute maximum at $x_0\in\Delta$, and it is $g''_0:=g''(x_0)<0$. We want to evaluate the integral 
\begin{align}  
I_M = \int_{\Delta}\d x\, e^{M g(x)+h(x)}
\end{align} 

The standard saddle point approximation is:

\begin{align}  
I^{\rm (std)}_M = e^{M g_0+h_0} \left(\frac{2\pi}{M |g_0''|}\right)^{1/2} 
\end{align} 
and one can easily proof that

\begin{align}  
\lim_{M\to\infty}\frac{I_M}{I^{\rm (std)}_M} = 1
.
\end{align} 

Let us consider the alternative approximation:

\begin{align}  
I^{\rm (alt)}_M =  e^{M g_0+h_0} \left(\frac{2\pi}{M|g_0''|-h_0''}\right)^{1/2} \exp{\left(\frac{1}{2}\frac{M^{-1}{h'_0}^2}{|g_0''|-M^{-1}h_0''}\right)}
.
\end{align}  

It is easy to proof that it also satisfies:

\begin{align}  
\lim_{M\to\infty}\frac{I_M}{I^{\rm (alt)}_M} = 1
\end{align} 
but it is less immediate to see that it is actually a better approximation than $I^{\rm (std)}_M$. Indeed, it is:

\begin{align}  
\lim_{M\to\infty}\frac{I_M-I^{\rm (alt)}_M}{I_M-I^{\rm (std)}_M} = 1
\end{align} 
and not $<1$, as one could initially have expected! The reason is that the alternative approximation is a correction for large, but not infinite values of $M$. Rather, we want to proof that the quotient above is $\le 1$ {\it for large enough $M$}. The proposition to be demonstrated is the following. Given $g,h,\Delta,x_0$, there is some $\tilde M$ above which it is

\begin{align}  \label{eq:proposition}
\left| \frac{I_M-I^{\rm (alt)}_M}{I_M-I^{\rm (std)}_M} \right| \le 1 \qquad \forall M>\tilde M
.
\end{align} 

Let us see how this could be proven. We need some preliminary propositions.\\

{\bf Lower and upper bounds.} We follow the demonstration in \cite{wikipediaLaplace}. First, we are able to write analytical expressions for lower and upper bounds to $I_M$, that will be referred to as $\Ilb{\epsilon}{M},\Iub{\epsilon}{M}$ respectively, both depending on a real parameter $\epsilon>0$. For the lower bound, one increases by $\epsilon$ the curvatures of $g$ and $f$, in such a way that $\forall \epsilon>0, \exists\, \delta>0$ such that $Mg(x)+h(x)\ge  Mg_0-M(|g''(x)|+\epsilon)\delta_x^2/2+h_0+h'_0\delta_x+(h''_0-\epsilon)\delta_x^2/2$ for all $x$ in the interval $|\delta_x|<\delta$. As a consequence:

\begin{align}  
I_M &\ge \int_{|\delta_x|<\delta}\d x\, e^{M g(x)+h(x)}   \ge \Ilb{\epsilon}{M}  \\
\Ilb{\epsilon}{M}  &:=   e^{M g_0 + h_0}\int_{-\delta M^{1/2}}^{\delta M^{1/2}}\d y\,e^{-y^2(|g_0''|+\epsilon)/2+y^2M^{-1}(h''_0-\epsilon)/2+y M^{-1/2} h_0'}
.
\end{align} 

For the upper bound, we decrease the curvature by a quantity $\epsilon>0$, which is small enough so that the increased curvature is still lower than zero, $-g''_0+\epsilon<0$. By continuity of $g''$ and the Taylor theorem, for each of such $\epsilon$'s $\exists\, \delta>0$ such that

\begin{align}  
M g(x) + h(x) \le M (g_0-(|g''_0|-\epsilon)\delta_x^2/2)+h_0+h'_0\delta_x+(h''_0+\epsilon)\delta_x^2/2
\end{align} 
for all $x$ such that $|\delta_x|<\delta$. Now, let $M^*$ be a value for which the integral 

\begin{align}  
\tilde I := \int_\Delta\d x\, e^{M^*g(x)+h(x)} <\infty
\end{align} 
is finite. Hence, given $\epsilon$ and $\delta$, there exist a $\eta>0$ such that $M g(x)\le M(g_0-\eta)$ in all the points $|\delta_x|>\delta$ outside the interval $(-\delta+x_0,x_0+\delta)$. Therefore, 
 
\begin{align}  
I_M &= &\int_{|\delta_x|>\delta}\d x\, e^{M g(x)+h(x)} + &\int_{|\delta_x|<\delta}\d x\, e^{M g(x)+h(x)} \le& \\
&\le &\int_{|\delta_x|>\delta}\d x\, e^{M^* g(x)} e^{(M-M^*) (g_0-\eta)+h(x)} + &\int_{|\delta_x|<\delta}\d x\, e^{M g(x)+h(x)} \le& \\
&\le  &e^{(M-M^*)(g_0-\eta)}\tilde I + &\int_{|\delta_x|<\delta} \d x\, e^{M g(x)+h(x)}  \le & \\
&\le  &e^{(M-M^*)(g_0-\eta)}\tilde I + &e^{Mg_0+h_0}\int_{|\delta_x|<\delta} \d x\,e^{ -M(|g''_0|-\epsilon)\delta_x^2/2+ \delta_x h'_0+(h''_0+\epsilon)\delta_x^2/2} \le & \\
&\le  &e^{(M-M^*)(g_0-\eta)}\tilde I + &e^{Mg_0+h_0}\int_{-\infty}^{\infty} \d x\,e^{-M(|g''_0|-\epsilon)\delta_x^2/2+ h_0+\delta_x h'_0+(h''_0+\epsilon)\delta_x^2/2}, &
\end{align} 
or, using Gaussian formulae:

\begin{align}  
I_M \le \Iub{\epsilon}{M} = e^{(M-M^*)(g_0-\eta)}\tilde I + e^{M g_0 + h_0} \left(\frac{2\pi}{M(|g''_0|-\epsilon)-(h_0''+\epsilon)}\right)^{1/2} \exp{\left(\frac{1}{2}\frac{{h'_0}^2}{M(|g_0''|-\epsilon)-(h''_0+\epsilon)}\right)}
.
\end{align}

The strategy to demonstrate the proposition in (\ref{eq:proposition}) is to examine, for large $M$, the following lower and upper bounds

\begin{align}  
q_M^{{\rm u}(\epsilon)} &:= \frac{\Iub{\epsilon}{M}-I^{\rm (alt)}_M}{\Ilb{\epsilon}{M}-I^{\rm (std)}_M} \\
q_M^{{\rm l}(\epsilon)} &:=\frac{\Ilb{\epsilon}{M}-I^{\rm (alt)}_M}{\Iub{\epsilon}{M}-I^{\rm (std)}_M} 
\end{align} 
 for the ratio: 

\begin{align}  
q_M^{{\rm l}(\epsilon)} \le \frac{I_M-I^{\rm (alt)}_M}{I_M-I^{\rm (std)}_M} \le q_M^{{\rm u}(\epsilon)}   \qquad \forall \epsilon
.
\end{align} 

We demonstrate that $q_M^{{\rm u}(\epsilon)}\le 1$ and that $q_M^{{\rm l}(\epsilon)}\ge -1$ for large enough $M$. Let us consider $q_M^{{\rm u}(\epsilon)}$. It is easy to see that:
 
\begin{align}  \label{eq:myratio}
q_M^{{\rm u}(\epsilon)}  = 
\frac{ e^{-\eta M}\, u\, \tilde I + \left(\frac{2\pi}{M(|g''_0|-\epsilon)-(h_0''+\epsilon)}\right)^{1/2} e^{\frac{1}{2}\frac{{h'_0}^2}{M(|g_0''|-\epsilon)-(h_0''+\epsilon)}} -  \left(\frac{2\pi}{M|g_0''|-h_0''}\right)^{1/2} e^{\frac{1}{2}\frac{{h'_0}^2}{M|g_0''|-h_0''}}
} 
{\int_{-\delta M^{1/2}}^{\delta M^{1/2}}\d y\,e^{-\frac{y^2}{2}(|g_0''|+\epsilon-M^{-1}(h_0''-\epsilon))+y M^{-1/2} h_0'} - 
\left(\frac{2\pi}{M |g_0''|}\right)^{1/2} 
.
}
\end{align} 
where $u:=\exp[-M^*(g_0-\eta)-h_0]$. It is instructive to see what happens if we take first the $M\to\infty$ limit of this expression:
 
\begin{align}  
\lim_{M\to\infty} \frac{\Iub{\epsilon}{M}-I^{\rm (alt)}_M}{\Ilb{\epsilon}{M}-I^{\rm (std)}_M} = 
\frac{ (|g_0''|-\epsilon)^{-1/2} - (|g_0''|)^{-1/2} }{ (|g_0''|+\epsilon)^{-1/2} - (|g_0''|)^{-1/2} } 
.
\end{align} 

Our best estimators for $I_M$ are recovered in the $\lim_{\epsilon\to 0}$, which gives

\begin{align}  
\lim_{\epsilon\to 0} \lim_{M\to\infty}\frac{\Iub{\epsilon}{M}-I^{\rm (alt)}_M}{\Ilb{\epsilon}{M}-I^{\rm (std)}_M} =  1
.
\end{align} 

Now, it is clear what is happening: in the $\lim_{M\to\infty}$ our corrections to the standard Laplace method (i.e., the differences between $I^{({\rm alt})}$ and $I^{({\rm std})}$) vanish. Let us now consider the ratio in (\ref{eq:myratio}), and suppose that $M$ is so large that one can neglect the exponentially decaying term $\exp{(-\eta M)}$ in the numerator, and that we can substitute the integral in the denominator by the same integral with the $-\infty,\infty$ limits (the differences between both decay very fast as the complementary error function $\rm erfc$ in $\delta M^{1/2}$, whose leading term is $\sim\delta M^{1/2}\exp(\delta^2 M)$). Indeed, these are the only two terms decaying exponentially in $M$. We have:

\begin{align}  \label{eq:myratioapp}
\frac{\Iub{\epsilon}{M}-I^{\rm (alt)}_M}{\Ilb{\epsilon}{M}-I^{\rm (std)}_M} \simeq 
\frac{ \left(\frac{2\pi}{M(|g''_0|-\epsilon)-(h_0''+\epsilon)}\right)^{1/2} e^{\frac{1}{2}\frac{{h'_0}^2}{M(|g_0''|-\epsilon)-(h_0''+\epsilon)}} -  \left(\frac{2\pi}{M|g_0''|-h_0''}\right)^{1/2} e^{\frac{1}{2}\frac{{h'_0}^2}{M|g_0''|-h_0''}}
} 
{\left(\frac{2\pi}{M(|g''_0|+\epsilon)-(h_0''-\epsilon)}\right)^{1/2} e^{\frac{1}{2}\frac{{h'_0}^2}{M(|g_0''|+\epsilon)-(h_0''-\epsilon)}} - 
\left(\frac{2\pi}{M |g_0''|}\right)^{1/2} 
}
\qquad \text{large $M$}
.
\end{align} 

Mind that, as far as $h'_0\ne 0$ or $h''_0\ne 0$, this ratio is lower than one in absolute value, because the difference between both terms in the numerator is lower in absolute value than the difference between both terms in the denominator (and it would vanish for $\epsilon=0$). Notice that, given $M$, $\epsilon$ cannot be arbitrary low for (\ref{eq:myratioapp}) to be a good approximation of (\ref{eq:myratio}): it has to be $\epsilon$ (and, consequently, $\delta$ and $\eta$ in (\ref{eq:myratio})) large enough so that we can neglect $\exp{(-\eta M)}$ and substitute $\int_{-\delta M^{1/2}}^{\delta M^{1/2}}$ by $\int_{-\infty}^{\infty}$. The larger $M$, the lower we can take $\epsilon$, and the better is the estimation in (\ref{eq:myratioapp}). However, it is not necessary that $\epsilon$ is very small. The only thing we need is $M$ large enough for the upper bound (\ref{eq:myratioapp}) to be $\le 1$. The same argument can be applied to $q_M^{{\rm l}(\epsilon)} \ge-1$, and this leads us to Eq. (\ref{eq:proposition}).

\subsection{Validity of the analytic expression for the mean readout rate in the quenched-noise approximation}

\begin{figure}
\includegraphics[width=0.75\textwidth]{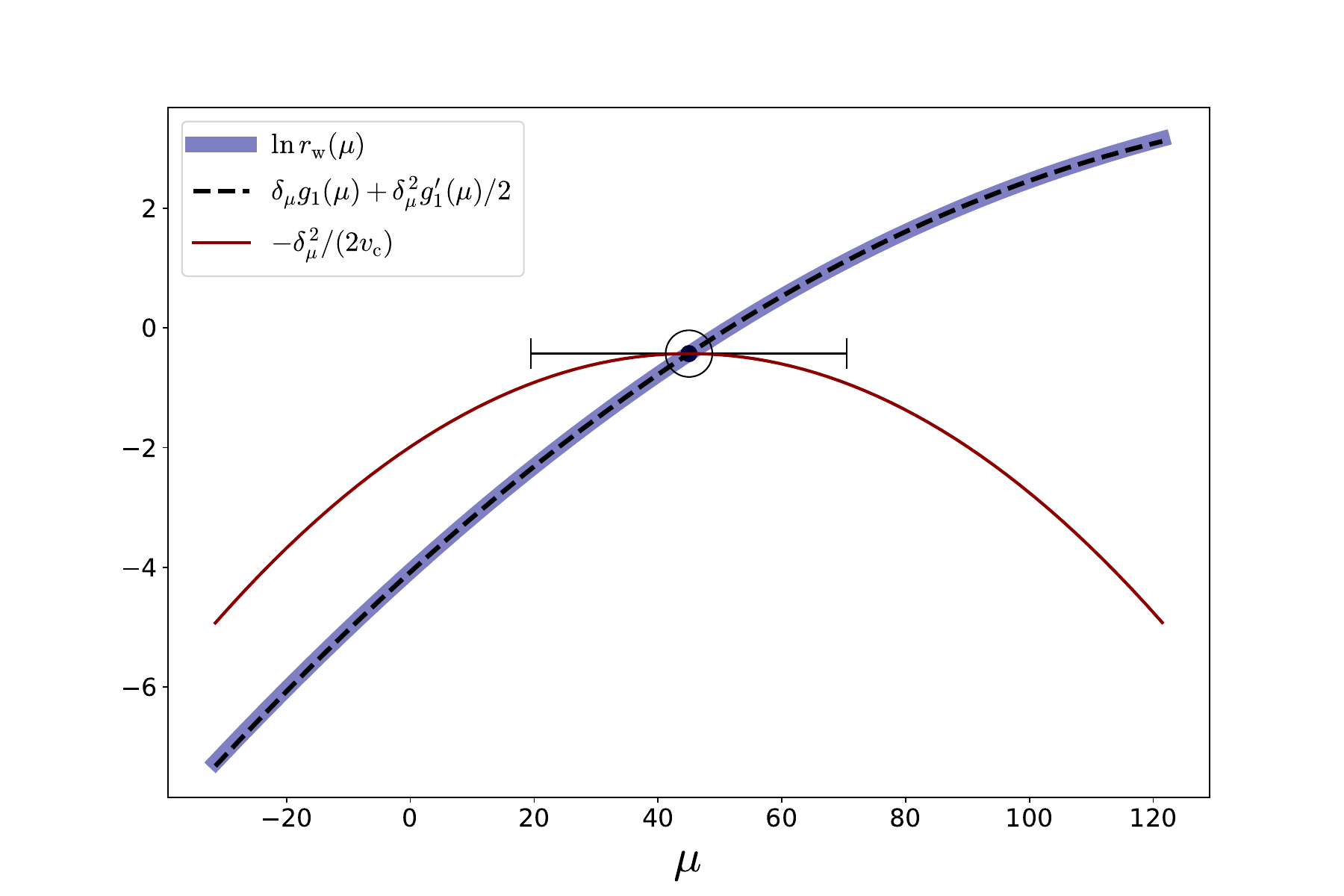}%
\caption{\label{fig:theoremvalidity} {\it Necessary conditions for the demonstration of the quenched-noise approximation for the rate mean in Eq. (10)}. The derivation holds in the extent to which terms of $O[(\mu-\mui)^3]$ of $\ln\rw$ in Eq. (\ref{eq:taylor}) are negligible in intervals of few standard deviations $\vc^{1/2}$ of $\mui$ around its mean $\mui$ (outside which the contribution of the integral (\ref{eq:integraltosolve}) is negligible). We plot the different terms in the exponent $f(\mu)$ in Eq. (\ref{eq:taylor}), with $\mui=45$ and the parameters of the article simulations for $\alphaV=0.9$. Indeed, within an interval of length $6\vc^{1/2}$ around $\mui$, the $\ln \rw$ is well approximated by its first two orders in $\delta_\mu$. The big circle and its error-bar indicate the mean $\mui$ and (twice the) standard deviation $\vc^{1/2}$ of the normal distribution.}
\end{figure}

We have seen that the condition in which this equation is derived is low enough $\vc$. This condition can be actually seen to be satisfied in the simulations corresponding to the main analysis of this article: the 1st simulation for constant $\vc$ presented in Secs. \ref{sec:auxiliary},\ref{sec:valentemodelsolution}. For our derivation to hold (see above in this appendix), we need to neglect powers of $O[(\delta_\mu)^3]$ of $\ln\rw$ in an interval of the order of few multiples of $\vc^{1/2}$ around $\mui$. Indeed, this is what happens even in the less favorable case (that of the higher $\vc(\alphaV=0.9)\simeq 25.4994$ for $\alphaV=0.9$) in the 1st simulation, see Fig. \ref{fig:theoremvalidity}. As a consequence, the analytic expression for $\rq$ in Eq. (\ref{eq:rqanalytical}) coincides with the numeric estimation of the integral in the general expression for $\rq$, Eq. (\ref{eq:quenchedmeandef}), as we illustrate in Fig. (\ref{fig:r_quenched}-B).

We remind that in the 3rd simulation (analysed in Appendix \ref{sec:varyingsigmav}), the RO neuron properties are analysed as a function of $\sigmaV$, for constant $\tauc$, which implies varying $\vc$. For the highest values of $\sigmaV$, the value of $\vc$ lies outside the validity limits of the derivation of Eq. (\ref{eq:rqanalytical}), and this equation begins to differ from the numerical integration of Eq. (\ref{eq:quenchedmeandef}) (see Fig. \ref{fig:SNR_varyingsigmac}). This example illustrates the fact that the validity limit of the analytical expression for the mean rate in the quenched-noise approximation, $\rq$ in Eq. (\ref{eq:rqanalytical}) has a validity limit (low $\vc$) that, in general, differs from that of the quenched approximation itself (Eq. (\ref{eq:quenchedmeandef}) for the mean and Eq. (\ref{eq:nestedintegrals}) for the temporal distribution function), which is $\tauc\gg\taum$. For the highest values of $\sigmaV$, hence of $\vc$ in Fig. \ref{fig:SNR_varyingsigmac}, the analytic expression for $\rq$ is no longer valid, while the quenched noise approximation still holds (since it is $\tauc=0.1$, $\taum=0.005$).

\section{SNR behavior varying $\sigmac$ at fixed $\tauc$ \label{sec:varyingsigmav}}

\begin{figure}
\includegraphics[width=0.7\textwidth]{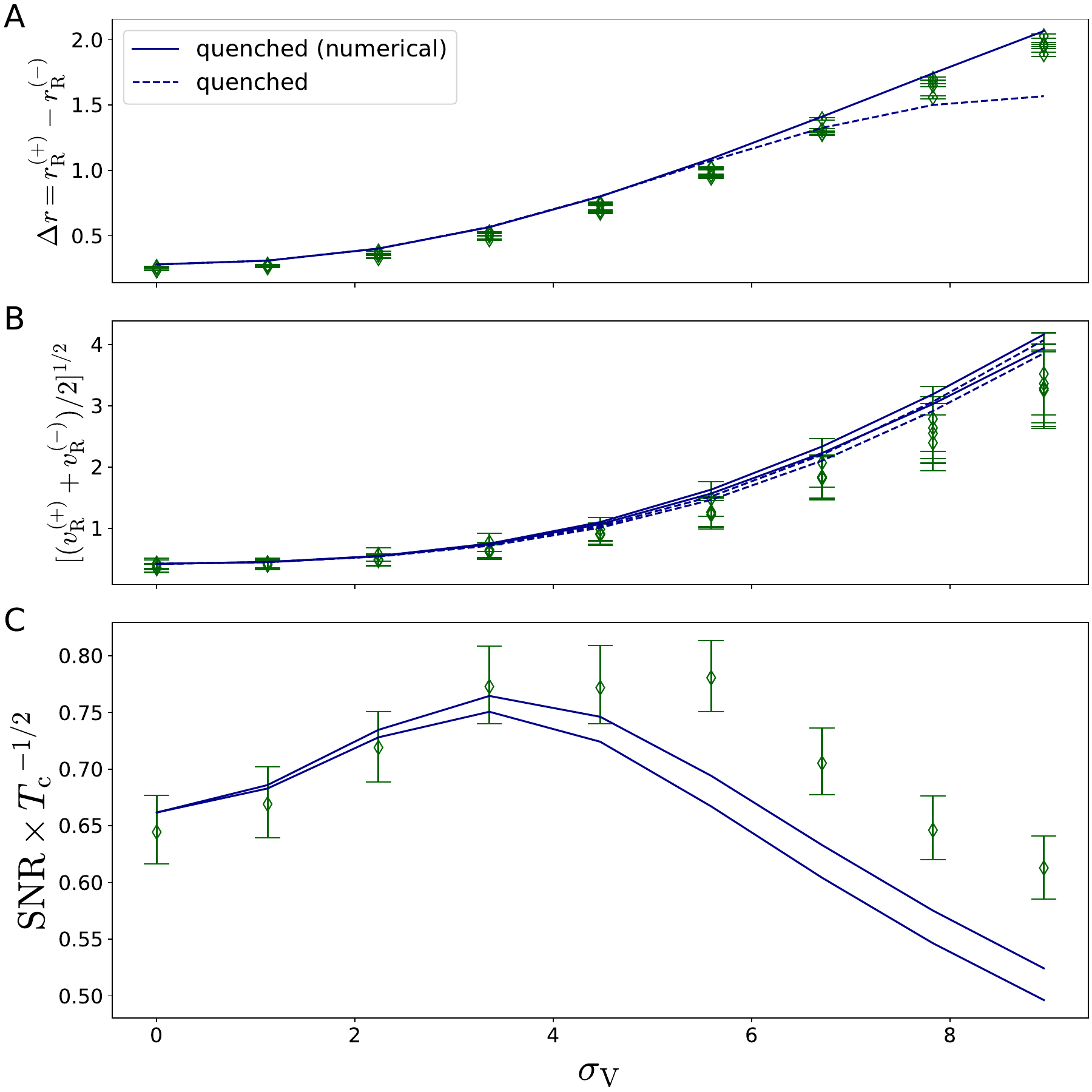} 
\caption{{\it Mean rate enhancement, rate standard deviation and SNR, as a function of $\sigmaV$.} As in Fig. \ref{fig:numSNRsqrt} but for a fixed value of $\tauc=0.1$, and varying $\sigmaV$. See the rest of model parameters in Table \ref{table:parameters3}.  \label{fig:SNR_varyingsigmac}}
\end{figure}

As a complementary analysis of that of Sec. \ref{sec:valentemodelsolution}, we have analysed the behavior of the SNR in the NERM versus the value of $\sigmaV$ (consequently $\sigmac$) for a fixed value of $\tauc$. The simulation parameters of the corresponding simulation (the 3rd simulation, see Appendix \ref{sec:numerical}) are specified in Table \ref{table:parameters3}. The results are summarised in Fig. \ref{fig:SNR_varyingsigmac}. 

The salient features of the figure are the following. First, the mean readout rate enhancement $\Delta\rro$ in panel A is, again statistically consistent, at least for low enough $\sigmaV$, with the quenched-noise approximation. The discrepancies for large $\sigmaV$ are attributed to the finite-NIS effect, as we observed in Sec. \ref{sec:valentemodelsolution}. The analytical expression for the quenched-noise effect, Eq. (\ref{eq:rquenched_analytical}) coincides with its numerical counterpart, Eq. (\ref{eq:rquenched}), for low enough $\sigmaV$ (i.e., low enough $\vc$), as expected from Sec. \ref{sec:demonstrationrq}. The same arguments apply to the stimulus-averaged rate standard deviation in panel B. As a consequence of the underestimation of the variance due to the finite-NIS effect, the SNR is overestimated for large $\sigmaV$, and departs from the analytical prediction. Overall, this case illustrates the non-monotonic dependence of the SNR on $\sigmac$, for which the analytical solution provides an interpretable explanation (see Sec. \ref{sec:valentemodelsolution}).  

In this case, and differently from the so called 1st simulation, the finite-NIS reduction algorithm does not help us to construct a better estimation of the SNR. The reason is that the underestimation of the rate variance $\vro$ due do the finite-NIS effect is more severe for this simulation parameters. As we wrote above, unfortunately the finite-NIS reduction algorithm does not provide us with an accurate estimation for the rate variance, given the used values of $\nis$ and $\Tc$, as it does for the rate mean. Therefore, although the finite-NIS estimation for the rate mean is closer to the theoretical $\rq$, the finite-NIS correction does not help much to construct a better numerical estimation of the SNR. We therefore conclude that it is the theoretical curve in Fig. \ref{fig:SNR_varyingsigmac}-C to be correct, and the points corresponding to the numerical simulations to be an overestimation of the SNR due to the finite-NIS effect. 

\section{Additional figures for alternative values of the model parameters \label{sec:alternativefigures}}

\begin{figure}
\includegraphics[width=1\textwidth]{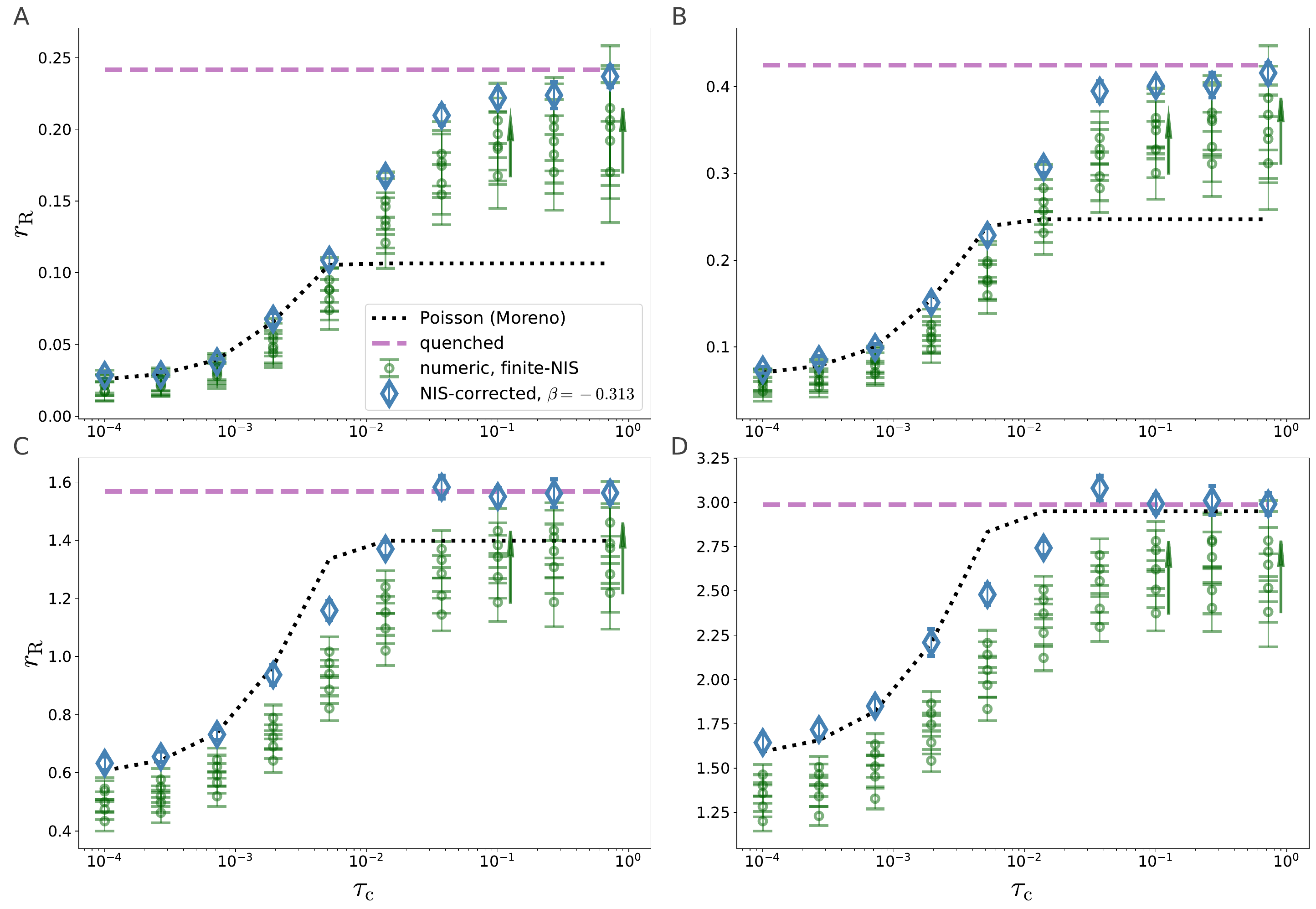} 
\caption{{\it Mean readout firing rate $\rro$ vs $\tauc$: comparison between theoretical values and numerical estimations.} Panels (A,B,C,D) correspond, respectively, to the different values of $\mui=32.65,35.0,44.45,50.35$. Green points and error-bars are fixed-NIS numerical estimations; blue diamonds and error-bars are the finite-NIS-corrected estimations; the dotted black line is the Moreno extrapolated function in \cite{moreno2002}, while the pink dashed line is the quenched-noise approximation, Eq. (\ref{eq:rquenched}). See the rest of model parameters in Table \ref{table:parameters1}.  \label{fig:rRvstauc}}
\end{figure}

In Fig. \ref{fig:rRvstauc} we show our numerical and analytical estimations of $\rro$ versus $\tauc$ (constant $\vc$), for four values of $\mui$. The finite-NIS corrected estimate is statistically compatible with the Moreno et al solution for low enough $\tauc$ and for all values of $\mui$ (except, perhaps, for the largest one $\mui=50.350$). Rather remarkably, for large enough $\tauc$ and {\it all the probed values of $\mui$}, the finite-NIS-corrected estimation $\tilde\rro$ is compatible with the quenched-noise approximation. We make notice that, as expected, for low values of $\mui=32.650,35.600,44.450$, the Moreno parameter $\alpha_{\rm M}=\sigmac^2/(w\mui)$ is large for the Moreno approximation to work well and, indeed, only the quenched-noise approximation is compatible with the (NIS-corrected) numerical data (see Appendix \ref{sec:validity} as well). For large enough $\mui=50.350$, $\alpha_{\rm M}$ is low enough, so that the Moreno solution approaches the quenched-noise solution for $\rro$, and becomes statistical compatible with the numerical $\tilde\rro$.

\begin{figure}
\includegraphics[width=1.\textwidth]{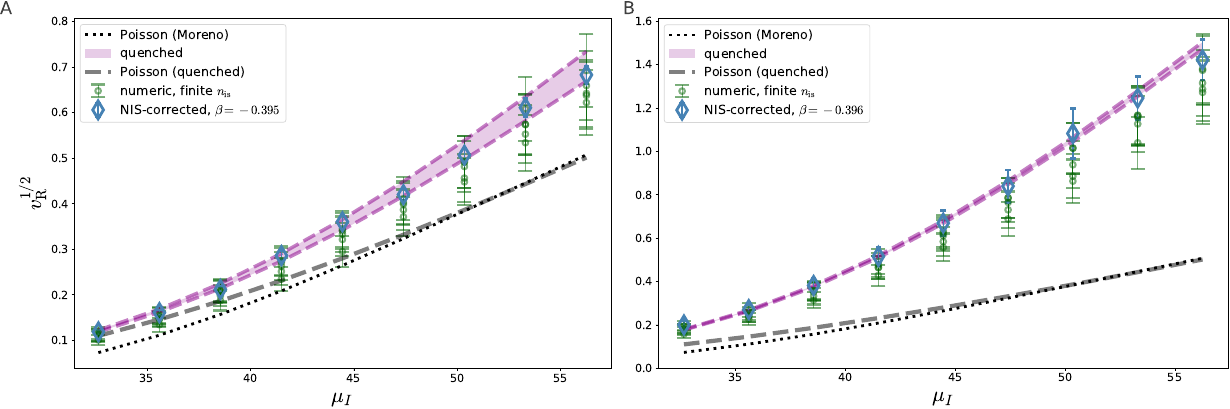} 
\caption{ {\it Standard deviation of the readout firing rate $\vro^{1/2}$ vs $\mui$: comparison between theoretical values and numerical estimations.} Panels A,B correspond, respectively, to the different values of $\tauc=0.1, 0.72$. Green points and error-bars are fixed-NIS numerical estimations; blue diamonds and error-bars are the finite-NIS-corrected estimations; the dotted black line is the Poisson-Moreno approximation $\vro=\rro_{\rm M}/\Tc$, while the gray dashed line is the Poisson-quenched approximation $\vro=\rro_{\text q}/\Tc$; the pink dashed line fork correspond to the upper and lower-bounds of the quenched-noise approximation, Eq. (\ref{eq:varrate_quenched}). See the rest of model parameters in Table \ref{table:parameters1}. 
\label{fig:sRvsmui}}
\end{figure}

\begin{figure}
\includegraphics[width=1.\textwidth]{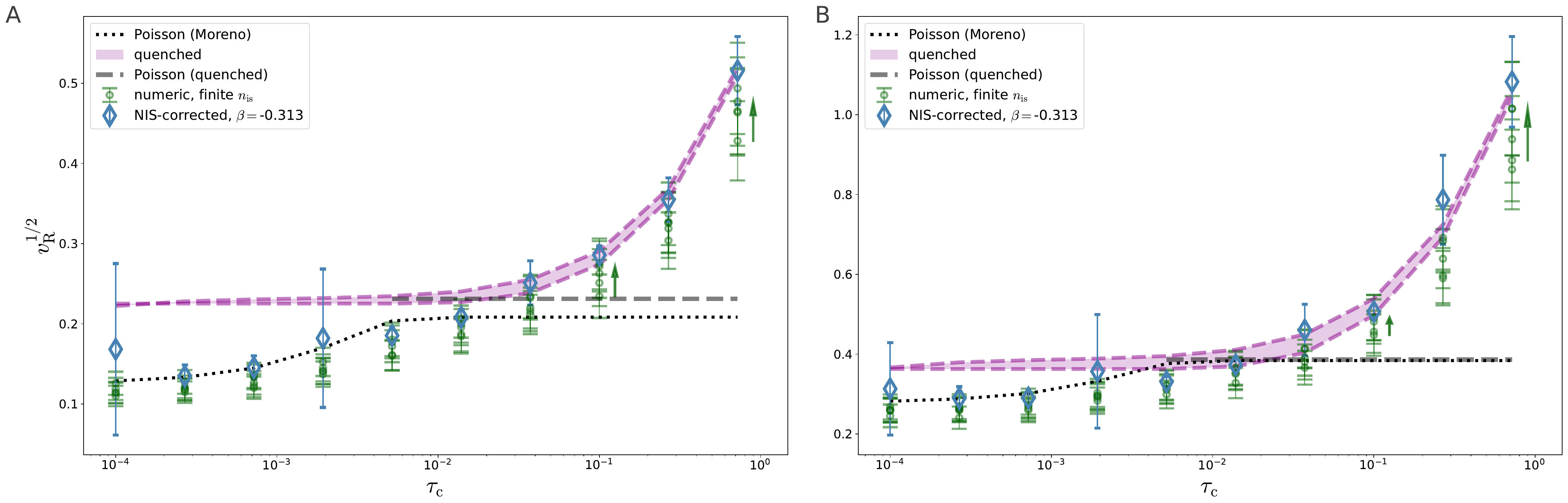} 
\caption{{\it Standard deviation of the readout firing rate $\vro^{1/2}$ vs $\tauc$: comparison between theoretical values and numerical estimations.} Panels A,B correspond, respectively, to two different values of $\mui=41.5,50.35$. See the rest of model parameters in Table \ref{table:parameters1}.  
\label{fig:sRvstauc}}
\end{figure}

In Figs. \ref{fig:sRvsmui},\ref{fig:sRvstauc} we show our numerical and analytical estimations of $\vro^{1/2}$ versus $\mui$ (for two values of $\tauc$) and versus $\tauc$ (for two values of $\mui$, with constant $\vc$), respectively. We there report as well our NIS-corrected estimations for the rate variance. We have not used such estimations, since they present a relatively larger statistical error (than their $\rro$ counterpart). We observe that, as happens for $\rro$, and rather remarkably, for low and high enough values of $\tauc$, the finite-NIS-corrected estimations $\tilde\vro^{1/2}$ become, respectively, compatible with the Poisson-Moreno and the quenched-noise approximations.

\begin{figure}
\includegraphics[width=0.7\textwidth]{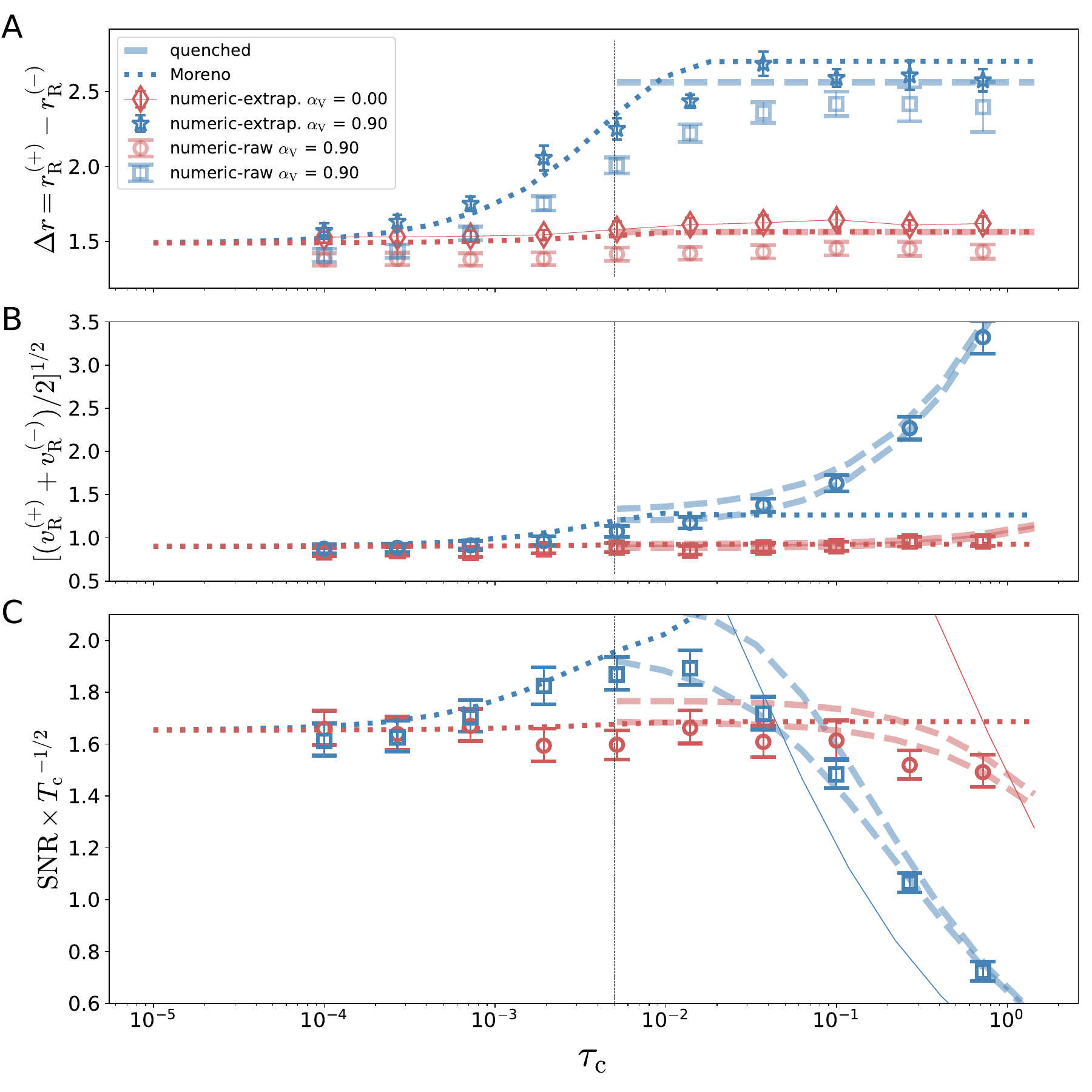}%
\caption{{\it Readout SNR in the unit time interval}, as in Fig. \ref{fig:numSNRsqrt} but for different values of the tuning curves $\mui^{(\pm)}$. The tuning curves are $(\mu_I^{(+)},\mu_I^{(-)}) = (35.6,50.35)$. The thin dotted lines are the {\it encoding} signal-to-noise ratio in the unit time interval, Eq. (\ref{eq:snrencoding}). See the rest of model parameters in Table \ref{table:parameters1}. 
\label{fig:numSNRsqrt_alt}}
\end{figure}

In Fig. \ref{fig:numSNRsqrt_alt} we present the equivalents of Fig. \ref{fig:numSNRsqrt} for an alternative value of the input averages, $\mui^{(\pm)}$. We observe a qualitatively identical behavior.

\begin{figure}[h!]
\includegraphics[width=0.7\textwidth]{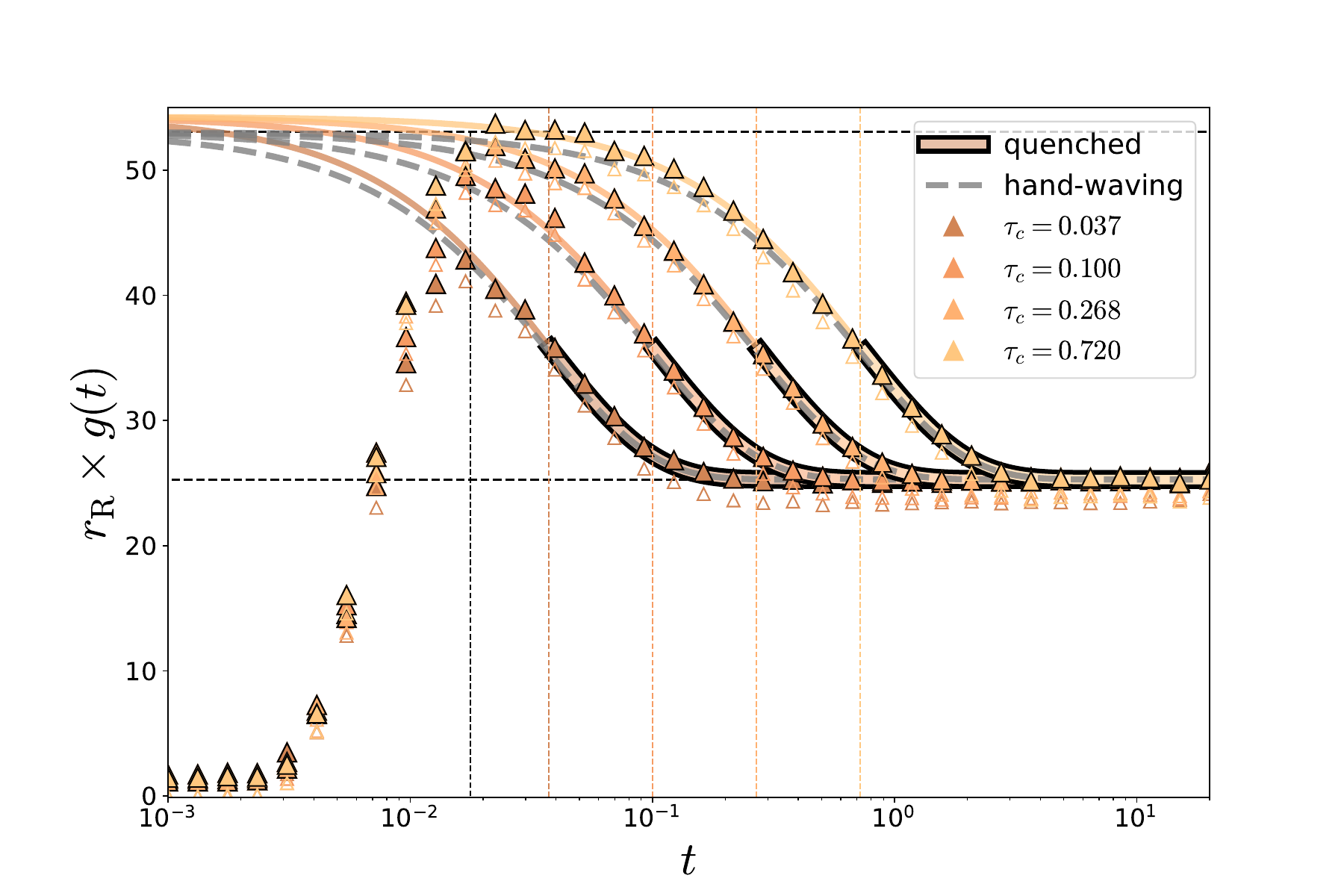}%
\caption{{\it Temporal distribution function $g(t)$}, as in Fig. \ref{fig:g_quenched} but for a different value of $\mui=56.25$. See the rest of model parameters in Table \ref{table:parameters1}. \label{fig:g_quenched_alt} }
\end{figure}

Finally, in Fig. \ref{fig:g_quenched_alt} we present a variant of Fig. \ref{fig:g_quenched} for a different value of $\mui$.

\section{Analytical expression for the rate variance in the quenched-noise approximation \label{sec:quenched_variance}}

\subsection{Sketch of the derivation of the temporal distribution function in the quenched-noise approximation}

The quenched-noise approximation to $g$ is as follows. Let $n_j(\ell)$ be the number of LIF spikes in the time interval $[t_j,t_{j+1}]$ where $t_j:=j\ell$ and $j$ is an integer in $[0,L={\rm int}(\Tc/\ell)]$. We define the {\it discrete-time temporal distribution function} in the quenched-noise approximation, $\gq$, as:

\begin{align} \label{eq:gq}
\rq\, \gq\left(t_j-t_{j'}\right) = \frac{1}{\ell^2}\<\< n_j(\ell) n_{j'}(\ell) \>_{{\bf n}|{\bm \mu}}\>_{{\bm \mu}}
\end{align} 
where the expected values are, respectively: with respect to the white-LIF neuron probability density $\text{prob}({\bf n}|{\bm{\tilde \mu}})$ of sequences $(n_1(\ell),\ldots,n_L(\ell))$ of $L$ spike counts, given the sequence of $L$ average input rates $(\tilde\mu_1,\ldots,\tilde\mu_L)$; and with respect to the probability of sequences of consecutive mean input rates, $\text{prob}({\bm{\tilde \mu}})$. We expect $\gq$ in Eq. (\ref{eq:gq}) to approximate well $g$ in resolution timescales $\ell$ satisfying $\taum\ll \ell\ll\tauc\ll\Tc$.   

The first hypothesis in our calculation consists in the factorisation of the inner correlation: $\< n_j n_{j'} \>_{{\bf n}|{\bm \mu}}=\<n_j\>_{n_j|\tilde\mu_j}\<n_{j'}\>_{n_{j'}|\tilde\mu_{j'}}$. Consistently with the quenched-noise approximation, $n_j(\ell)$ can be thought as exhibiting the statistics of the white-LIF neuron, with an input average $\tilde\mu\sim {\cal N}(\tilde\mu;\mui,\vc)$. We consequently write $\<n_j\>_{n_j|\tilde\mu_j} \ell^{-1}=\rw(\tilde\mu_j)$, or: $\rro\, \gq\left( t_j-t_{j'} \right) \simeq \< \rw(\mu_j) \rw(\mu_{j'}) \>_{{\bm \mu}}$, which in its turn leads to:

\begin{align}
\label{eq:nestedintegrals}
\rq\, \gq\left( t-t' \right) \simeq \int \d\mu\int\d\mu'\, \rw(\mu)\rw(\mu')P(\mu,t|\mu',t')P(\mu')
\end{align} 
where $t,t' := \ell j,\ell j'$, $P(\cdot)={\cal N}(\cdot;\mui,\vc)$ is the stationary distribution of input noises, and $P(\mu,t|\mu',t')$ is the Green's function of an Ornstein-Uhlenbeck process: (mind $\xi$ in Eq. (\ref{eq:I_diffusion})):

\begin{align}
P(\mu,t|\mu',t') &= {\cal N}\left( \mu ; \mui + \omega\mu' , \vc(1-\omega^2)   \right) \\
\omega&:=e^{-|t-t'|/\tauc} \label{eq:omegadef}
.
\end{align} 

An important remark on the validity limits of this derivation is that we expect Eq. \ref{eq:nestedintegrals} to be valid for $\taum\ll t-t' \ll\tauc$, i.e., to exhibit the validity limits of the general quenched-noise approximation. Afterwards, the analytical approximations to both nested integrations, that in $\<\rw(\mu_t)\>_{P(\mu_t|\mu_{t'})}$ (compacting the notation) and, afterwards, that in 

\begin{align}
\<\rw(\mu_{t'}) \<\rw(\mu_t)\>_{P(\mu_t|\mu_{t'})}\>_{P(\mu_{t'})}
,
\end{align}
can be performed in the saddle-point approximation, for large $\vc$, using, as in Appendix \ref{sec:demonstrationrq}, the variant of the Laplace integration method in Eq. (\ref{eq:there}). It is straightforward to arrive to an expression for $\gq$ in such a quenched-noise approximation, for low enough $\vc$. Interestingly, in the low- and large- $|t-t'|$ limits, the $\rro \gq(t-t')$ take, respectively, the values $\<\rw^2(\mu)\>_{P(\mu)}$ and $\<\rw(\mu)\>^2_{P(\mu)}$ (as expected, since for large $|t-t'|$ we expect $\mu,\mu'$ to be uncorrelated, and so $\rw(\mu)$ and $\rw(\mu')$). In particular, for large enough $|t-t'|$ (or, more precisely, neglecting terms of order $\omega^2$ in front of first-order terms in $\omega$), we get a simplified expression for $\gq$:

\begin{align} 
\rro \gq^{(\text{lo-}\omega)}(t-t')  &\simeq \rq^2 \left( 1 + e^{-|t-t'|/\tauc}\, Q\right) \qquad \text{large $|t-t'|/\tauc$} 
\end{align}
where $Q$ is a constant in time, depending on the white-LIF properties $\mui,\sigmai,\taum$, and on $\vc$ (see the explicit expression in Eq. (\ref{eq:Q}) below in Sec. \ref{sec:nested}).

\subsection{Evaluation of the nested integrals \label{sec:nested}}

We want to evaluate the nested integral in Eq. (\ref{eq:nestedintegrals}). In the Laplace approximation around $\mu_\omega:=\mui + \omega\mu'$, and for large $M=\tauc/\taum$ it is:

\begin{subequations}
\label{eq:gqinnerintegral}
\begin{align}
\<\rw(\mu)\>_{P(\mu|\mu')} &\simeq e^{-M \ffunc(\mu_\omega)} \left( \frac{2\pi}{M \ffunc''(\mu_\omega)} \right)^{1/2} e^{\frac{M \ffunc'(\mu_\omega)^2}{2\ffunc''(\mu_\omega)}} \\
\ffunc(y) &:= \frac{\taum}{\sigma_\omega^2} ( y -\mu_\omega )^2 - \frac{\taum}{\tauc}\ln(\rw(y))\\
\sigma_\omega &:=\sigmac(1-\omega^2)^{1/2}  
\end{align}
\end{subequations}
This amounts to:  

\begin{align}
\<\rw(\mu)\>_{P(\mu|\mu')} &\simeq \rw(\mu_\omega) \left[ \frac{1}{1-v_\omega g'_1(\mu_\omega)} \right]^{1/2}  
	\exp{\left( \frac{v_\omega}{2} \frac{g_1(\mu_\omega)^2}{1 - v_\omega g_1'(\mu_\omega)}\right)} \\
v_\omega &:= \sigma_\omega^2/2\tauc
.
\end{align}

We notice that, as required, $\<\rw(\mu)\>_{P(\mu|\mu')}$ coincides with $\rw(\mu')$ for $t=t'$, and with our quenched-noise approximation, $\rro = \<\rw(\tilde\mu)\>_{P(\tilde\mu)}$ for $|t-t'|=\infty$ (or $\omega=0$). The final step is to perform the integral $\int\d\mu'\, \rw(\mu')\,\<\rw(\mu)\>_{P(\mu|\mu')}$. It is, in saddle-point approximation around $\mui$:

\begin{align} \label{eq:g_quenched}
\rro \gq(t-t')  &\simeq \frac{2^{1/2}}{\sigma_\omega} \left(\frac{\tauc}{\pi\sigmac^2}\right)^{1/2} \left( \frac{2\pi}{M \hfunc''(\mui)} \right)^{1/2} e^{\frac{M \hfunc'^2(\mui)}{2\hfunc''(\mu_i)}} \\
\hfunc(y) &:= \frac{\taum}{\sigmac^2} (y-\mui)^2 -\frac{\taum}{2\tauc^2}\frac{g_1(y_\omega)^2}{\ffunc''(y_\omega)} + \frac{\taum}{2\tauc}\ln(\ffunc''(y_\omega)) - \frac{\taum}{\tauc}\ln\left(\rw(y)\rw(y_\omega) \right) \\
y_\omega &:= \mui+\omega y
\end{align}
where $\ffunc''(\mu_\omega)=(2\taum/\sigma_\omega^2)-(\taum/\tauc)g_1'(\mu\omega)$ (see Eq. (\ref{eq:gqinnerintegral})). 
%

We will omit the explicit expression of $\rro g(t-t')$, that can nevertheless be straightforwardly unfolded in terms of the derivatives of $\ffunc$ and $\hfunc$, that in their turn depend on the first, second and third-order derivatives of $\rw$. We will actually write an explicit expression for $\rro g(t-t')$ for large values of $|t-t'|/\tauc$. Neglecting terms of order $\omega^2$ in front of zeroth- and first-order terms in $\omega$, in $\hfunc'^2$ and in $\hfunc''$, we can evaluate Eq. (\ref{eq:g_quenched}):

\begin{align} 
\rro g(t-t')  &\simeq \<\rw(\mu)\>_{P(\mu)}^2 \left( 1 + \omega\, Q\right) \qquad \text{low $\omega$} 
\end{align}
where:

\begin{align}
\label{eq:Q}
Q &:= \frac{1}{2}\frac{g_1(\mui) \left[2g_1(\mui) + \tauc^{-1} (F_2 + F_3) \right]}{\vc^{-1} - g_1'(\mui)} \\
F_2 &: = \left.\left(\frac{g_1^2}{\gfunc''}\right)'\right|_{\mui}  \\
F_3 &: = \left.\frac{g_1''}{\gfunc''}\right|_{\mui}
\end{align}

The explicit expressions for $F_2,F_3$ are:

\begin{align}
F_2 & = \left(\frac{g_1^2}{\gfunc''}\right)' = \frac{2g_1g_1'\gfunc'' + \frac{1}{\tauc} g_1^2g_1''}{{\gfunc''}^2} \\
\gfunc'' & = \frac{1}{2\sigmac^2}-\frac{g_1}{\tauc} \\
g_1' & = -\frac{\tw'' \tw-{\tw'}^2}{\tw^2} \\
g_1'' & = 3\frac{ \tw'' \tw'}{\tw^2} -2 \left(\frac{\tw'}{\tw}\right)^3 - \frac{\tw'''}{\tw}
\end{align}
where $\tw',\tw''$ are in Eq. (\ref{eq:twprimeprime}), and where:

\begin{align}
\tw'''(y) &= 2\sqrt{\pi}\frac{\taum^{5/2}}{\sigmai^3} \left( 2\Theta(y) [\Theta(y)\phi({\Theta(y)}) + \pi^{1/2}] - 
							    2H(y)      [H(y)\phi({H(y)})           + \pi^{1/2}] \right) 
.
\end{align}

\def\Tm{{T_{\rm s}}}
\def\pgivenmu{{{\rm P}|\mu}}

\section{Hand-waving estimation for the rate variance in the quenched-noise approximation \label{sec:handwaving}}

\subsection{Hand-waving ansatz for the temporal distribution function}

We here present a hand-waving argument, consistent with the quenched-noise approximation, that leads to an alternative analytic expression for the rate variance of the LIF model with white+colored noise input. Interestingly, the hand-waving approximation for $\vro$ and $g$ captures relatively well the behavior of the numeric estimation of these quantities, but {\it worst than the full quenched-noise calculation in Sec. \ref{sec:quenched_variance}, of which the hand-waving approximation is a further simplification} (please, see Figs. \ref{fig:g_quenched},\ref{fig:rstd_quenched}).

The approximation stands on a simple ansatz for the form of the temporal distribution function $g(t)=\<\rhoro(t')\rhoro(t'+t)\>/\rro$ of the white+colored noise LIF. We suppose it to be of the same functional form $a+b\,e^{-|t|/\tauc}$ of the {\it encoding temporal distribution function} (see Eq. (\ref{eq:currentstatistics}) in Sec. \ref{sec:diffusion}): $g_{\rm en}(t)=\<I(t')I(t'+t)\>/\<I(t')\>=(\vc/\mui) e^{-|t|/\tauc} +\mui$. The ansatz is:

\begin{align}
\label{eq:gansatz}
\gq^{(\text{hand})}(t) = \frac{\uq}{\rq} e^{-|t|/\tauc} + \rq 
,
\end{align}
where the coefficients $a=\rq$, $b=\uq/\rq$ are fixed so that the ansatz is consistent with the quenched-noise approximation. The emerging stationary variance $C_-(0) = \rq \gq^{(\text{hand})}(0)-\rq^2$ is, indeed, the stationary variance according to the quenched-noise approximation (in analogy with $\rq$ in Eq. (\ref{eq:quenchedmeandef}) in Sec. \ref{sec:quenched_mean}):

\begin{align}
\uq(\mui) :=& \<\rw(\tilde\mu)^2\>_{{\cal N}(\tilde\mu|\mu_I,\vc)} -\rq^2
.
\end{align}

A straightforward calculation in the saddle-point approximation, completely analogous to that in Sec. \ref{sec:quenched_mean}, leads to:

\begin{align} \label{eq:uq}
\uq(\mui) +\rq^2(\mui) &= \rw(\mui)^2 \left[\frac{1}{1-2\vc g'_1(\mui)}\right]^{1/2} \exp\left( \frac{2\vc g_1^2(\mui)}{1-2\vc g'_1(\mui)} \right)
.
\end{align}

We can now compute the $\text{Var}(\text{rate})$ associated to our $\gq^{(\text{hand})}$ in Eq. (\ref{eq:gansatz}), using the relation between $\text{Var}(\text{rate})$ and $C_-$ in Eq. (\ref{eq:variancecorrection}), and taking $\rq$ as the amplitude of the singular part of $C(t)=\rq \delta(t)+ C_-(t)$, with $C_-(t)=\rq\gq^{(\text{hand})}(t)-\rq^2$. The result is our hand-waving expression for the rate variance:

\begin{align}
\label{eq:ratevariance_quenched}
{\rm Var}\left[\frac{\nro(\Tc)}{\Tc}\right] &\simeq \frac{\rq}{\Tc} + \frac{2\tauc}{\Tc}\, \uq
.
\end{align} 

\subsection{The variance of step-inhomogeneous Poisson processes}
 
We will now see that Eq. (\ref{eq:ratevariance_quenched}) can be found in an alternative, easily interpretable manner. In Eq.  (\ref{eq:ratevariance_quenched}), the first term corresponds to the variance of the rate {\it of a homogeneous Poisson process whose firing rate is equal to the firing rate in the quenched-noise approximation}. The second term in Eq. (\ref{eq:ratevariance_quenched}) accounts for the inhomogeneity of the rate of the Poisson process. Overall, Eq. (\ref{eq:ratevariance_quenched}) corresponds to the variance of {\it an inhomogeneous Poisson process such that the firing rate is constant in time windows of length $2\tauc$ and, within each window, it assumes a constant, stochastic value $r_\mu:=\rw(\mu)$, where $\mu$ is independently sampled from the stationary probability distribution ${\cal N}(\mu;\mu_I,v_{\rm c})$ in the quenched approximation}. 

In other words, our hand-waving approximation to the variance of our white+colored-noise input LIF neuron for $\tauc\gg\taum$ in Eq. \ref{eq:ratevariance_quenched}, can be seen to emerge as well under the following set of assumptions: 

\begin{enumerate}
\item The rate variance ${\rm Var}[\nro(\Tc)/\Tc]$ {\it of the white-noise LIF neuron with $\sigmac=0$} and input mean $\mu$ can be well approximated with the variance of a Poisson process with the corresponding firing rate ${\rm Var}[\nro(\Tc)/\Tc]\simeq \rw(\mu)/\Tc$. This assumption (also adopted in the above derivation of Eq. (\ref{eq:ratevariance_quenched})) is approximately satisfied in the parameter region considered in our analyses. Indeed, for null or low enough $\sigmac$, the variance deviations from Poisson are of the same order of the numerical estimation uncertainty of $\vro$, or lower, see Fig. \ref{fig:rstd_quenched}. 

\item The dynamics of the input current $I$ can be approximated with a process in which $I$ is constant in time intervals of size $\Tm=2\tauc$, and independently sampled across different intervals, from its stationary distribution. We assume, in other words, that the encoding current correlation functions can be approximated as: $\text{Cov}\left(I(t'),I(t'+t)\right)=\vc\,\text{Heav}(2\tauc-t)$, where $\text{Heav}$ is the Heaviside function.

\item Within time windows of size as large as $2\tauc$, the LIF effectively behaves as a white-noise LIF with {\it constant input current} $\mu$, sampled from its stationary distribution according to the quenched-noise approximation. 
\end{enumerate}

Let us derive Eq. (\ref{eq:ratevariance_quenched}) under these assumptions. Let the time interval of length $T$ be divided into $M$ small intervals of length $\Tm=T/M$ so that $r_\mu$ is constant in such smaller intervals. The number of spikes $n(T)$ is, in the large interval of length $T$, a stochastic variable that can be expressed as $n(T)=\sum_{m=1}^M n_m(\Tm)$, where $n_m(\Tm):=n\left(m\Tm\right) - n\left((m-1)\Tm\right)$ is the number of spikes in the $m$-th small interval. Now, let the $n_m(\Tm)$'s be independent and identically distributed, and that their average and variance be finite. Then, the Central Limit Theorem wants, for sufficiently large $M$, the average rate $n(T)/T$ to exhibit mean and variance as follows:

\begin{align}
\<\frac{n(T)}{M}\> &\simeq \<n_m(\Tm)\>\\
{\rm Var}\left[\frac{n(T)}{M}\right] &\simeq \frac{1}{M}  {\rm Var}\left[ n_m(\Tm) \right] \label{eq:CLTargument}
\end{align} 
where expectation values and variances refer to the annealed probability distributions of both the realizations of the Poisson spikes given $r_\mu$, and of the different instances of $r_\mu$. This is equivalent to suppose that one first samples $M$ independent $\mu\sim{\cal N}(\cdot;\mu_I,v_{\rm c})$ and, for each of such $\mu$'s, one samples $n_m(\Tm)\sim {\rm Poisson}(\cdot;\Tm r_\mu)$. 

The average of the single $n_m$'s is:

\begin{align}
\label{eq:avcount}
\<n_m(\Tm)\> &= \<\< n_m(\Tm) \>_\pgivenmu \>_\mu = \Tm \bar r \\
\bar r &:= \< r_\mu \>_\mu
\end{align} 
where we have used that the average over the joint probabilities on $n_m$'s and $\mu$'s is $\<\cdot\> =  \<\<\cdot\>_{\pgivenmu}\>_\mu$, where $\<\cdot\>_\pgivenmu$ refers to the expectation over the conditional probability ${\rm Poisson}(\cdot;\Tm r_\mu)$, and $\<\cdot\>_\mu$ the expectation according to ${\cal N}(\cdot;\mu_I,v_{\rm c})$. In the above online equation, the identity $\< n_m(\Tm) \>_\pgivenmu  = r_\mu$ is just the equation for the average of ${\rm Poisson}(\cdot;\Tm r_\mu)$. \\

The variance of the single $n_m$'s is:

\begin{align}
{\rm Var}\left[ n_m(\Tm) \right] &=  \<\< \left( n_m(\Tm) - \Tm \bar r \right)^2  \>_\pgivenmu \>_\mu 
\end{align} 
where we have used Eq. (\ref{eq:avcount}) $\<n_m(\Tm)\>= \Tm \bar r$. Now, sum and subtract $\Tm r_\mu$ within the parentheses, to get:

\begin{align}
{\rm Var}\left[ n_m(\Tm) \right] &=  \<\< \left( n_m(\Tm) - \Tm  r_\mu \right)^2  \>_\pgivenmu \>_\mu + \Tm^2 \<\left( \bar r -r_\mu \right)^2\>_\mu
\end{align} 
where we have used again Eq. (\ref{eq:avcount}). Using the equation for the variance of the Poisson distribution $\< \left( n_m(\Tm) - \Tm  r_\mu \right)^2  \>_\pgivenmu =  \Tm  r_\mu$, we finally get:

\begin{align}
{\rm Var}\left[ n_m(\Tm) \right] &=   \Tm  \bar r + \Tm^2\, {\rm Var}_\mu\left[ r_\mu\right] 
.
\end{align} 

Now, we are interested in the variance of the rate $n(T)/T$. It is ${\rm Var}[ n(T)/T ]  = (M^2/T^2) {\rm Var}[ n(T)/M ]$ and, using the Central Limit Theorem argument in Eq. (\ref{eq:CLTargument}):

\begin{align}
\label{eq:ratevariance_quenched_var}
{\rm Var}\left[\frac{n(T)}{T}\right] &\simeq \frac{\bar r}{T} + \frac{\Tm}{T}\, {\rm Var}_\mu\left[ r_\mu \right]
\end{align} 

Now, taking $\Tm=2\tauc$ we get Eq. (\ref{eq:ratevariance_quenched}).

\section{Necessary conditions for the SNR enhancement \label{sec:necessaryconditions}}

{\bf An upper bound in $\muip$ for the SNR enhancement to occur.} We will now argue that, since the SNR numerator is $\Delta\rro\le \rro(\mui^{(+)})$, a necessary condition for the SNR to increase with the noise correlation strength is $\mui^{(+)}<\mui^*$. We will adopt the following double delta notation: $\delta$ for the increment when introducing the colored noise, $\alphaV>0$, while $\Delta$ for the $s=\pm$ change, with $\Delta\mui=\muip-\muim>0$. We focus in the regime when the white-noise LIF response curve $\rw$ is convex, for low enough $\mui<\mui^*$, since it is the only one in which one observes a $\delta({\rm SNR})>0$. As we have argued above, in this regime it is $\delta\rro>0$ (we can understand this fact thanks to the quenched-noise approximation, at least when $\tauc\gtrsim \taum$). Moreover, we also know (again thanks to the quenched-noise hypothesis) that the increment in the SNR denominator in the presence of colored noise is positive $\delta \<\vro^{(s)}\>_s$, since both terms of $\vro$ in Eq. (\ref{eq:variancecorrection}) increase with $\alphaV$. We conclude that, in the regime where one numerically observes $\delta({\rm SNR})>0$ ($\mu$ sufficiently lower than $\mu^*$), the condition $\delta({\rm SNR})>0$ implies $\delta\Delta\rro>0$. Indeed, $\delta({\rm SNR})>0$ can be written as:

\begin{align}
\frac{\Delta \rro}{\<\vro\>} > \frac{\Delta \rw}{\<\vw\>}
\end{align}
where $\rro,\rw$ refer, again, to the readout rate in the presence and absence of noise correlations, respectively, and where $\<\vro\>:=(\vro^{(+)}+\vro^{(-)})/2$. This translates into $\Delta\rro>\Delta\rw \<\vro\>/\<\vw\>>\Delta\rw$ (because, as we said before, $\delta\vro>0$). We conclude that $\delta\Delta\rro>0$. Thus, having $\delta\Delta\rro>0$ is a necessary condition for $\delta({\rm SNR})>0$. Moreover, since also $\delta\rrom>0$ (because, in general, $\delta\rro>0$), we also have that $\delta\rrop>0$ is a necessary condition (since $\delta\rrop=\delta\rrom+\delta\Delta\rro \ge \delta\Delta\rro\ge 0$). Therefore, $\delta\Delta\rro>0$ and $\delta\rrop>0$ are necessary conditions for $\delta({\rm SNR})>0$ (which are not sufficient, since the denominator could increase too much). As a matter of fact, $\delta\rrop>0$ implies $\muip<\mui^*$. First, because, in the light of the quenched-noise approximation, it must be $\mui<\mui^*$ for $\delta\rro>0$. Second, if we expand $\rq$ in Eq. (\ref{eq:rquenched_analytical}) up to first order in $\vc$, and we impose $\rq=\rw$, we obtain, again, that the solution is $\mu^*$ satisfying $g_1^2+g_1'=0$. For larger values of $\vc$, for which the $O[\vc^2]$ is non-negligible, the solution of $\rq=\rw$ occurs for lower values of $\mui<\mui^*$. We conclude that $\mui<\mui^*$ is a necessary condition for the SNR to be enhanced. Indeed, this is what we observe in all the simulations in the article: the SNR enhancement occurs for $\mui<\mui^*$. 

{\bf A lower upper bound in $\muip$ for the SNR enhancement to occur.} We can make the condition $\muip\le\mui^*$ more stringent by imposing that, beyond the numerator, the SNR itself should increase. This lower upper bound for $\muip$ is found by imposing the further condition that not only it has to be $\delta\Delta\rro>0$, but $\delta({\rm SNR})>0$. First, we assume firing rates so low that, in the white-noise case for $\alphaV=0$, we can neglect the over-Poissonian deviations of $\rro$ (mind that, for the white-noise model, this is verified in almost all cases probed in this article in which $\alphaV=0$, c.f. Fig. \ref{fig:rstd_quenched}). Neglecting $\rrom$, and taking $\vro\Tc\simeq\rw$, the white-noise SNR is (we omit the $\Tc$ proportionality factors) ${\rm SNR}_{\rm w}\simeq (2\rw)^{1/2}$. Now, take ${\rm SNR}_{\rm q}\simeq \rq^{(+)}(\vq^{(+)}/2)^{-1/2}$ (again neglecting $\rrom$), Taylor-expand our expression for $\rq^{(+)}/{(\vq^{(+)})}^{1/2}$ (taking $\vq$ from Eq. (\ref{eq:uq}) in Appendix \ref{sec:handwaving}), and stop at first order in $\vc$. Equating $\text{SNR}_{\rm q} = \text{SNR}_{\rm w}$ results in the transcendental equation $g_1'+g_1^2 = 2\tauc\rw g_1^2$, whose solution is dubbed $\hat\mui$. The numerator $\rro^{(+)}$ and the denominator ${\vro^{(+)}}^{1/2}$ both increase with $\sigmac$, so that both effects cancel at first order in $\vc$. In fact, to order $o[\vc]$, the condition $\text{SNR}_{\rm q} = \text{SNR}_{\rm w}$ translates into $g_1'+g_1^2 = 2\tauc\rw g_1^2$, and equation that {\it does not depend on $\vc$}. The same happens to its solution $\hat\mui$ (green line in Fig. \ref{fig:snr_vs_muI}). 

We conclude that $\delta(\text{SNR})>0$ implies $\mui^{(+)}<\hat\mui$ for low enough $\vc$ and $\muim$ (since we have neglected $\rrom$). This condition can therefore be taken as a necessary condition on $\mui^{(+)}$ for the SNR to increase in the presence of noise correlations, in the NERM, and for large enough $\tauc\gtrsim \taum$, and low enough $\vc$ and $\muim$. 

While we expect the upper bound $\mui^*$ to hold always, the lower upper bound $\hat\mui$ is therefore expected to hold for low enough values of $\muim$ and $\vc$. In Sec. \ref{sec:conditions} we have mentioned that not only $\mui<\mui^*$, but even the more stringent condition $\mui<\hat\mui$ is respected in all the figures in which we have evaluated the SNR. This is not very surprising, in the light of the fact that, for high enough values of $\muim$ and of $\vc$, also the condition $\delta(\text{SNR})>0$ tends to not be satisfied any more (as explained in Sec. \ref{sec:conditions}). In practice, there are no cases, among those probed here, in which $\delta(\text{SNR})>0$ and, at the same time, the conditions low $\muim$ and low $\vc$ are not satisfied. Therefore, the upper bound $\hat\mui$ tends to hold in all cases.

\section{The encoding SNR
\label{sec:encodingsnr}}

In Fig. \ref{fig:numSNRsqrt} we compared the readout SNR with the \emph{encoding} SNR, and noticed that the later may become lower than the former. This fact is consistent with the data processing inequality \cite{cover1999elements}, stating that, whenever $X\to Y\to Z$ is a Markov chain, or (overloading the symbol $P$) whenever $P(X,Y,Z)=P(X)P(Y|X)P(Z|Y)$, their mutual information satisfy $I(\hat X;\hat Z)\le I(\hat X;\hat Y)$. Let $X$ be the stimulus $s$, $Y$ be \emph{the number of encoding spikes} (from any encoding neuron) \emph{in the time interval $[0,\Tc]$, $n(\Tc)$}, and $Z$ be the number of readout spikes in $[0,\Tc]$, $\nro(\Tc)$. First, the quantity that we have chosen to quantify the information carried by $X$ and $Y$ or $Z$ is not the mutual information, but the encoding and readout signal-to-noise ratios (that, indeed, depend on the statistics of $Y|X$ and of $Z|X$, respectively). But, most importantly, the readout variable $Z = \nro(\Tc)$  does not depend on the encoding number of spikes only, but on the whole history of encoding density of spikes, ${\cal R}=(\rho(t))_{t=0}^\Tc$. In other words, if as variables $X,Y,Z$ we chose $s,n(\Tc),\nro(\Tc)$, the data processing inequality does not apply, since $P(Z)$ is not a function of $Y$ only: $p(Z|Y)$ does depend on $X$. Some information regarding the stimulus is actually encoded in the temporal fluctuations of the current $(\rho(t))_{t=0}^\Tc$, not only through the number of spikes $n(\Tc)=\int_0^\Tc\d t\,\rho(t)$.

\section{Dependence on $\tauc$ of the colored noise amplitude \label{sec:constantvc}}

We are interested in the dependence of the readout rate mean and variance, $\rro^{(s)},\vro^{(s)}$, on the encoding and decoding properties in the Valente model. In order to study the dependence on the correlation time $\tauc$, there are two possibilities (see references in \cite{schwalger2008}). The first one is to consider that {\it the stationary noise amplitude $\vc$ is constant} (where, throughout this section, by {\it constant} we mean  {\it independent of $\tauc$}), being $\sigmac=(2\tauc\vc)^{1/2}$ consequently a function of $\tauc$. The second one is taking $\sigmac^2$ constant (and, hence, $\tauc$-dependent stationary noise amplitude $\vc=\sigmac^2/(2\tauc)$). The first scenario is suitable to address the $\tauc\to \infty$ limit of our problem since, if $\vc$ is constant, the colored noise does have an influence for infinitely large $\tauc$ (indeed, $\text{Cov}\left(I(t),I(t')\right)=\vc$, see Eq. (\ref{Icorrelations1RON})), while for $\tauc\to 0$ the coloured noise has no influence (indeed, $\text{Cov}\left(I(t),I(t')\right)=0$). Oppositely, the  constant-$\sigmac$ scenario is the suitable one to consider the $\tauc\to 0$ regime of the encoding-readout model, since for $\tauc\to 0$ the colored noise term $\sigmac\xi(t)$ reduces to a delta-correlated noise ($\text{Cov}\left(I(t)I(t')\right)=(\sigmai^2+\sigmac^2)\delta(t-t')$, see Eq. (\ref{Icorrelations1RON})), while for infinite $\tauc$ it vanishes (see Table \ref{table:tauc}).

In this work, we analyze the behavior of the readout neuron SNR, both according to the analytical and numerical solutions of the stochastic differential Eq. (\ref{eq:VI_diffusion}), for several values of the parameters. For the $\tauc$-dependency, \emph{we consider constant stationary variance $\vc$}, since we are specially interested in the behavior of the SNR in the moderate and large-$\tauc$ regimes. 
The $\tauc\to 0$ limit is less interesting in what, even in the case of constant $\sigmac$, the colored-noise LIF neuron behaves for $\tauc=0$ {\it as a white-noise LIF neuron} with white noise amplitude equal to $\sigmai^2+\sigmac^2$. In this situation, the readout SNR could increase in the presence of small-$\tauc$ noise correlations, but not if one chooses, as a reference, the white-LIF readout with white-noise amplitude $=\sigmai^2+\sigmac^2$. Oppositely, and as show in this article, for constant $\vc$ and large values of $\tauc$, the behavior of the colored-noise LIF is non-trivially different from that of the white-noise LIF. 

We stress that setting a  constant value for $\vc$  does not lead to a loss of generality, as it only determines the value of $\sigmac$ that we probe when varying $\tauc$. Equation (\ref{Icorrelations1RON}) implies that, the larger the value of $\tauc$, the larger must be $\sigmac$ in order for the colored noise term to have a significant impact in $I(t)$, hence in the readout neuron statistics. Setting a constant value for $\vc$ accounts for this issue. The constant-$\sigmac$ scenario is left to future studies.  


\begin{table}[h] 
\caption{Two-time correlator of the input current, ${\sf c}({|t-t'|}) := \< (I(t) - {\mu_I})(I(t') - {\mu_I}) \>$ in the two different scenarios: constant instantaneous colored noise amplitude $\sigmac$ (and $\tauc$-dependent stationary noise amplitude $\vc$); constant $\vc$ (and $\tauc$-dependent $\sigmac$) (from equation (\ref{Icorrelations1RON})). \label{table:tauc}}
\begin{tabular}{l|l|l}
 & $\tauc = 0$ &  $\tauc = \infty$ \\
\hline
constant $\vc=\sigmac^2/(2\tauc)$ &  ${\sf c}({|t-t'|})= \sigmai^2\,\delta(t-t')$  & ${\sf c}({|t-t'|})= \sigmai^2\, \delta(t-t')+ \vc$ \\
constant $\sigmac$ &  ${\sf c}({|t-t'|})= (\sigmai^2+\sigmac^2)\,\delta(t-t')$  & ${\sf c}({|t-t'|})= \sigmai^2\,\delta(t-t')$ \\
\end{tabular}
\end{table}


\section{Underestimation of the rate variance \label{sec:varunderestimation}}

\begin{figure}
\includegraphics[width=0.7\textwidth]{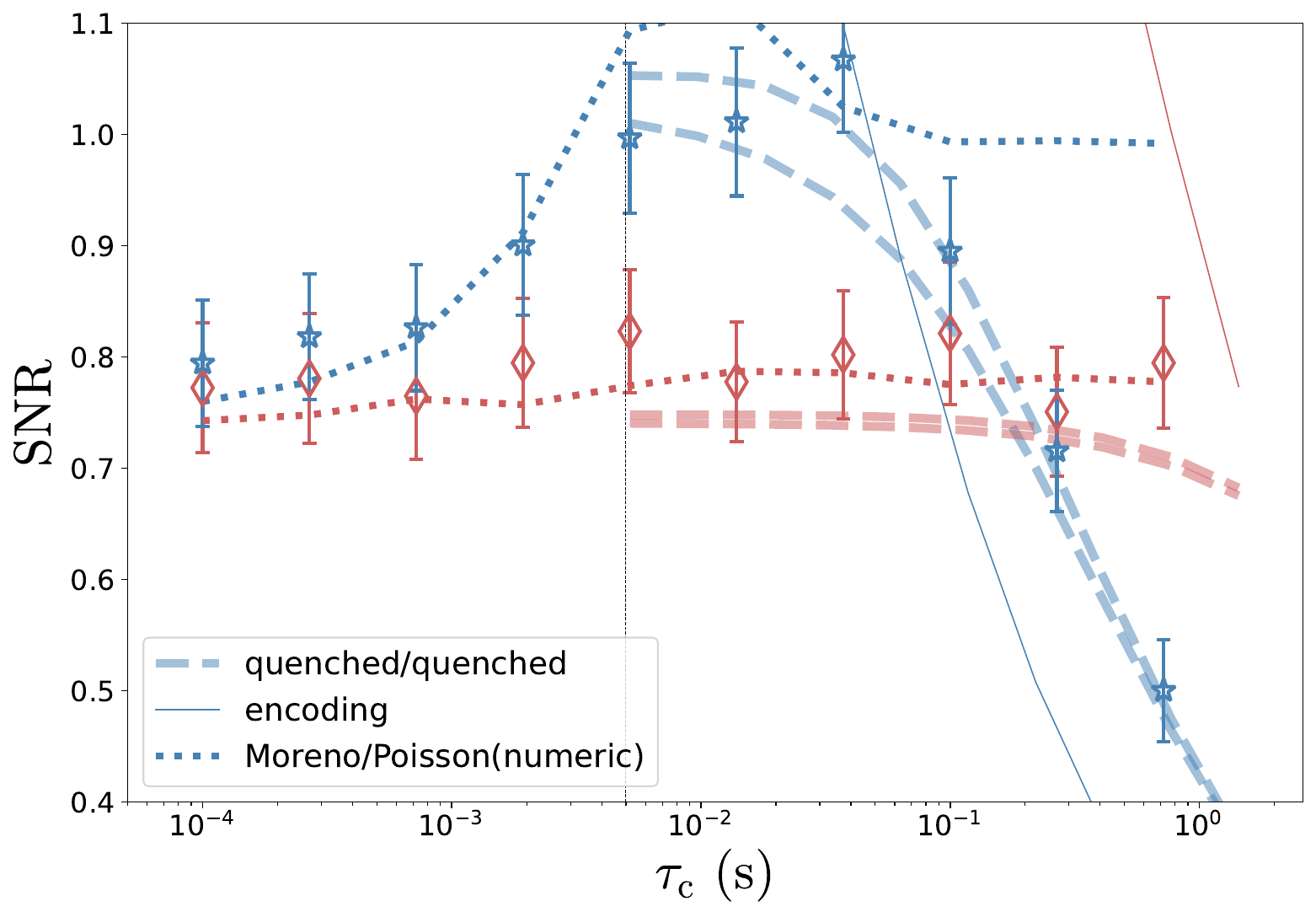}%
\caption{{\it SNR vs $\tauc$}, as in Fig. \ref{fig:numSNRsqrt}, but using the NIS-corrected estimator for the numerator, $\Delta\rro$. The thin dotted lines are the {\it encoding} signal-to-noise ratio in the unit time interval, Eq. (\ref{eq:snrencoding}). See the rest of model parameters in Table \ref{table:parameters1}. \label{fig:numSNRsqrt_semicorrected}}
\end{figure}

As already mentioned, we do not employ, in the main text, the finite-NIS correction algorithm to the variance, since it leads to large errors in the SNR. Consequently, our estimations for $\vro$ are slightly underestimated. This is, however, not a serious issue since, as we said, the overestimations in the numerator and denominator of the SNR in Fig. \ref{fig:numSNRsqrt} partially cancel. In Fig. \ref{fig:numSNRsqrt_semicorrected} we report the SNR but, differently from Fig. \ref{fig:numSNRsqrt}, we use the NIS-corrected estimation for the numerator, $\Delta\rro$. In this situation, for our approximations to work well in the $\tauc\ll\taum$ limit, one has to use the Poisson-Moreno approximation, but taking {\it the numerical estimation for $\rro$} (not the Moreno theoretical one) {\it in the Poisson hypothesis: $\vro=\tilde\rro/\Tc$}. This works, because the Poisson hypothesis $\tilde\rro\simeq\tilde\vro/\Tc$ still holds for finite $\nis$, when both $\rro$, $\vro$ are underestimated. We consequently expect that the SNR in Fig. \ref{fig:numSNRsqrt_semicorrected} is an overestimation of the SNR (since the numerator is NIS-corrected, while the denominator is underestimated), while the one in Fig. \ref{fig:numSNRsqrt} is a more accurate estimation. 


\section{SNR versus $\mui^{(-)}$\label{sec:SNRvsmuImin}}

\begin{figure}[h!]
\includegraphics[width=0.7\textwidth]{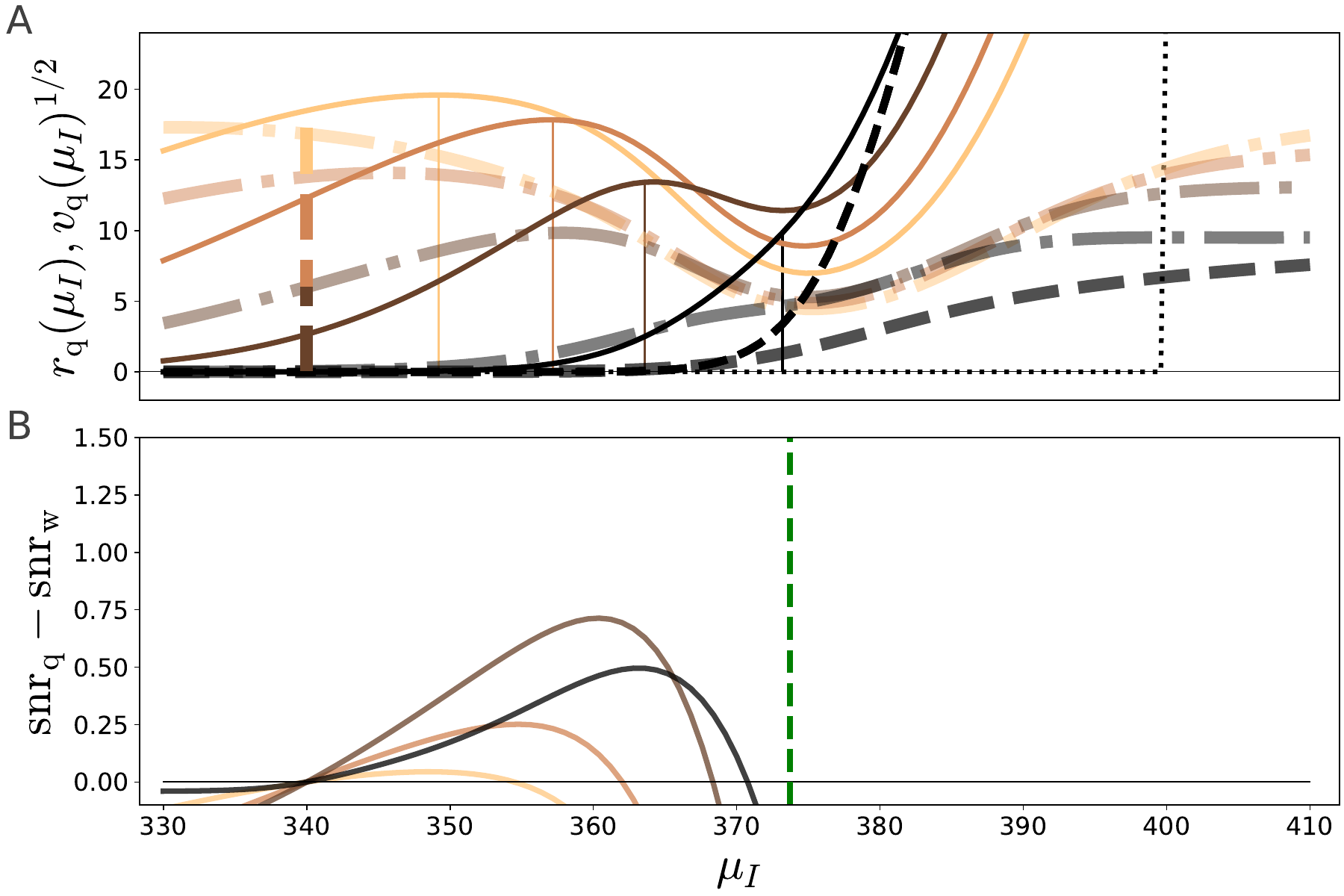}\\%
\caption{{\it Readout rate mean, variance, and excess of SNR}, as in Fig. \ref{fig:snr_vs_muI}, but for a higher $\mui^{(-)}=340$ (signaled as a dashed vertical line). The rest of the model parameters are: $\tauc = 0.1$, $\taum = 0.0025$, $w = 0.001 $. \label{fig:snr_vs_muImin}}
\end{figure}

\begin{figure}[h!]
\includegraphics[width=0.7\textwidth]{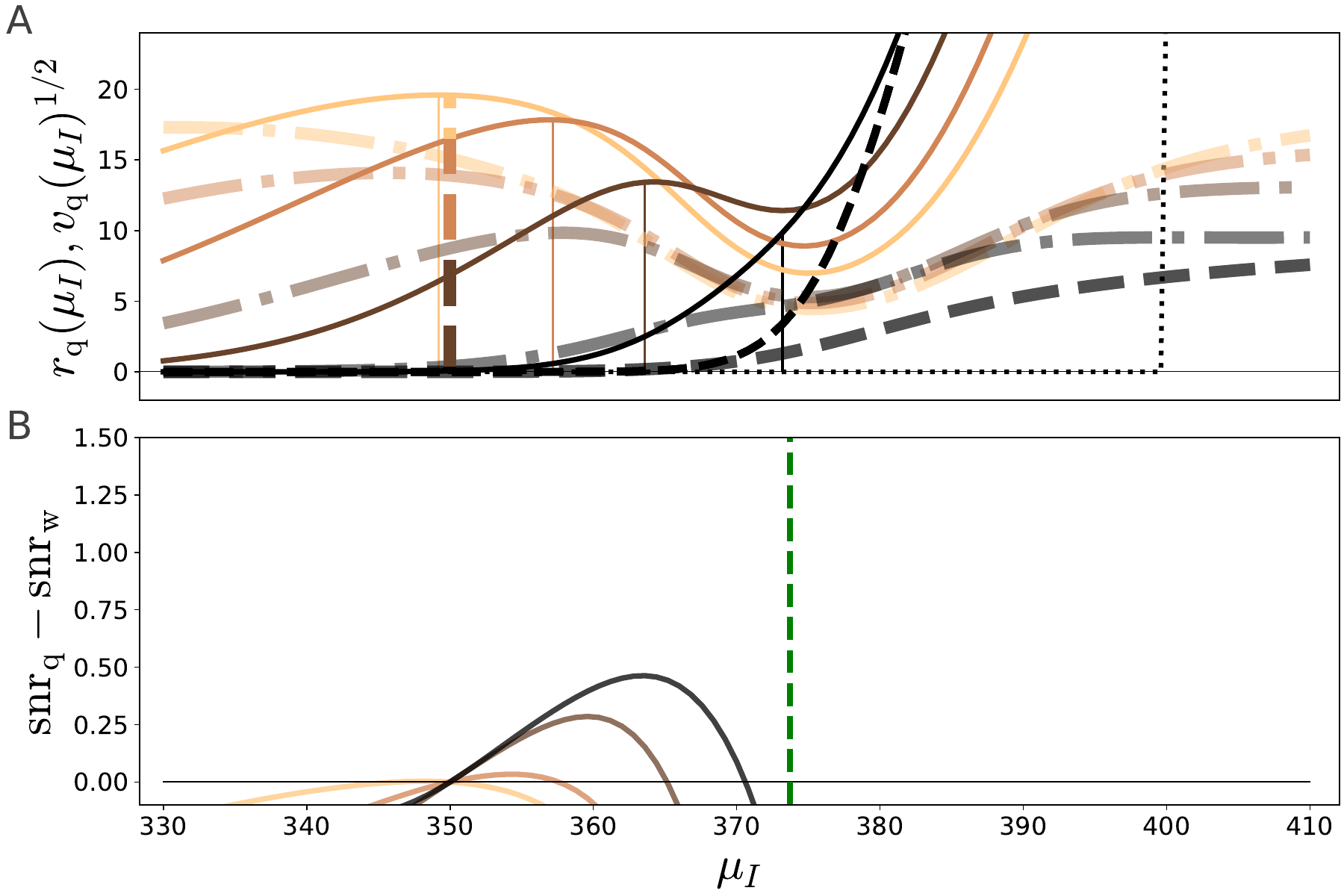}%
\caption{{\it Readout rate mean, variance, and excess of SNR}, as in Fig. \ref{fig:snr_vs_muImin}, but for an even higher $\mui^{(-)}=350$ (signaled as a dashed vertical line). \label{fig:snr_vs_muImin350}}
\end{figure}

See Figs. \ref{fig:snr_vs_muImin},\ref{fig:snr_vs_muImin350} for an analysis of how the SNR depends on $\mui^{(\pm)}$ and on $\sigmac$. The figure illustrates how the behavior in Fig. \ref{fig:snr_vs_muI} (in which $\mui^{(-)}$ is so low that $\rro^{(-)}$ is negligible) changes for non-vanishing values of $\rro^{(-)}$.

\section{Theoretical SNR dependence on other variables \label{sec:othervariables}}

For completeness, we report the behavior of the SNR per unit time in as a function of the parameters $\taum$, $w$, $\alphaV$, according to our Moreno et al and quenched-noise theories. 

\begin{figure}[h!]
\includegraphics[width=1.\textwidth]{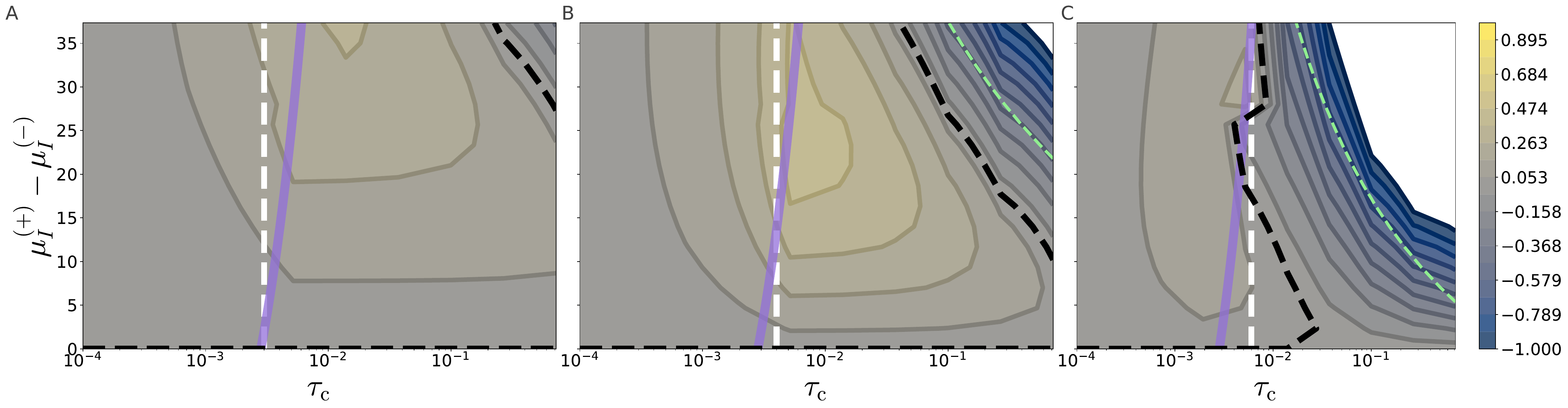}%
\caption{{\it Contour plot of the excess of SNR with respect to the absence of noise correlations}, as in Fig. \ref{fig:theoreticalSNR_vs_muI}, except that panels A,B,C correspond, respectively, to different values of $\taum=0.03,0.04,0.06$. The rest of the model parameters are: $\mui^{(-)}= 32.65$, $\sigmaV= 1.104$, $\alphaV=0.9$, $w= 0.45$. \label{fig:theoreticalSNR_vs_taum} }
\end{figure}

\begin{figure}
\includegraphics[width=1.\textwidth]{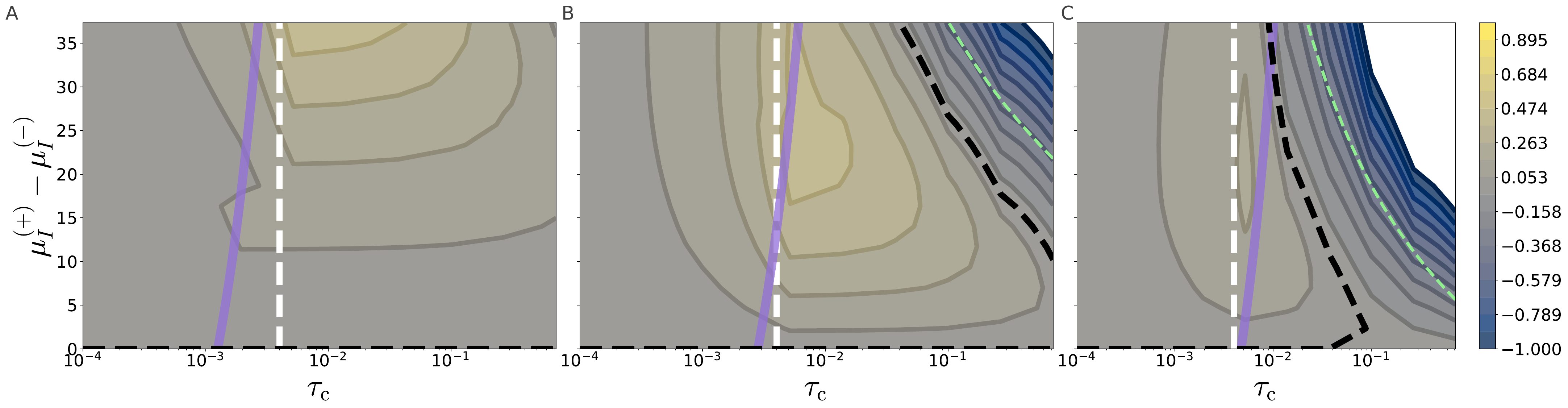}%
\caption{{\it Contour plot of the excess of SNR with respect to the absence of noise correlations}, as in Fig. \ref{fig:theoreticalSNR_vs_muI}, except that panels A,B,C correspond, respectively, to different values of $w=0.2,0.45,0.75$.  The rest of the model parameters are: $\mui^{(-)}= 32.65$, $\taum= 0.04$, $\sigmaV= 1.104$, $\alphaV=0.9$. \label{fig:theoreticalSNR_vs_wgain}}
\end{figure}

\begin{figure}
\includegraphics[width=1.\textwidth]{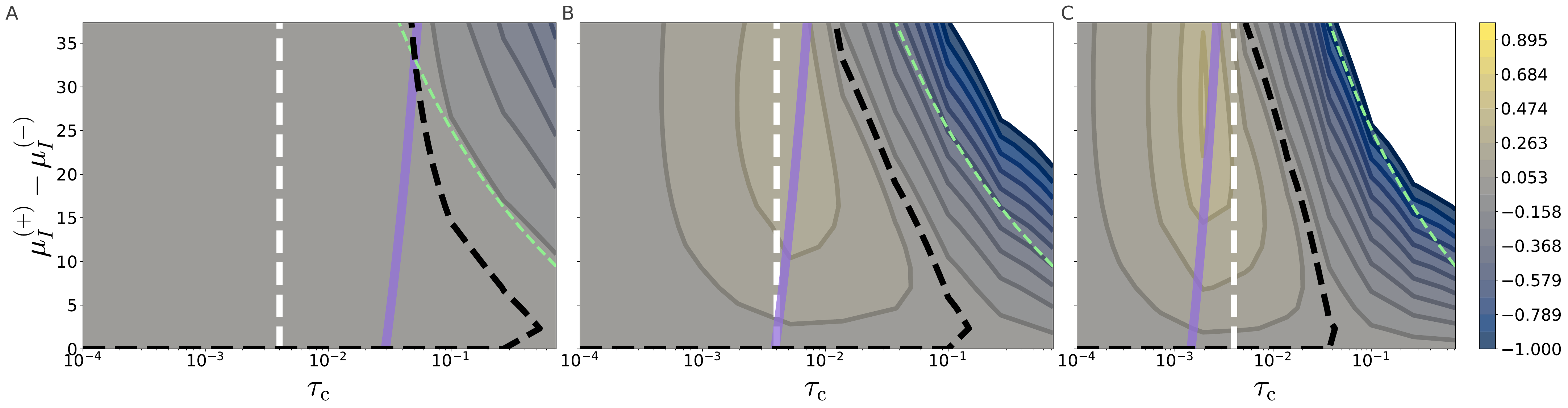}%
\caption{{\it Contour plot of the excess of SNR with respect to the absence of noise correlations}, as in Fig. \ref{fig:theoreticalSNR_vs_muI}, except that panels A,B,C correspond, respectively, to different values of $\alphaV=0.25,0.9,1.5$. The rest of the model parameters are: $\mui^{(-)}= 45$, $\taum= 0.04$, $\sigmaV= 1.104$, $\alphaV=0.9$, $w= 0.45$. \label{fig:theoreticalSNR_vs_alphaV}}
\end{figure}

In Figs. \ref{fig:theoreticalSNR_vs_taum},\ref{fig:theoreticalSNR_vs_wgain},\ref{fig:theoreticalSNR_vs_alphaV}, the SNR has been estimated according to our analytical solutions: for low $\alpha_{\rm M}$, we use the Moreno et al solution for $\Delta\rro$ (interpolated between the $\tauc\ll\taum$ and $\tauc\gg\taum$ solutions, as in \cite{moreno2002}), and a Poisson-Moreno hypothesis for the denominator $\<\vro^{(s)}\>_s$ (i.e., supposing $\vro=\rro/\Tc$, and the Moreno expression for $\rro$); for large $\alpha_{\rm M}$, we use the quenched-noise approximation for both the numerator and the denominator of the SNR. In practice, whenever $\alpha_{\rm M}$ is larger than a threshold $\hat \alpha_{\rm M}=0.75$, we use the quenched-noise approximation. The green dashed line in Figs. \ref{fig:theoreticalSNR_vs_muI},\ref{fig:theoreticalSNR_vs_taum},\ref{fig:theoreticalSNR_vs_wgain},\ref{fig:theoreticalSNR_vs_alphaV} signals the $\alpha_{\rm M}=\hat \alpha_{\rm M}$ transitions in the $(\Delta\mui,\tauc)$ plane. Although the Moreno and quenched-noise approximations are expected to hold only in the $\tauc\ll\taum$ and $\tauc\gg\taum$ limits, respectively, in practice they are relatively congruent around the $\alpha_{\rm M}=\hat \alpha_{\rm M}$ line for intermediate values of $\tauc$. 

{\bf Dependence on $\taum$.} In Fig. \ref{fig:theoreticalSNR_vs_taum} we see how the SNR contour plots change with the value of $\taum$. We observe a qualitatively identical effect: reducing $\taum$ has the effect of shifting the white-noise LIF activation function $\rw$ to the right (mind that $\rw$ depends on $\mui$ through $\mui\taum^{1/2}$ in Eq. (\ref{eq:rwhiteLIF})), which has a similar effect, qualitatively, as lowering $\mui$, and vice-versa.

{\bf Dependence on $w$.} In Fig. \ref{fig:theoreticalSNR_vs_wgain} we see the SNR as a function of $w$. The dependence on $w$ is qualitatively equal to that on $\mui^{(-)}$ in Fig. \ref{fig:theoreticalSNR_vs_taum}: low $w$ implies a shift of the activation functions to the right (see Fig. \ref{fig:r_interpretation}): for low enough $w$, reducing $w$ has the same qualitative effect than increasing $\mu^{(+)}$ (panels A,B). Increasing $w$ beyond a certain threshold, $\mui^{(+)}$ approaches $\hat\mui$, and eventually also $\mui^{(-)}$ approaches $\hat\mui^{(-)}$: hence, the SNR decreases everywhere in the $\Delta\mui,\tauc$ plane. 

{\bf Dependence on $\alphaV$.} We report the dependence with $\sigmac$ (equivalently, $\alphaV$) in \ref{fig:theoreticalSNR_vs_alphaV}, for $\alphaV=0.25,0.9,1.5$. The largest probed value of $\alphaV$ is larger than one. While this is not, in principle, possible according to the definition of encoding noise correlations, in this section it is an effective way of increasing $\sigmac(\alphaV)=w\sigmaV[N(1+(N-1)\alphaV^2)]^{1/2}$, above the value that it would have for $\alphaV=1$ (see Sec. \ref{sec:numerical}), and that could correspond to an equivalent increment in $N$, $w$ or $\sigmaV$, keeping $\alphaV\le 1$. First, for low enough $\alphaV$ (panel A), the amplitude of noise correlations is too low to imply a significant increment of SNR. For larger values of $\alphaV$, the non-monotonic dependence along both axes that we described in Sec. \ref{sec:conditions} emerge (panel B). For larger and larger $\alphaV$, {\it and for a high enough value of $\mui^{(-)}$, near enough $\hat\mui^{(-)}$}, the maximum of the SNR moves to the left, in the x-axis (panels B,C). This means that, at a fixed value of $\mui^{(\pm)}$, increasing the value of $\sigmac$ may imply a lower $\hat\mui^{(-)}$ and, consequently, a lower SNR, as we explained in Fig. \ref{fig:snr_vs_muImin}.

\bibliography{biblio.bib}

\end{document}